\documentclass[a4paper,10pt]{article}
\pdfoutput=1 

\usepackage{jcappub} 
\usepackage[T1]{fontenc} 

\usepackage{subcaption}
\usepackage{mathtools,amssymb,lipsum}

\title{\boldmath A direct detection method of galaxy intrinsic ellipticity-gravitational shear correlation in non-linear regimes using self-calibration}

\usepackage{amsmath}
\usepackage[utf8]{inputenc}
\usepackage[export]{adjustbox}
\usepackage{placeins}
\usepackage[T1]{fontenc}
\usepackage{hyperref}
\usepackage{float}
\restylefloat{table}
\usepackage{longtable}
\usepackage{graphicx}	

\graphicspath{{figures/}}

\DeclareRobustCommand{\VAN}[3]{#2}
\let\VANthebibliography\thebibliography
\def\thebibliography{\DeclareRobustCommand{\VAN}[3]{##3}\VANthebibliography}

\author[a]{Avijit Bera,\thanks{E-mail: Avijit.Bera@utdallas.edu}}
\author[a]{Leonel Medina Varela,}
\author[a]{Vinu Sooriyaarachchi,}
\author[a]{Mustapha Ishak,}
\author[b]{Carter Williams,}
\author{and the LSST Dark Energy Science Collaboration}

\affiliation[a]{Department of Physics, The University of Texas at Dallas, Dallas, TX 75080, USA}
\affiliation[b]{Department of Physics, Northeastern University, Boston, MA 02115, USA}
\emailAdd{Avijit.Bera@utdallas.edu}

\abstract{
\par\noindent

Intrinsic alignment (IA) of galaxies is a challenging source of contamination in the Cosmic shear (GG) signals. The galaxy intrinsic ellipticity-gravitational shear (IG) correlation is generally the most dominant component of such contamination for cross-correlating redshift bins. One of the most effective techniques to mitigate such contamination is the self-calibration (SC) method which extracts the IG correlation and allows for its removal from the GG signal. In a photometric survey, the SC method first extracts the galaxy number density-galaxy intrinsic ellipticity (gI) correlation from the observed galaxy-galaxy lensing correlation using the redshift dependence of lens-source pairs. The IG correlation is computed through a scaling relation using the gI correlation and other lensing observables. The applicability of the SC method has so far been focused on the linear IA scales and the linear galaxy bias. We extend the SC method beyond the linear regime by modifying its scaling relation which can account for the non-linear galaxy bias model and various IA models. In this study, we provide a framework to detect the IG correlation for the redshift bins for source galaxies for the proposed year 1 survey of the Rubin Legacy  Survey of Space and Time (LSST Y1). We tested the method for the tidal alignment and tidal torquing (TATT) model of IA and we found that the scaling relation is accurate within 10$\%$ and 20$\%$ for cross-correlating and auto-correlating redshift bins, respectively. Hence the suppression of IG contamination in observed GG correlation can be accomplished with a factor of 10 and 5, for cross-correlating and auto-correlating redshift bins, respectively. We tested the method's robustness and found that the suppression of IG contamination by a factor of 5 is still achievable for all combinations of cross-correlating bins even with the inclusion of a moderate amount of uncertainties on IA and bias parameters, respectively. We also make available, a branch of the code \texttt{FAST-PT} to provide gI correlations up to 1-loop order term used by the new SC method.     
}

\keywords{
Intrinsic alignment -- TATT -- Self-calibration -- \texttt{FAST-PT} -- LSST DESC Y1 
}

\begin{document}
\maketitle
\flushbottom

\newpage

\section{Introduction}\label{sec:intro}
\par\noindent

Cosmic shear which is an outcome of weak gravitational lensing due to the large-scale structure of the universe, serves as an important tool for exploring the dark sector of universe. It not only provides a means to probe the spread of dark matter across the cosmos but also enhances our understanding of the dark energy equation of state and the matter fluctuation amplitude parameter improving constraints by a factor of 2 to 4. This has been supported by numerous studies, and some of the recent ones are \cite{Fu_2010}, \cite{Joudaki_2009}, \cite{Schrabback_2010}.


Systematic effects in cosmic shear measurements pose significant challenges to precision cosmology. To fully harness the potential of this probe and accurately determine cosmological parameters, it is crucial to comprehend and manage them. 
One of the primary systematic effects of weak lensing measurement is the correlated IA of galaxies. This alignment contaminates the lensing signal and acts as a nuisance factor. For reviews and tutorials on IA of galaxies, see e.g. \cite{Troxel_2015}, \cite{Kirk_2015}, \cite{Joachimi_2015}.

For example, \cite{Bridle_2007} and \cite{Joachimi_2010} demonstrated that ignoring IA could bias the determination of the dark energy equation of state by as much as $50\%$. \cite{Hirata_2007} found that the matter power spectrum amplitude could be influenced by IA up to $30\%$, underscoring the need for methods to identify and eliminate the IA from the cosmic shear signal. In \cite{Krause_2015}, forecasts of the impact of IA on cosmic shear measurements for future surveys (DES, Euclid, LSST, WFIRST) using simulated likelihood analyses and realistic covariances that include higher-order moments of the density field in the computation can be found. 


In two-point shear-shear correlations, there are two types of IA contamination. The first, known as the II correlation, originates from the correlation between the intrinsic ellipticity of two galaxies. If these galaxies are spatially close, they may align due to the tidal force field of the same nearby matter structure. This correlation is generally weak and often ignored if the galaxies are spatially distant. The second type, known as the IG correlation, arises when a galaxy is aligned by a nearby matter structure which also contributes to the lensing of a different galaxy in the background. This correlation, first identified by \cite{Hirata2004IntrinsicAI}, resulted in an anti-correlation between cosmic shear and intrinsic ellipticity, as the tidal force and gravitational lensing tend to align galaxy shapes in orthogonal directions. This holds true for the non-linear alignment (NLA) model with a positive IA amplitude or when the tidal alignment (TA) contribution is stronger than the tidal torquing (TT) for the TATT model. This observation can vary based on the strength of the TA/TT mechanism for a particular catalog of galaxies. IG is generally the most dominant component of IA contamination in the weak lensing signal for cross-correlating bins but it also depends on the IA amplitude parameters and the redshift bins.

The IG correlation has been measured in various subsets of the SDSS spectroscopic and imaging samples for low redshift samples of galaxies by different groups. \cite{Mandelbaum_2006}  reported the detection of the large-scale IG correlation, and \cite{Hirata_2007} found an even stronger IG correlation for luminous red galaxies (LRGs). These papers demonstrated that IG contamination can affect the lensing measurement and can bias the cosmological parameters up to $10-30\%$ in some cases for the matter fluctuation amplitude. This was confirmed by numerical simulations, where a contamination level of $10\%$ was found by \cite{Heymans_2006}. Further measurements of the IG correlation were made in the SDSS dataset by other groups, including \cite{Okumura_2009} and \cite{Joachimi_2010} who measured strong two-point IA correlations in various SDSS and MegaZ-LRG samples.

Despite the maturity of the IA studies in the field for nearly three decades, progress on deriving estimators to measure the IG components at higher redshifts remains a challenge due to the difficulties in separating the intrinsic ellipticity and cosmic shear contributions from the observed ellipticities of the galaxies and technical challenges to get good quality redshift information for typical lensing sources, (\cite{Troxel_2015}, \cite{Kirk_2015}, \cite{Joachimi_2015}, \cite{Lamman_2024}). Current efforts have focused on mitigation techniques using modelling and marginalization over IA parameters using photometric surveys (\cite{Samuroff_2019}, \cite{Samuroff_2023}) or estimators efficient at small redshifts using spectroscopic surveys (\cite{Singh_2016}, \cite{singh2023increasing}). Extraction and direct measurements of the IG signal at high redshifts remain an area of needed development. 

The SC method, a key technique for extracting the IG correlation, has been effectively used to measure such signals for two-point correlations\cite{Zhang_2010b}. Specifically, besides the gravitational shear-gravitational shear (GG) correlation (cosmic shear), one should also extract the gI correlations from the galaxy-galaxy lensing signal from the same lensing survey. The IG correlation is then computed and removed from the lensing signal. In essence, SC works by utilizing correlations from the same lensing survey to isolate the gI correlation. It then derives the IG signal using a scaling relation, which we elaborate on further in this paper. This technique is called SC because it employs correlations and other necessary information that can be extracted from the same gravitational lensing survey. 

One of the usefulness of the SC method is in extracting the galaxy intrinsic ellipticity-cosmic microwave background (CMB) lensing ($\mathrm{I} \phi$) cross-correlation from the gravitational shear-CMB lensing cross-correlation \cite{Troxel_2014}. \cite{Yao_2023} investigated the impact of IA in the Kilo-Degree Survey (KiDS) gravitational shear-Planck CMB lensing convergence cross-correlation and used the SC method to extract the IA contamination.  

In the last decade, a series of studies
brought the SC from a theoretical method to a practical method that is applied to forecasted measurements or real data from current lensing surveys, (see e.g. \cite{Troxel_2012a}, \cite{Troxel_2012b}, \cite{Troxel_2012c}, \cite{Troxel_2015}, \cite{Yao_2017}, \cite{Yao_2018}, \cite{Yao_2019}, \cite{Pedersen_2020}, \cite{Yao_2020}). Notably, \cite{Pedersen_2020} and \cite{Yao_2020} made the first detection of the IG signal in the KIDS survey. Later on, \cite{Yao_2020b} made a detection in the LEGACY survey.   

The choice of scales in astrophysical studies is crucial. To address this, \cite{SRD_V1_LSST} adopted a maximum wavenumber, denoted as $k_{\mathrm{max}}$, approximately equal to $0.3 h$/MPc. This specific scale corresponds to the point where nonlinear bias leads to deviations of around $10\%$ from the linear bias for galaxies at redshift $ z \sim 0.5$. This value should serve as an effective upper limit for wavenumbers. Although the statistical precision of the LSST will exceed $10\%$, we anticipate that any differences will be accounted for by fitting beyond-linear bias parameters using slightly larger scales than $k_{\mathrm{max}}$. To facilitate combined analyses across different probes, a common set of logarithmically spaced angular bins ($\ell$ bins) have been defined for large-scale structure analyses, and weak gravitational lensing (WLSS). These bins cover the range $20 \leq \ell \leq 15000$, with the value of 15000 chosen to accommodate galaxy cluster lensing profiles in the 1-halo regime. Importantly, these bin limits represent the edges of the bins, not their centers. For the WLSS analysis, $\ell_{\mathrm{max}}$ is defined for each redshift bin based on its corresponding $k_{\mathrm{max}}$ and redshift distribution, following the relation $k_{\mathrm{max}} =  k_{\mathrm{max}} \chi(\bar{z}) - 0.5$. Any $\ell$ bins above this $\ell_{\mathrm{max}}$ value in our standardized binning scheme are discarded.

The existing SC approach has primarily concentrated on linear scales for IA correlations. However, with the advent of upcoming stage IV high-precision surveys such as the LSST and the Roman Space Telescope, it is essential to extract and mitigate the IA signal on nonlinear scales. This will enable these surveys to fully realize their potential in measuring a pure cosmic shear signal, which could subsequently allow us to place tighter constraints on cosmological parameters.

In this paper, we introduce a SC technique capable of mitigating the IG correlation in nonlinear regimes. We have considered $k_{\mathrm{max}}=1.0 h$/MPc for all the redshift bins where the $k_{\mathrm{max}} \sim 0.3 h$/MPc has been proposed for LSST Y1 survey. The method that we propose is applicable beyond the proposed range of wavenumbers where the nonlinear effects of galaxy bias and IA should be taken into account for the LSST Y1 survey. This method will tackle IA contamination at levels typically depicted in literature by the TATT model for the intrinsic ellipticity of galaxies, and a galaxy bias expanded up to the second order in both the matter density field and the tidal field.

In this study, we aim to mitigate the IG correlation for the proposed redshift bins for source galaxies which will be observed by the LSST Y1 survey. The proposed SC method is accurate approximately within $10\%$ for cross-correlating bin pairs and within $20\%$ for auto-correlating bin pairs, allowing for a significant reduction of IG contamination. The method is more precise for larger bin separations and bin pairs at higher redshifts. Residual errors after SC are negligible compared to the minimum error in measurement of the GG correlation, ensuring minimal loss of cosmological information. Even with a moderate amount of uncertainties in the IA parameters and galaxy bias parameters, the scaling relation remains accurate within $20\%$ for all combinations of redshift bins, indicating that these uncertainties have minimal impact on the SC technique and the accuracy of the scaling relation.

The organization of the paper is described as follows. In section \ref{sec:2}, we first describe the observables relevant to the SC method in the weak lensing surveys. We briefly review the TATT model for the IA of galaxies and the nonlinear expansion of galaxy bias using the standard perturbation theory approach in subsections \ref{sec:2.1} and \ref{sec:2.2}, respectively. In subsection \ref{sec:2.3}, we provide the final expression for the matter density-galaxy intrinsic ellipticity (mI) and galaxy number density-galaxy intrinsic ellipticity (gI) power spectra. The choices of IA model parameters, galaxy bias parameter, fiducial cosmology and the photo-$z$ model are described in subsection \ref{sec:2.4}. In section \ref{sec:3} we develop the formalism of a SC method which has the potential to recover the pure lensing signal after extracting the IG contamination in the nonlinear regime. We describe the method of separating the gI correlation and gG correlation from lensing observables in subsection \ref{sec:3.1} and we then describe the detailed derivation of the scaling relation in subsection \ref{sec:3.2}. In section \ref{sec:4}, we investigate the performance of the SC by calculating the scaling relation's accuracy and determining the errors associated with the entire method. The contribution of catastrophic photo-$z$ errors and the magnification bias on the overall error budget have been discussed in sections \ref{sec:5.1}, \ref{sec:5.2}, respectively. We study the effects of the uncertainties in the cosmological and IA parameters in section \ref{sec:6}. We provide a summary of the errors associated with the SC method in section \ref{sec:7}. We present the conclusions and discussions in section \ref{sec:8}. The detailed calculation and the \texttt{FASTPT} implementation of the gI power spectrum are shown in Appendix \ref{AppA} and Appendix \ref{AppB}, respectively.

\section{Background}\label{sec:2}
\par\noindent

In this section, we provide a brief overview of the IA of galaxies and how it contaminates the weak gravitational signal at the power spectrum level.

\subsection{Observables in weak-lensing surveys}\label{sec:2.1}
\par\noindent

The SC method, first introduced by \cite{Zhang_2010b}, is designed to calculate and remove the primary contamination from the weak gravitational lensing signal, which is caused by the IA of galaxies. The key information required for this technique from weak gravitational lensing surveys \cite{Bernstein_2009} includes each galaxy’s shape, angular position, and photometric redshift. More specifically, the SC technique typically requires two basic observables from the weak lensing surveys: (1) the galaxy number density ($\delta_{\mathrm{g}}$) of a photo-$z$ bin, and (2) the galaxy shape, which can be expressed in terms of the galaxy’s shear - a direct measure of the cosmic shear induced by weak gravitational lensing. The observed shear of each galaxy is significantly contaminated due to the intrinsic ellipticity of galaxies, which is caused by the gravitational tidal force associated with the local matter density and the universe’s large-scale structure. The shot noise, created by the random orientation of the intrinsic ellipticity, can be straightforwardly corrected and is not relevant to the SC technique. The observed shear of a galaxy $\gamma^{\mathrm{obs}}$ can be expressed as a combined effect of the gravitational shear $\gamma^{\mathrm{G}}$ and the galaxy's intrinsic ellipticity $\gamma^{\mathrm{I}}$ as $\gamma^{\mathrm{obs}} = \gamma^{\mathrm{G}} + \gamma^{\mathrm{I}}$. $\gamma^{\mathrm{I}}$ represents the E-mode contribution of the correlated part of the galaxy's intrinsic ellipticity due to the IA of the galaxy. Though the intrinsic ellipticity has a non-vanishing B-mode contribution through higher order corrections,  their amplitude is generally much too small to be observed by current surveys \cite{Krause_2010}. In this work, we can safely ignore the B-mode as we are only interested in the E-mode. It is convenient to work with the lensing convergence $\kappa$ instead of $\gamma$ as we are concerned only with the weak limit. $\kappa$ is the projection of the matter density along the line of sight. From $\gamma^{\mathrm{obs}}$ it is straightforward to obtain $\kappa^{\mathrm{obs}} = \kappa^{\mathrm{G}} + \kappa^{\mathrm{I}}$. For a flat universe and under the Born approximation, $\kappa$ of a galaxy (source) at redshift $z_{\mathrm{G}}$ and direction $\hat{\theta}$ is 
\begin{eqnarray}\label{eq:1}
    \kappa(\hat{\theta}) = \int_{0}^{\chi_{\mathrm{G}}} \delta_{\mathrm{m}}(\chi_{\mathrm{L}}, \hat{\theta}) W_{\mathrm{L}}(z_{\mathrm{L}}, z_{\mathrm{G}}) d\chi_{\mathrm{L}}, 
\end{eqnarray}
where $\hat{\theta}$ is the direction on the sky and $W_{\mathrm{L}}(z_{\mathrm{L}}, z_{\mathrm{G}})$ is the lensing kernel.  $\chi_{\mathrm{L}} \equiv \chi(z_{\mathrm{L}})$ and $\chi_{\mathrm{G}} \equiv \chi(z_{\mathrm{G}})$ are the comoving angular diameter distances\footnotemark{} to the lens and the source, respectively. $\delta_{\mathrm{m}}(\chi_{\mathrm{L}}, \hat{\theta})$ is the matter density at direction $\hat{\theta}$ and at $\chi_{\mathrm{L}}$. The units of the comoving angular diameter distance $\chi$ is $c/H_0$, where $H_0$ is the present-day Hubble constant. The lensing kernel is 
\begin{eqnarray}\label{eq:2}
    W_{\mathrm{L}}(z_{\mathrm{L}}, z_{\mathrm{G}}) = \begin{cases} \frac{3}{2} \frac{H_0^2}{c^2}\Omega_{\mathrm{m}} (1+z_{\mathrm{L}}) \chi_{\mathrm{L}} \Big( 1-\frac{\chi_{\mathrm{L}}}{\chi_{\mathrm{G}}}\Big) & \text{if $z_{\mathrm{L}} < z_{\mathrm{G}}$}\\ 0 & \text{otherwise} \end{cases}
\end{eqnarray}
where $\Omega_{\mathrm{m}}$\footnotemark[\value{footnote}] is the present-day matter density in units of the critical density. Following the Limber equation, the two-dimensional angular lensing power spectrum between two cross-correlating bins ($i\ \&\ j$) is 
\begin{eqnarray}\label{eq:3}
     C_{ij}^{\mathrm{GG}}(\ell) = \int_0^{\infty} \frac{W_i(z_{\mathrm{L}}) W_j(z_{\mathrm{L}})}{\chi^2} {P}_{\mathrm{m}}\Big(k=\frac{\ell}{\chi}, z \Big)  d\chi,
\end{eqnarray}
where ${P}_{\mathrm{m}}$\footnotemark[\value{footnote}] is the non-linear matter power spectrum and $W_i$ is the weighted lensing efficiency for the $i^{\mathrm{th}}$ redshift bin defined as the lensing kernel averaged over the $i^{\mathrm{th}}$ redshift bin 
\begin{eqnarray}\label{eq:4}
    W_i(\chi_{\mathrm{L}}) = \int_0^\infty W_{\mathrm{L}}(z_{\mathrm{L}}, z_{\mathrm{G}}) \bar{n}_i(\chi_{\mathrm{G}}) d\chi_{\mathrm{G}},
\end{eqnarray}
where $\bar{n}_i(z)$ is the normalized true redshift distribution of galaxies in the $i^{\mathrm{th}}$ redshift bin and follows $\bar{n}_i(z) dz = \bar{n}_i(\chi) d\chi$. $W_i$ for the 5 source-redshift bins for the proposed LSST Y1 survey have been shown in Figure \ref{fig:3}. 

\footnotetext{Calculated using Core Cosmology Library (\texttt{CCL}) \cite{Chisari_2019}.}

For two-point statistics, photometric surveys provide three observable correlations which come from the measurements of their positions, their shapes, and a combined use of both. The first observable is the angular power spectrum of the observed shapes of the galaxies in the $i^{\mathrm{th}}$ and $j^{\mathrm{th}}$ redshift bins ($C_{ij}^{\gamma\gamma}(\ell)$) which is 
\begin{eqnarray}\label{eq:5}
    C_{ij}^{\gamma\gamma}(\ell) = C_{ij}^{\mathrm{GG}}(\ell) + C_{ij}^{\mathrm{IG}}(\ell) + C_{ij}^{\mathrm{GI}}(\ell) + C_{ij}^{\mathrm{II}}(\ell),
\end{eqnarray}
where $C_{ij}^{\alpha\beta}$ represents the two-dimensional angular cross-correlation power spectrum between the quantities $\alpha$ in the $i^{\mathrm{th}}$ redshift bin and $\beta$ in the $j^{\mathrm{th}}$ redshift bin. I and G represent the intrinsic ellipticity and the gravitational shear of the galaxy, respectively. In this paper, $i<j$ represents $\Bar{z}_i<\Bar{z}_j$. For $i<j$, $C_{ij}^{\mathrm{IG}}>>C_{ij}^{\mathrm{GI}}$ holds if the catastrophic error is reasonably small. For cross-correlating bins, $C_{ij}^{\mathrm{GI}}$ is negligible and can be ignored safely. The II correlation only exists at a small line-of-sight separation. Though $C_{ij}^{\mathrm{II}}$ is negligible for non-adjacent redshift bins, it is non-negligible for adjacent redshift bins. We tested the fractional strength (II/IG) of these contaminations using the angular power spectrum for the proposed redshift bins of source galaxies for LSST Y1 survey, and we observed that $C_{ij}^{\mathrm{II}}/C_{ij}^{\mathrm{IG}}$ is of the order of $1$, $0.1$ and $<0.01$ for the same,  adjacent and non-adjacent bin pairs, respectively. A similar study on the fractional contributions of the IG and II correlation functions in cosmic shear correlation function i.e., (IG/GG and II/GG) can be found in Figure 9 of \cite{samuroff2024joint}. For the sake of simplicity, we consider that $C_{ij}^{\mathrm{II}}$ is only comparable to $C_{ij}^{\mathrm{IG}}$ for the same bins. So, for $i<j$, $C_{ij}^{\gamma\gamma}(\ell)$ can be approximated as 
\begin{eqnarray}\label{eq:6}
    C_{ij}^{\gamma\gamma}(\ell) \approx C_{ij}^{\mathrm{GG}}(\ell) + C_{ij}^{\mathrm{IG}}(\ell). 
\end{eqnarray}

The second observable which arises from the correlation between the ellipticities of the galaxies in the $i^{\mathrm{th}}$ redshift bin and the galaxy number density in the $j^{\mathrm{th}}$ redshift bin, is known as the galaxy-galaxy lensing and is denoted by $C_{ij}^{\mathrm{g} \gamma}(\ell)$  
\begin{eqnarray}\label{eq:7}
    C_{ij}^{\mathrm{g}\gamma }(\ell) = C_{ij}^{\mathrm{gG}}(\ell) + C_{ij}^{\mathrm{gI}}(\ell).
\end{eqnarray}
  
The third observable which arises from the correlation of the number densities of the galaxies in the $i^{\mathrm{th}}$ and $j^{\mathrm{th}}$ redshift bins, is known as galaxy clustering and is denoted by  $C_{ij}^{\mathrm{gg}}(\ell)$.

\subsection{Nonlinear alignment of galaxies}\label{sec:2.2}
\par\noindent

To calculate the IA power spectra (e.g. $P_{\mathrm{gI}}$, $P_{\mathrm{mI}}$) in a non-linear regime, we consider a non-linear model for galaxy bias and an astrophysically motivated non-linear model for IA of galaxies. In this subsection, we briefly discuss a perturbative expansion of galaxy bias using the standard perturbation theory (SPT) approach. Then we provide the basics of the TATT model for the intrinsic ellipticities of galaxies \cite{Blazek_2019}. For details, we refer to the reader to look at Appendix \ref{AppA} and Appendix \ref{AppB}.

\subsubsection{Galaxy bias and Standard Perturbation Theory}\label{sec:2.2.1}
\par\noindent

Galaxy density is usually modelled and expressed in terms of the local matter density field and the tidal field associated with the large-scale structure. Ignoring higher-order terms and functions of velocity divergence $\theta$, which appear at the higher order; following the notation of \cite{Baldauf_2012}, we consider the perturbative expansion of the galaxy bias, which includes the second-order contribution from matter density field and tidal fields. The expression for such a model of galaxy bias is   
\begin{eqnarray}\label{eq:8}
    \delta_{\mathrm{g}}(\mathbf{x})= b_1\delta_{\mathrm{m}}(\mathbf{x}) + \frac{b_2}{2}\big( \delta_{\mathrm{m}}(\mathbf{x}) \big)^2 + \frac{b_{\mathrm{s}}}{2}\big( s(\mathbf{x}) \big)^2 + ... ,
\end{eqnarray}
where $\delta_{\mathrm{m}}(\mathbf{x})$ and $s(\mathbf{x})$ are the nonlinear matter density field and tidal field, respectively, in configuration space. We have not considered $b_{3\rm NL}$ in our study. The matter density field can be decomposed in terms of linear matter density and relevant gravity kernels in the SPT \cite{Bernardeau_2002} is  
\begin{eqnarray}\label{eq:9}
    \delta_{\mathrm{m}} = \delta_{\mathrm{m}}^{(1)} + \delta_{\mathrm{m}}^{(2)} + \delta_{\mathrm{m}}^{(3)} + ...,
\end{eqnarray}
where $\delta_{\mathrm{m}}^{(1)}$ is the linear matter density field; $\delta_{\mathrm{m}}^{(2)}$ and $\delta_{\mathrm{m}}^{(3)}$ are the second and third-order contributions. In configuration space, a tidal tensor $s_{ij}(\mathbf{x})$ which is defined in terms of $\delta_{\mathrm{m}}(\mathbf{x})$ is
\begin{eqnarray}\label{eq:10}
    s_{ij}(\mathbf{x}) = \Big[\nabla_i \nabla_j \nabla^{-2} - \frac{1}{3}\delta_{ij}\Big]\delta_{\mathrm{m}}(\mathbf{x}).
\end{eqnarray}

For more details, we refer the readers to follow Appendix \ref{AppA} and \cite{Desjacques_2018}. 

\subsubsection{Tidal alignment and tidal torquing }\label{sec:2.2.2}
\par\noindent

According to the tidal alignment (TA) or linear alignment (\cite{Catelan_2001}, \cite{PhysRevD.70.063526}) mechanism, the axes of a triaxial galaxy are preferentially aligned with the axes of the tidal field. The second-order contribution in the alignment model comes from tidal torquing (TT) (\cite{Mackey_2002}, \cite{Hui_2002}, \cite{Schafer_2012}). In the TT mechanism, the first and second-order tidal fields are accounted for through the formation of the angular momentum axis and the resulting torque on this axis, respectively. In general, the TA model is applicable for pressure-supported large elliptical galaxies, while the TT model is more suitable for the galaxies that are dominated by angular momentum i.e., for the spin-supported disc galaxies. Where the TA model can correctly predict the large-scale behaviour of IA in the linear regime \cite{Hirata_2007}, the TT model incorporates the small-scale feature of IA in the nonlinear regime. The most complete model that encapsulates both the TA and TT mechanisms is the TATT \cite{Blazek_2019}. The TATT model can successfully explain both the linear and non-linear features of the intrinsic ellipticities of galaxies which can be expanded up to the second order in terms of the linear matter density field is 
\begin{eqnarray}\label{eq:11}
    \gamma^{\mathrm{I}}_{ij}(\mathbf{x}) = C_1s_{ij} + C_2\bigg(s_{ik}s_{kj} - \frac{1}{3}\delta_{ij} s^2 \bigg) + C_{1\delta}\delta_{\mathrm{m}} s_{ij} + 
    C_{2\delta} \delta_{\mathrm{m}}\bigg(s_{ik}s_{kj} - \frac{1}{3}\delta_{ij} s^2 \bigg)  + ...,
\end{eqnarray}
where all the fields are evaluated at \textbf{x} and the summation over repeated indices are considered. $C_i$ parameters, which are analogous to galaxy bias parameters, capture each term's effective strength and include the contributions from small-scale physics. The intrinsic ellipticity of galaxies can not be uniquely defined but depends on the measurement techniques that influence the values of $C_i$. 
So, Equation \ref{eq:11} can be considered as the ``intrinsic ellipticity'' of the galaxies . We will use this model in the Eulerian perturbation theory to evaluate all the quantities at the observed position of the galaxy. In Fourier space, $\gamma$ can be decomposed into curl-free (E) and divergence-free (B) components and those are defined as the standard practice in weak gravitational lensing and CMB measurements (\cite{Mackey_2002}, \cite{Kamionkowski_1998}). E and B mode decomposition of $\gamma$ are 
\begin{eqnarray}\label{eq:12}
    \gamma_{\mathrm{(E,B)}}(\mathbf{k}) &=& C_1 f_{\mathrm{(E,B)}}(\hat{k})\delta_{\mathrm{m}}(k)
    + C_{1\delta} \int \frac{d^3\mathbf{k_1}}{(2\pi)^3} f_{\mathrm{(E,B)}}(\hat{k}_1) \delta_{\mathrm{m}}(\mathbf{k}_1) \delta_{\mathrm{m}}(\mathbf{k}_2) \\\nonumber
    && +\ C_2 \int \frac{d^3\mathbf{k_1}}{(2\pi)^3} h_{\mathrm{(E,B)}}(\hat{k}_1,\hat{k}_2) \delta_{\mathrm{m}}(\mathbf{k}_1) \delta_{\mathrm{m}}(\mathbf{k}_2)\\\nonumber
    && +\ C_{2\delta} \int \frac{d^3\mathbf{k_1}}{(2\pi)^3} h_{\mathrm{(E,B)}}(\hat{k}_1,\hat{k}_2) \delta_{\mathrm{m}}(\mathbf{k}_1) \delta_{\mathrm{m}}(\mathbf{k}_2)\delta_{\mathrm{m}}(\mathbf{k}_3).
\end{eqnarray}
where $\mathbf{k}_1+\mathbf{k}_2+\mathbf{k}_3 = \mathbf{k}$, with $\mathbf{k}_3 = 0$ except in the last term. The angular operators $f_{\mathrm{(E,B)}}(\hat{k})$ and $h_{\mathrm{(E,B)}}(\hat{k}_1,\hat{k}_2)$ are defined in Appendix \ref{AppA} and \cite{Blazek_2019}. 

\begin{figure}[hbt!]
\begin{subfigure}{.48\textwidth}
    \centering
    \includegraphics[width=.8\linewidth, height=5cm]{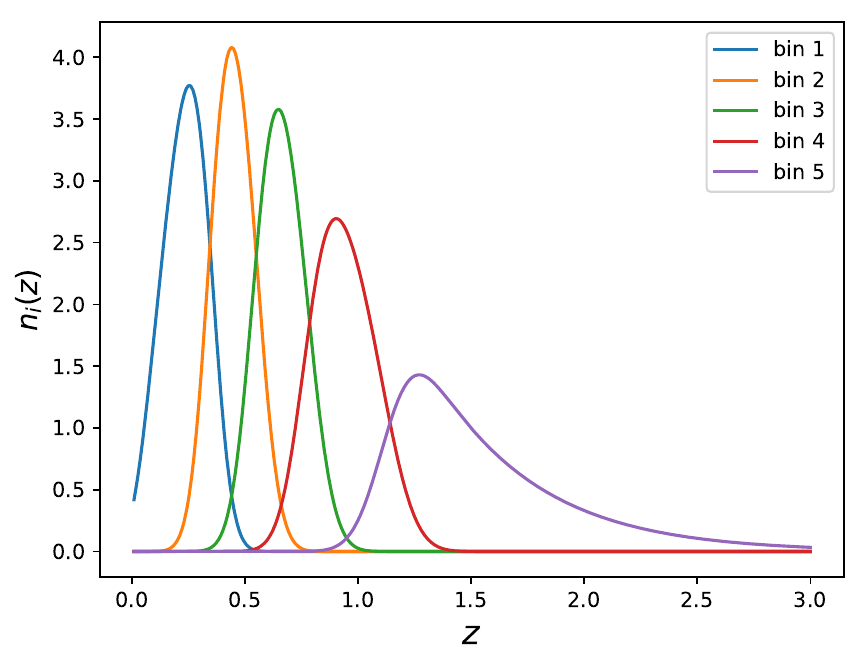}
    \caption{Photometric bins}
    \label{fig:1a}
\end{subfigure}
\begin{subfigure}{.48\textwidth}
    \centering
    \includegraphics[width=0.9\linewidth, height=5cm]{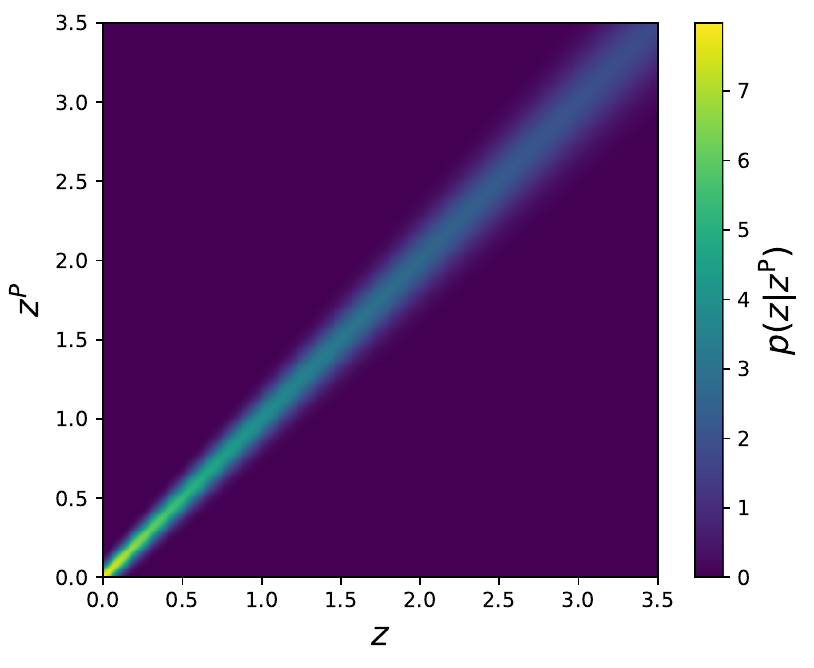} 
    \caption{$P(z|z^P)$}
    \label{fig:1b}
    \end{subfigure}
    \caption{(a). Redshift distribution of the $5$ tomographic bins for source galaxies for the proposed LSST Y1 survey. In this plot, $n_i(z)$s are normalized. (b).The photo-$z$ PDF for the redshift distribution of the source galaxies for the proposed LSST Y1 survey is described by a Gaussian probability distribution function.}
    \label{fig_1}
\end{figure}


\subsection{Intrinsic alignment power spectra}\label{sec:2.3}
\par\noindent

We provide the expression of the IA correlations up to one loop order for the TATT model. We have only considered the E-mode contribution of intrinsic ellipticity to calculate the IA power spectra as the B-mode contribution is suppressed in the leading order \cite{Krause_2010}. Considering the B-mode contribution for II correlations is mandatory as it is not suppressed in the smallest scale. We follow the method developed by \cite{Blazek_2019} and the convention provided in Appendix \ref{AppA}.

The matter density-galaxy intrinsic ellipticity power spectrum $P_{\mathrm{mI}}$ is
\begin{eqnarray}\label{eq:13}
    P_{\mathrm{mI}}(k,\mu_k) &=&  C_1 p(\hat{k})P_\delta(k) + C_{1\delta} P_{\mathrm{0|0E}}(k,\mu_k) + C_2 P_{\mathrm{0|E2}}(k,\mu_k).
\end{eqnarray}

The galaxy density-galaxy intrinsic ellipticity power spectrum $P_{\mathrm{gI}}$ is
\begin{eqnarray}\label{eq:14}
   P_{\mathrm{gI}}(k,\mu_k) &=&  b_1 C_1 p(\hat{k}) P_\delta(k) + 
     b_1 C_{1\delta} P_{\mathrm{0|0E}}(k,\mu_k) + 
     \frac{b_2}{2}C_1 p(\hat{k}) P_{\mathrm{00|E}}(k,\mu_k) \\\nonumber  && +\ \frac{b_2}{2} C_{1\delta} P_{\mathrm{00|0E}}(k,\mu_k) +
     \frac{b_{\mathrm{s}}}{2}C_{1\delta} P_{\mathrm{SS|E}}(k,\mu_k) + 
     \frac{b_{\mathrm{s}}}{2}C_{1\delta} P_{\mathrm{SS|0E}}(k,\mu_k) \\\nonumber
     && +\ b_1 C_2 P_{\mathrm{0|E2}}(k,\mu_k) +  \frac{b_2}{2}C_2  P_{\mathrm{00|E2}}(k,\mu_k) + \frac{b_{\mathrm{s}}}{2}C_{1\delta}  P_{\mathrm{SS|E2}}(k,\mu_k).
\end{eqnarray}

 The detailed calculation and implementation of these power spectra in \texttt{FAST-PT}, can be found in Appendix \ref{AppA}, and Appendix \ref{AppB}, respectively.

\subsection{Models and parameters}\label{sec:2.4}
\par\noindent


This subsection introduces the photometric model, fiducial cosmology and IA parameters, and galaxy bias parameters chosen for this study.  

\begin{figure}[hbt!]
\begin{subfigure}{.48\textwidth}
    \centering
    \includegraphics[width=.9\linewidth, height=5cm]{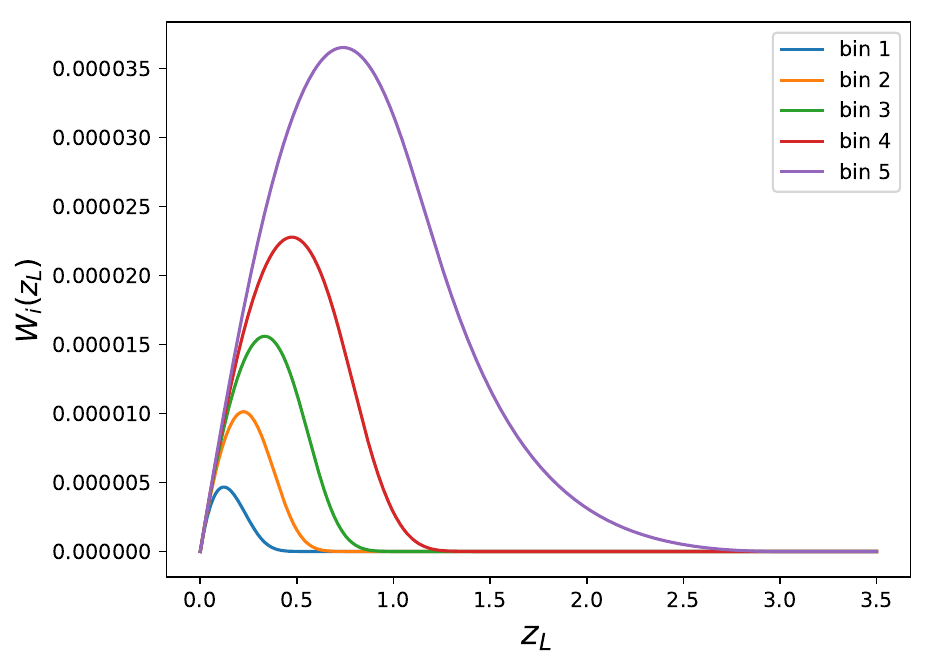}
    \caption{Weak lensing kernel}
    \label{fig:2a}
\end{subfigure}
\begin{subfigure}{.48\textwidth}
    \centering
    \includegraphics[width=0.9\linewidth, height=5cm]{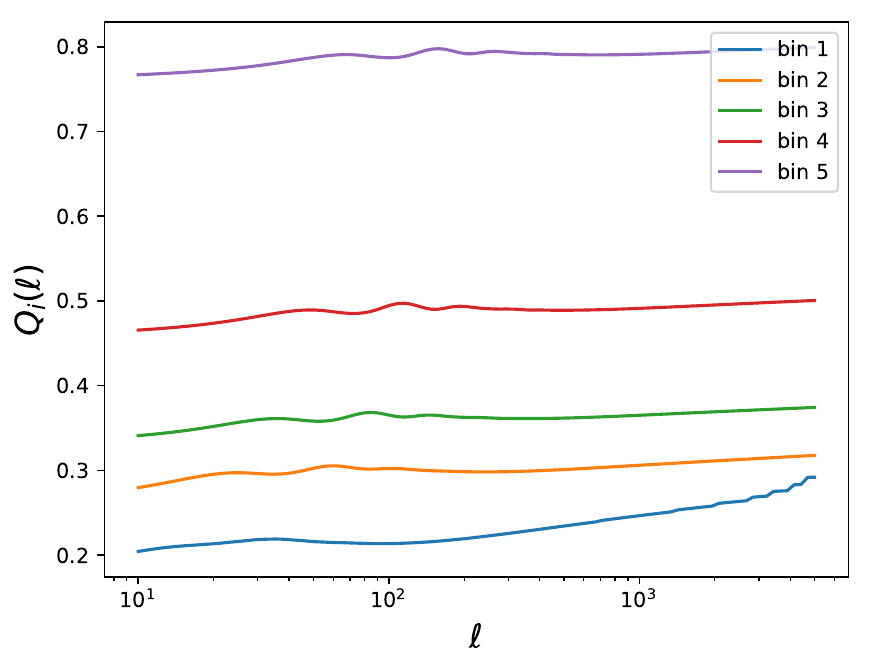} 
    \caption{Suppression factor}
    \label{fig:2b}
    \end{subfigure}
    \caption{(a). $W_i(z)$ which is the weighted lensing efficiency for the $i$th redshift bin is shown in Equation \ref{eq:4}. $W_i(z)$ for the 5 tomographic bins for source galaxies for the proposed LSST Y1 survey in the figure. (b). $Q_i(\ell)$ describes the suppression of the galaxy-galaxy correlation in the same redshift bin after discarding those pairs with source redshift higher than their lens redshift. $Q_i(\ell)$ for the 5 tomographic bins for source galaxies for the proposed LSST Y1 survey have been shown in the figure.}
    \label{fig:2}
\end{figure}



\subsubsection{IA parameters}\label{sec:2.4.1}
\par\noindent

The IA strength parameters $C_1$ and $C_2$ can be parameterized as a function of redshift $z$ following the convention of (\cite{Blazek_2019}, \cite{Samuroff_2019}) 
\begin{eqnarray}\label{eq:15}
    C_1(z) &=& -A_1 (\bar{C_1} \rho_{\text{crit}}) \frac{\Omega_{\mathrm{m}}}{D(z)} \bigg( \frac{1+z}{1+z_0}\bigg)^{\eta_1},\\\label{eq:16}
    C_2(z) &=& 5A_2 (\bar{C_1} \rho_{\text{crit}}) \frac{\Omega_{\mathrm{m}}}{D^2(z)} \bigg( \frac{1+z}{1+z_0}\bigg)^{\eta_2},
\end{eqnarray}
with the four IA parameters ($A_1$, $A_2$, $\eta_1$, $\eta_2$), where $\eta_1$ and $\eta_2$ which are considered as free parameters are set to a fiducial value of 1 for each of them. We consider the IA parameters $A_1 = 1$ and $A_2 = -1$. The number $\bar{C_1} = 5\times 10^{-14} h^{-2} M_{\text{sun}}^{-1}$ Mpc$^3$ was determined from the windowed shear variance in SuperCOSMOS and assuming the NLA model (i.e. the nonlinear matter power spectrum used with the linear IA model) \cite{Bridle_2007}. $\bar{C_1} \rho_{\text{crit}} \approx 0.014$. $\Omega_{\mathrm{m}}$ and $D(z)$\footnotemark{} are the matter density of the universe and growth factor, respectively, which are computed assuming the fiducial cosmology. We consider the pivot redshift $z_0 = 0.62$ following the convention of \cite{Samuroff_2019}. Following \cite{Blazek_2019}, we consider $C_{1\delta} = b_{\mathrm{ta}} C_1$, and we choose $b_{\mathrm{ta}}=1$ \cite{Samuroff_2019}. Similiarly, we have chosen $C_{2\delta} = b_{\mathrm{ta}} C_2$.

\footnotetext{Calculated using Core Cosmology Library (\texttt{CCL}) (\cite{Chisari_2019}).}

\subsubsection{Galaxy bias parameters}\label{sec:2.4.2}
\par\noindent

The galaxy bias parameter $b_1$ is a free parameter and we choose $b_1=2$ for our entire study. We also consider $b_{\mathrm{s}} = (-4/7)\times(b_1-1)$, obtained from the Lagrangian co-evolution (\cite{Saito_2014}; \cite{Pandey_2020}) and $b_2 = 0.9\times(b_1-1)^2-0.5$, motivated from the spherical collapse picture \cite{Desjacques_2018}.

\subsubsection{Fiducial cosmology}\label{sec:2.4.3}
\par\noindent

We consider ($\Omega_c$=0.2726, $\Omega_b$=0.03964, h=0.6766, $n_s$=1.021, $\sigma_8$=0.836) as the fiducial cosmology for our study. We use the same cosmology to calculate the power spectra using \texttt{CCL} \cite{Chisari_2019} $\&$ \texttt{FAST-PT} \cite{Fang_2017} and to validate the SC method in the next sections.

\subsection{\texorpdfstring{Photo-$z$ model}{Photo-z model}}\label{sec:2.4.4}
\par\noindent

The galaxy redshift distribution over the $i^{\mathrm{th}}$ redshift bin is $n_i^{\mathrm{P}}(z^{\mathrm{P}})$ and $n_i(z)$ is a function of photo-$z$ and true redshift, respectively, which are related by the photo-$z$ probability distribution function $p(z|z^{\mathrm{P}})$. We apply the SC method for the redshift distribution of source galaxies proposed for the LSST Y1 survey, and we follow the information provided in the Science Requirements Document (SRD) of LSST Dark Energy Science Collaboration \cite{SRD_V1_LSST}. We assume a survey coverage of half the sky ($f_{\text{sky}} \sim 0.5$) with a survey depth of $\sim 20$ galaxies per arcmin$^2$ and a redshift distribution similar to \cite{SRD_V1_LSST}
\begin{eqnarray}\label{eq:17}
    n(z) \propto z^2 \text{exp}[-(z/z_0)^{\alpha}],
\end{eqnarray}
with $(z_0, \alpha) = (0.62, 0.94)$. $n(z)$ gives the overall redshift distribution of the survey. The shear shape noise is described by $\gamma_{\mathrm{rms}} = 0.18 + 0.042z$, and the photo-$z$ error is described by a Gaussian probability distribution function (PDF) is
\begin{eqnarray}\label{eq:18}
    p(z|z^{\mathrm{P}}) = \frac{1}{\sqrt{2 \pi \sigma_z^2}} \text{exp} \bigg( -\frac{(z-z^{\mathrm{P}}-\Delta_z^i)^2}{2\sigma_{\mathrm{z}}^2} \bigg),
\end{eqnarray}
with the photo-$z$ scatter of $\sigma_{\mathrm{z}} = 0.03(1+z)$, and the redshift bias is chosen as $\Delta_z = 0$ for each bin. 

The normalized redshift distribution for each tomographic redshift bin is expressed as $\Bar{n}_i(z)$ \cite{Yao_2019} 
\begin{eqnarray}\label{eq:19}
    \Bar{n}_i(z) = \frac{\int_{z^{\mathrm{P}}_{\mathrm{min}}}^{z^{\mathrm{P}}_{\mathrm{max}}} n(z) p(z|z^{\mathrm{P}}) dz^{\mathrm{P}}}{\int_0^\infty \Big[ \int_{z^{\mathrm{P}}_{\mathrm{min}}}^{z^{\mathrm{P}}_ {\mathrm{max}}} n(z) p(z|z^{\mathrm{P}}) dz^{\mathrm{P}}\Big] dz}.
\end{eqnarray}

Figure \ref{fig:1a} and \ref{fig:1b} represent the normalized redshift distribution of 5 tomographic bins and the photo-$z$ PDF for the source galaxies for the proposed LSST Y1 survey, respectively. 

\begin{figure}
    \centering
    \includegraphics[width=0.45\linewidth, height=5cm]{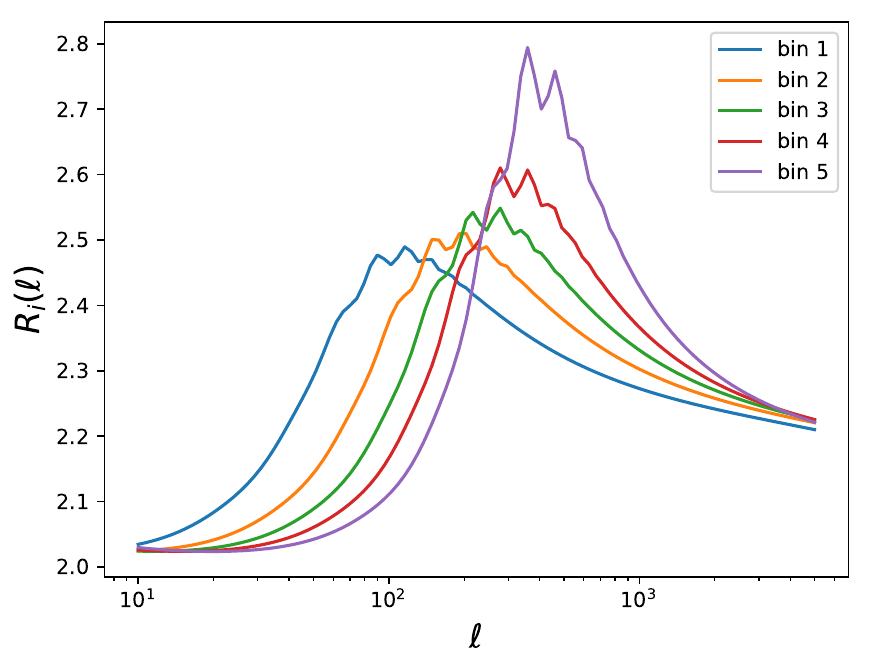} 
    \caption{$R_i(\ell)$ is defined as the ratio $P_{\mathrm{gI}}/P_{\mathrm{mI}}$ in the middle of the $i$th redshift bin.}
    \label{fig:3}
\end{figure}

\section{Self-calibration method for extraction of IG correlation in nonlinear regime}\label{sec:3}
\par\noindent

In a linear regime, it is appropriate to work with the linear matter density field and a linear galaxy bias, i.e., galaxy number density $\delta_{\mathrm{g}}$ is linearly proportional to matter density $\delta_{\mathrm{m}}$. In a nonlinear regime, $\delta_{\mathrm{g}}$ is no longer proportional to $\delta_{\mathrm{m}}$ and perturbative expansions become necessary for both $\delta_{\mathrm{g}}$ (Equation \ref{eq:9}) and $\delta_{\mathrm{m}}$ (Equation \ref{eq:10}) to correctly incorporate the small-scale behaviour. The scaling relation associated with the former SC technique assumes the ratio of $P_{\mathrm{gI}}$ and $P_{\mathrm{mI}}$ to be the linear galaxy bias which is no longer appropriate in the nonlinear regime. The nonlinear feature of IA of galaxies is considered through the TATT model and the corresponding IA power spectra. To apply SC in a nonlinear regime, we need to modify the scaling relation and we provide the necessary details in this section. We will test the performance and applicability of SC in the following section.  

The purpose of this SC method is to mitigate the IG contamination in the observed shear-shear correlation $C_{ij}^{\gamma \gamma}(\ell)$ mainly for cross-correlating bins, using the measurement $C_{ii}^{\mathrm{g}\gamma}(\ell)$ and $C_{ii}^{\mathrm{gg}}(\ell)$ from the surveys. The steps for the calculation of the IG correlation are described in detail in the following subsections. 

The measured IG contamination is in general an anti-correlation because the lensing-induced shear is tangential to the gradient of the gravitational potential, while the intrinsic ellipticity is parallel to the gradient. This observation can vary for different sets of IA parameters and redshift bins. For simplicity, we avoid the negative sign of the IG correlation for the rest of our study. The fraction of the IG contamination in the observed shear-shear signal can be quantified by $f_{ij}^{\mathrm{I}}(\ell)$ which is defined as the ratio of the two-dimensional angular IG power spectrum to two-dimensional angular GG power spectrum 
\begin{eqnarray}\label{eq:20}
    f_{ij}^{\mathrm{I}}(\ell) \equiv  \frac{C_{ij}^{\mathrm{IG}}(\ell)}{C_{ij}^{\mathrm{GG}}(\ell)},
\end{eqnarray}

After the SC, the residual IG contamination is expressed as the residual fractional errors on the lensing measurement, in which $\Delta f_{ij}^{(\mathrm{a})}$, $\Delta f_{ij}^{(\mathrm{b})}$ and $\delta f_{ij}$ denote the statistical errors and the systematic error, respectively. So, the performance of the SC is quantified by $\Delta f_{ij}^{(\mathrm{a})}$, $\Delta f_{ij}^{(\mathrm{b})}$ and $\delta f_{ij}$.

\begin{figure*}
    \centering
    \includegraphics[width=1.0\linewidth, height=4.5cm]{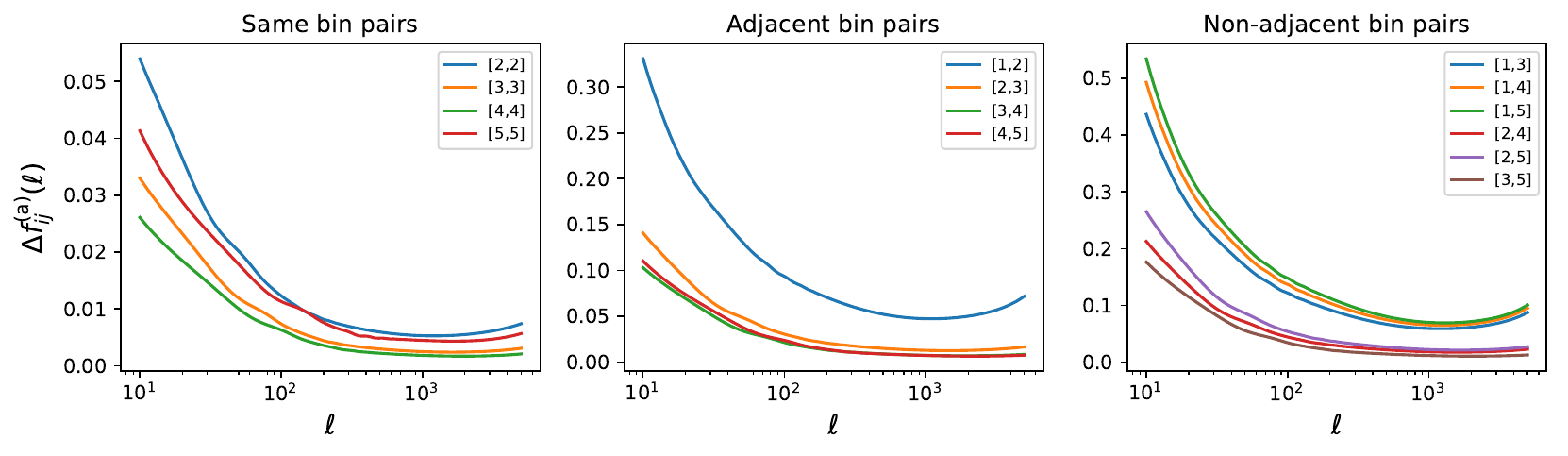}
    \caption{These curves represent residual statistical error $\Delta f_{ij}^{\mathrm(a)}$ which is equal to the threshold of the applicability of the SC technique $f_{ij}^{\text{thresh}}$. $\Delta f_{ij}^{(\mathrm{a})}$ are shown for all possible combinations of redshift bin pairs.}
    \label{fig:4}
\end{figure*}

\subsection{Separating the gG \& gI signals from lensing observables}\label{sec:3.1}
\par\noindent

The first step of the SC is to separate the galaxy number density-gravitational shear correlation $C_{ii}^{\mathrm{gG}}(\ell)$ and the galaxy number density-galaxy intrinsic ellipticity correlation $C_{ii}^{\mathrm{gI}}(\ell)$ from the observed galaxy-galaxy lensing correlation in the same redshift bin $C_{ii}^{\mathrm{g}\gamma}(\ell)$ based on the lensing geometry or the positions of the source-galaxies and the lens-galaxies. 


Using the lensing geometry, $C_{ii}^{\mathrm{gI}}(\ell)$ and $C_{ii}^{\mathrm{gG}}(\ell)$ (\cite{Zhang_2010b}, \cite{Troxel_2012a}, \cite{Yao_2020}) can be measured from $C_{ii}^{\mathrm{g}\gamma}(\ell)$, using the estimators 
\begin{eqnarray}\label{eq:21}
    \hat{C}_{ii}^{\mathrm{gI}}(\ell) &=& \frac{C_{ii}^{\mathrm{g}\gamma}|_{\mathrm{S}}(\ell) - Q_i(\ell)C_{ii}^{\mathrm{g}\gamma}(\ell)} {1-Q_i(\ell)},\\\label{eq:22}
    \hat{C}_{ii}^{\mathrm{gG}}(\ell) &=& \frac{C_{ii}^{\mathrm{g}\gamma}(\ell) - C_{ii}^{\mathrm{g}\gamma}|_{\mathrm{S}}(\ell)} {1-Q_i(\ell)},
\end{eqnarray}
with the suppression factor for the gG correlation in the $i^{\mathrm{th}}$ redshift bin $Q_i(\ell)$ defined as 
\begin{eqnarray}\label{eq:23}
    Q_i(\ell) = \frac{C_{ii}^{\mathrm{gG}}|_{\mathrm{s}}(\ell)}{C_{ii}^{\mathrm{gG}}(\ell)},
\end{eqnarray}
where the subscript $S$ in both of the equations, denotes the selection function defined as - 
\begin{eqnarray}\label{eq:24}
    S(z_{\mathrm{G}}^{\mathrm{P}}, z_{\mathrm{g}}^{\mathrm{P}}) = \begin{cases} 1 & \text{if $z_{\mathrm{G}}^{\mathrm{P}} < z_{\mathrm{g}}^{\mathrm{P}}$,}\\ 0 & \text{otherwise,} \end{cases}
\end{eqnarray}

Thus, $Q_i(\ell)$ measures the relative suppression of the gG signal, in the $i^{\mathrm{th}}$ redshift bin, due to the orientation dependence of the lens-source pairs,
where $C_{ii}^{\mathrm{gG}}|_{\mathrm{s}}(\ell) < C_{ii}^{\mathrm{gG}}(\ell)$, since those $z^{\mathrm{P}}_{\mathrm{G}} > z^{\mathrm{P}}_{\mathrm{g}}$ pairs that are disregarded contribute more to the galaxy-galaxy lensing correlation. 

The two-dimensional gG angular power spectrum is 
\begin{eqnarray}\label{eq:C1}
    C_{ii}^{\mathrm{gG}}(\ell) = \int_0^{\infty} \frac{W_i(\chi) \bar{n}_i(\chi)}{\chi^2} {P}_{\mathrm{mg}}\Big(k=\frac{\ell}{\chi}, z \Big)  d\chi,
\end{eqnarray}

The two-dimensional gG angular power spectrum with the suppression factor is
\begin{eqnarray}\label{eq:C2}
    C_{ii}^{\mathrm{gG}}|_{\mathrm{s}}(\ell) = \int_0^{\infty} \frac{W_i(\chi) \bar{n}_i(\chi)}{\chi^2} {P}_{\mathrm{mg}}\Big(k=\frac{\ell}{\chi}, z \Big) \eta_i (z, z_{\mathrm{g}}=z) d\chi, 
\end{eqnarray}
where $\eta(z_{\mathrm{L}},z_{\mathrm{g}})$ is defined as
\begin{eqnarray}\label{eq:C3}\nonumber
    \eta(z_{\mathrm{L}},z_{\mathrm{g}}) = \frac{2\int_{\Bar{z}_i-\Delta z_i/2}^{\Bar{z}_i+\Delta z_i/2} dz_{\mathrm{g}}^{\mathrm{P}} \int_{\Bar{z}_i-\Delta z_i/2}^{\Bar{z}_i+\Delta z_i/2} dz_{\mathrm{G}}^{\mathrm{P}} \int_0^\infty dz_{\mathrm{G}} W_{\mathrm{L}}(z_{\mathrm{L}},z_{\mathrm{G}}) p(z_{\mathrm{G}}|z_{\mathrm{G}}^{\mathrm{P}}) p(z_{\mathrm{g}}|z_{\mathrm{g}}^{\mathrm{P}}) S(z_{\mathrm{G}}^{\mathrm{P}}, z_{\mathrm{g}}^{\mathrm{P}}) n_i^{\mathrm{P}}(z_{\mathrm{G}}^{\mathrm{P}}) n_i^{\mathrm{P}}(z_{\mathrm{g}}^{\mathrm{P}})}{\int_{\Bar{z}_i-\Delta z_i/2}^{\Bar{z}_i+\Delta z_i/2} dz_{\mathrm{g}}^{\mathrm{P}} \int_{\Bar{z}_i-\Delta z_i/2}^{\Bar{z}_i+\Delta z_i/2} dz_{\mathrm{G}}^{\mathrm{P}} \int_0^\infty dz_{\mathrm{G}} W_{\mathrm{L}}(z_{\mathrm{L}},z_{\mathrm{G}}) p(z_{\mathrm{g}}|z_{\mathrm{g}}^{\mathrm{P}}) n_i^{\mathrm{P}}(z_{\mathrm{G}}^{\mathrm{P}}) n_i^{\mathrm{P}}(z_{\mathrm{g}}^{\mathrm{P}})},\\
\end{eqnarray}
where the normalization factor 2 comes from the relation  
\begin{eqnarray}\label{eq:C4}
    2 = \frac{\int_{\Bar{z}_i-\Delta z_i/2}^{\Bar{z}_i+\Delta z_i/2} dz_{\mathrm{g}}^{\mathrm{P}} \int_{\Bar{z}_i-\Delta z_i/2}^{\Bar{z}_i+\Delta z_i/2} dz_{\mathrm{G}}^{\mathrm{P}} p(z_{\mathrm{G}}|z_{\mathrm{G}}^{\mathrm{P}}) p(z_{\mathrm{g}}|z_{\mathrm{g}}^{\mathrm{P}})  n_i^{\mathrm{P}}(z_{\mathrm{G}}^{\mathrm{P}}) n_i^{\mathrm{P}}(z_{\mathrm{g}}^{\mathrm{P}})}{\int_{\Bar{z}_i-\Delta z_i/2}^{\Bar{z}_i+\Delta z_i/2} dz_{\mathrm{g}}^{\mathrm{P}} \int_{\Bar{z}_i-\Delta z_i/2}^{\Bar{z}_i+\Delta z_i/2} dz_{\mathrm{G}}^{\mathrm{P}} p(z_{\mathrm{G}}|z_{\mathrm{G}}^{\mathrm{P}}) p(z_{\mathrm{g}}|z_{\mathrm{g}}^{\mathrm{P}})
    S(z_{\mathrm{G}}^{\mathrm{P}}, z_{\mathrm{g}}^{\mathrm{P}}) n_i^{\mathrm{P}}(z_{\mathrm{G}}^{\mathrm{P}}) n_i^{\mathrm{P}}(z_{\mathrm{g}}^{\mathrm{P}})}.
\end{eqnarray}

Based on the definition, the range is $0 < Q_i < 1$. $Q_i \approx 1$ stands for very bad photo-$z$ quality, and $Q_i \approx 0$  means spectroscopic-level photo-$z$ quality. $Q_i$ is observable in the lensing surveys and can be calculated from the redshift distribution of galaxies. For details, we refer to \cite{Zhang_2010b}. 

$Q_i(\ell)$ for all 5 redshift bins are plotted in Figure \ref{fig:2b}. It can be observed that $Q_i(\ell)$ is approximately constant throughout the individual bins. $Q_i$ is almost scale-independent, as it is defined as the ratio of two power spectra of similar shapes. It increases with redshift which is expected for large photo-$z$ rms error at higher redshift and hence larger effective redshift width. Our results agree with the results of \cite{Zhang_2010b}, and \cite{Troxel_2012b}.



\subsection{Scaling relation}\label{sec:3.2}
\par\noindent

Under the Limber approximation, the two-dimensional galaxy intrinsic ellipticity-gravitational shear (IG) angular power spectrum between the $i^{\mathrm{th}}$ and $j^{\mathrm{th}}$ redshift bins is related to the three-dimensional non-linear matter density-galaxy intrinsic ellipticity cross-correlation power spectrum ${P}_{\mathrm{mI}}(k,z)$ through
\begin{eqnarray}\label{eq:25}
    C_{ij}^{\mathrm{IG}}(\ell) = \int_0^{\infty} \frac{\bar{n}_i(\chi) W_j(\chi)}{\chi^2} {P}_{\mathrm{mI}}\Big(k=\frac{\ell}{\chi}, z \Big)  d\chi,
\end{eqnarray}
where the weighted lensing efficiency $W_j(z)$ is defined in Equation \ref{eq:4}. The two-dimensional galaxy number density-galaxy intrinsic ellipticity (gI) angular power spectrum between the $i^{\mathrm{th}}$ and the $j^{\mathrm{th}}$ redshift bins is related to three-dimensional non-linear galaxy density-galaxy intrinsic ellipticity cross-correlation power spectrum ${P}_{\mathrm{gI}}(k,z)$ through
\begin{eqnarray}\label{eq:26}
    C_{ij}^{\mathrm{Ig}}(\ell) = \int_0^{\infty} \frac{\bar{n}_i(\chi) \bar{n}_j(\chi)}{\chi^2} {P}_{\mathrm{gI}} \Big(k=\frac{\ell}{\chi}, z \Big) d\chi.
\end{eqnarray}

The two-dimensional gI angular auto-correlation power spectrum for the $i^{\mathrm{th}}$ redshift bin is
\begin{eqnarray}\label{eq:27}
     C_{ii}^{\mathrm{Ig}}(\ell) = C_{ii}^{\mathrm{gI}}(\ell) = \int_0^{\infty} \frac{\bar{n}^2_i(z)}{\chi^2} {P}_{\mathrm{gI}}\Big(k=\frac{\ell}{\chi(z)}, z \Big)  \frac{dz}{d\chi}dz.
\end{eqnarray}

We define $R_i(\ell)$ as the ratio of $P_{\mathrm{gI}}(k=\ell/\chi, z)$ and $P_{\mathrm{mI}}(k=\ell/\chi, z)$ in the middle of the $i^{\mathrm{th}}$ redshift bin is 
\begin{eqnarray}\label{eq:28}
    R_i(\ell) = \frac{P_{\mathrm{gI}}(k=\ell/\chi_i, \bar{z}_i)}{P_{\mathrm{mI}}(k=\ell/\chi_i, \bar{z}_i)}.
\end{eqnarray}

Clearly, in the case of the linear galaxy bias $R_i(\ell)$ is the linear galaxy bias, but in the non-linear regimes, this is a combination of galaxy bias, IA and matter power spectrum. $R_i(\ell)$ is shown for all 5 redshift bins in Figure \ref{fig:3}. 

Under the thin redshift bin approximation for the $i^{\mathrm{th}}$ redshift bin $P_{\mathrm{mI}}$ and $P_{\mathrm{gI}}$ change slowly over the bin and can be approximated as $P_{\mathrm{mI}}(k=\ell/\chi_i, \bar{z}_i)$ and $P_{\mathrm{gI}}(k=\ell/\chi_i, \bar{z}_i)$, respectively. $C_{ij}^{\mathrm{IG}}(\ell)$ and $C_{ii}^{\mathrm{Ig}}(\ell)$ can be approximated as 
\begin{eqnarray}\label{eq:29}
     C_{ij}^{\mathrm{IG}}(\ell) &\simeq&  {P}_{\mathrm{mI}}\Big(k=\frac{\ell}{\chi_i},\bar{z}_i\Big) W_{ij}(z) \chi_i^{-2},\\\label{eq:30}
     C_{ii}^{\mathrm{Ig}}(\ell) &\simeq&  R_i(\ell) {P}_{\mathrm{mI}} \Big(k=\frac{\ell}{\chi_i},\bar{z}_i\Big) \frac{\chi_i^{-2}}{\Delta_i},
\end{eqnarray}
where $\chi_i = \chi(\bar{z}_i)$, and the weighted lensing kernel $W_{ij}$ is 
\begin{eqnarray}\label{eq:31}
    W_{ij} \equiv \int_0^\infty dz_{\mathrm{L}} \int_0^\infty dz_{\mathrm{G}} W_{\mathrm{L}}(z_{\mathrm{L}},z_{\mathrm{G}}) \bar{n}_i(z_{\mathrm{L}}) \bar{n}_j(z_{\mathrm{G}}),
\end{eqnarray} 
and the effective width of the $i^{\mathrm{th}}$ redshift bin $\Delta_i$ is 
\begin{eqnarray}\label{eq:32}
        \Delta_i^{-1} = \int_0^\infty \bar{n}^2_i(z) \frac{dz}{d\chi}dz.
\end{eqnarray}

Combining Equations \ref{eq:29} - \ref{eq:32} lead us to the SC scaling relation for IG correlation 
\begin{eqnarray}\label{eq:33}
    C_{ij}^{\mathrm{IG}}(\ell) \simeq \bigg[ \frac{W_{ij} \Delta_i}{R_i(\ell)} \bigg]C_{ii}^{\mathrm{Ig}}(\ell).
\end{eqnarray}

\begin{figure*}
    \centering
    \begin{subfigure}[h!]{1\textwidth}
    \includegraphics[width=1\linewidth, height=2.75cm]{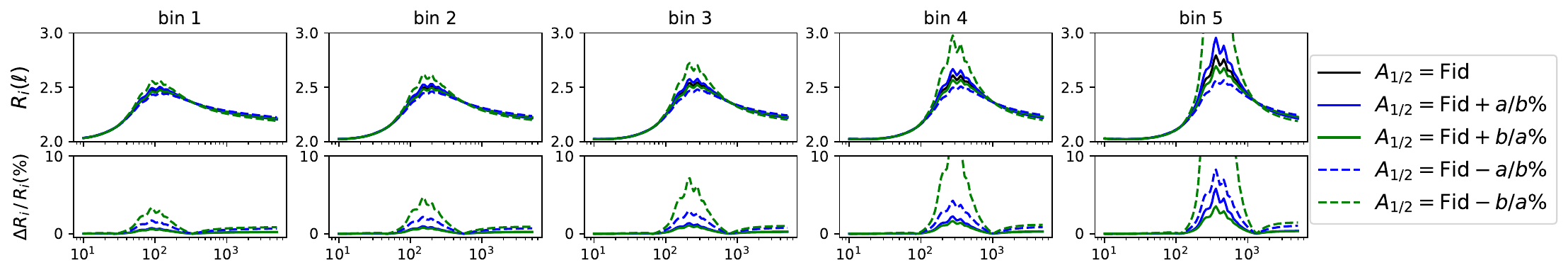} 
    \end{subfigure}
    \begin{subfigure}[h!]{1\textwidth}
    \includegraphics[width=1\linewidth, height=2.75cm]{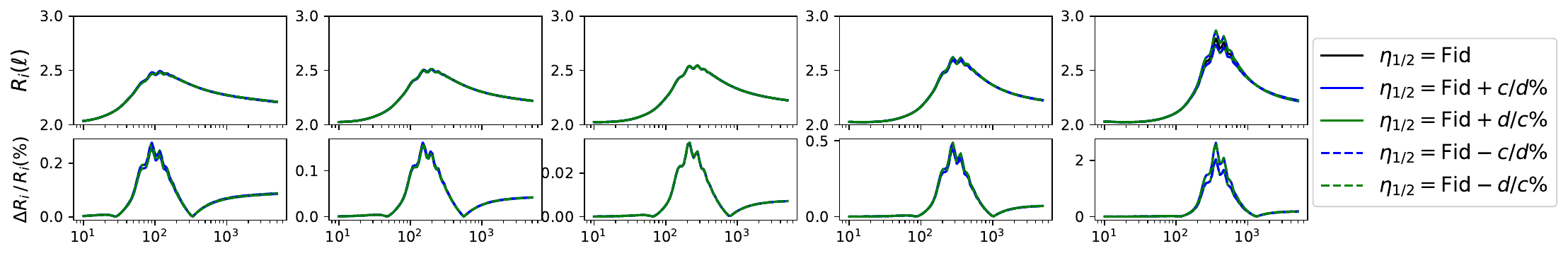}
    \end{subfigure}
    \begin{subfigure}[h!]{1\textwidth}
    \includegraphics[width=1\linewidth, height=2.75cm]{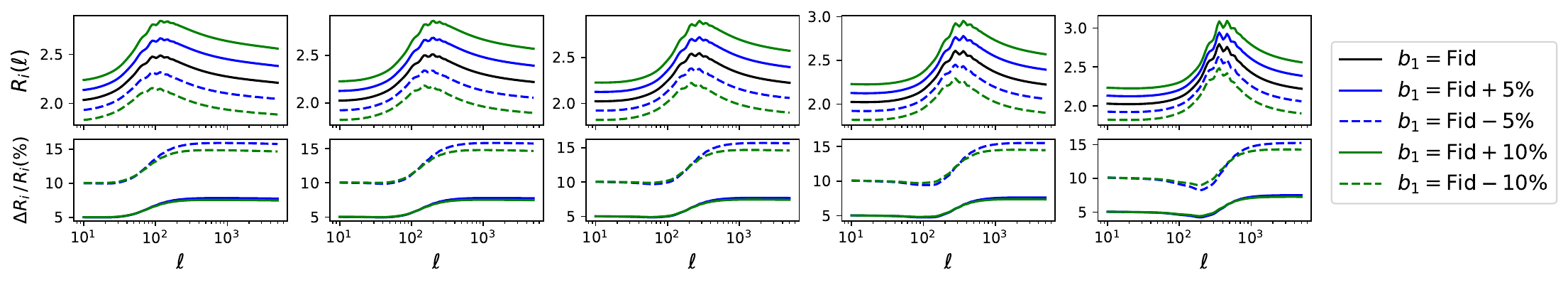}
    \end{subfigure}
    \caption{Variation of $R_i$ due to the uncertainties in IA and bias parameters are shown for the 5 redshift bins. The parameters are varied one at a time, shown in the box just right to the plots, keeping others fixed at their fiducial values. The first two rows represent different sets of $R_i$s when the uncertainties in $A_{1/2}$ are considered and the deviation from the $R_i$ which is calculated using the fiducial values of the IA parameters. The rest of the plots can be read similarly. $A_{1/2} = \mathrm{Fid} + a/b \%$ can be read as $A_1 = A_1^{\mathrm{Fid}} + a\% \ \&\ A_2 = A_2^{\mathrm{Fid}} + b\%$, where $a\ \&\ b$ refer to $50\ \&\ 60$, respectively. $\eta_{1/2} = \mathrm{Fid} + c/d \%$ can be read as $\eta_1 = \eta_1^{\mathrm{Fid}} + c\% \ \&\ \eta_2 = \eta_2^{\mathrm{Fid}} + d\%$, where $c\ \&\ d$ refer to $80\ \&\ 90$, respectively. }
    \label{fig:5}
\end{figure*}

To correctly predict the IG correlation in a nonlinear regime, one must use this scaling relation. $R_i(\ell)$ which incorporates the non-linear feature of galaxy bias and the IA model, can be determined through proper choices of the IA parameters ($A_1$, $A_2$, $\eta_1$, $\eta_2$) and galaxy bias parameters ($b_1$, $b_2$, $b_s$). The variations in $R_i(\ell)$ due to the uncertainties in the IA  parameters are shown in Figure \ref{fig:5}. We study the impact of these uncertainties on the performance of the SC method in the subsequent sections.

\section{Performance of the new SC method in the nonlinear regime}\label{sec:4}
\par\noindent

In this section, we investigate the accuracy of the SC technique by analyzing the errors associated with the measurement of $C_{ii}^{\mathrm{Ig}}$, uncertainties in the IA parameters and the accuracy of the scaling relation.

\subsection{\texorpdfstring{Measurement error in $C_{ii}^{\mathrm{Ig}}$ and threshold for IA detection}{Measurement error in Cii Ig and threshold for IA detection}}\label{sec:4.1}
\par\noindent

Measurements of $C_{ii}^{\mathrm{g}\gamma}$ and $C_{ii}^{\mathrm{gG}}$ are contaminated by cosmic variance and shot noise errors which propagate into $C_{ii}^{\mathrm{Ig}}(\ell)$. The error in measuring $C_{ii}^{\mathrm{Ig}}$ using equation \ref{eq:21} (see appendix C of \cite{Zhang_2010b}) for a bin with width $\Delta \ell$ is  
\begin{eqnarray}\label{eq:34}
    \big(\Delta C_{ii}^{\mathrm{Ig}}(\ell)\big)^2 &=& \frac{1}{2 \ell \Delta \ell} \Bigg( C_{ii}^{\mathrm{gg}}C_{ii}^{\mathrm{GG}} + \Bigg[ 1+ \frac{1}{3(1-Q)^2} \Bigg] \times \Big[ C_{ii}^{\mathrm{gg}} C_{ii}^{\mathrm{GG,N}} + C_{ii}^{\mathrm{gg, N}} \Big( C_{ii}^{\mathrm{GG}} + C_{ii}^{\mathrm{II}} \Big) \Big] \\\nonumber
    && +\ C_{ii}^{\mathrm{gg,N}}C_{ii}^{\mathrm{GG,N}} \Bigg[ 1+ \frac{1}{(1-Q)^2} \Bigg]\Bigg) 
\end{eqnarray}
where the superscript `$N$' denotes the measurement noise (e.g., random shape noise, shot noise in galaxy distribution). $C_{ii}^{\mathrm{gg, N}} = 4\pi f_{\text{sky}}/N_i$ and $C_{ii}^{\mathrm{GG, N}} = 4\pi f_{\text{sky}} \gamma_{\text{rms}}^2/N_i$, where $N_i$ is the total number of galaxies in the $i^{\mathrm{th}}$ redshift bin. If the strength of the IA signal is too small, the measurement error $\Delta C_{ii}^{\mathrm{Ig}}$ will be larger than $C_{ii}^{\mathrm{Ig}}$. So, $\Delta C_{ii}^{\mathrm{Ig}} = C_{ii}^{\mathrm{Ig}}$ sets a threshold for $f_{ij}^{\mathrm{I}}$. Combining Equations \ref{eq:20} and \ref{eq:33}, we get the threshold as  
\begin{eqnarray}\label{eq:35}
    f_{ij}^{\text{thresh}} = \Bigg( \frac{\Delta C_{ii}^{\mathrm{Ig}}}{C_{ij}^{\mathrm{GG}}}\Bigg) \Bigg( \frac{W_{ij} \Delta_i}{R_i}\Bigg).
\end{eqnarray}

$f_{ij}^{\text{thresh}}$ describes the minimum IA that can be detected through the SC method for $S/N =1$. Thus it defines the lower limit beyond which the SC is no longer applicable. It also describes the accuracy of the SC resulting from the measurement error in $C_{ii}^{\mathrm{Ig}}$. $\Delta C_{ii}^{\mathrm{Ig}}$ propagates into an error in determining $C_{ij}^{\mathrm{IG}}$ through the scaling relation and hence leaves a residual statistical error $\Delta f_{ij}^{(\mathrm{a})}$ in the GG measurement. Since $\Delta f_{ij}^{(\mathrm{a})}/ f_{ij}^{(\mathrm{I})} = \Delta C_{ii}^{\mathrm{Ig}}/C_{ii}^{\mathrm{Ig}}$, and combining Equation \ref{eq:33}, we get 
\begin{eqnarray}\label{eq:36}
    \Delta f_{ij}^{(\mathrm{a})} = \Bigg( \frac{\Delta C_{ii}^{\mathrm{Ig}}}{C_{ij}^{\mathrm{GG}}}\Bigg) \Bigg( \frac{W_{ij} \Delta_i}{R_i}\Bigg),
\end{eqnarray}

\begin{figure}
    \centering
    \includegraphics[width=0.8\linewidth]{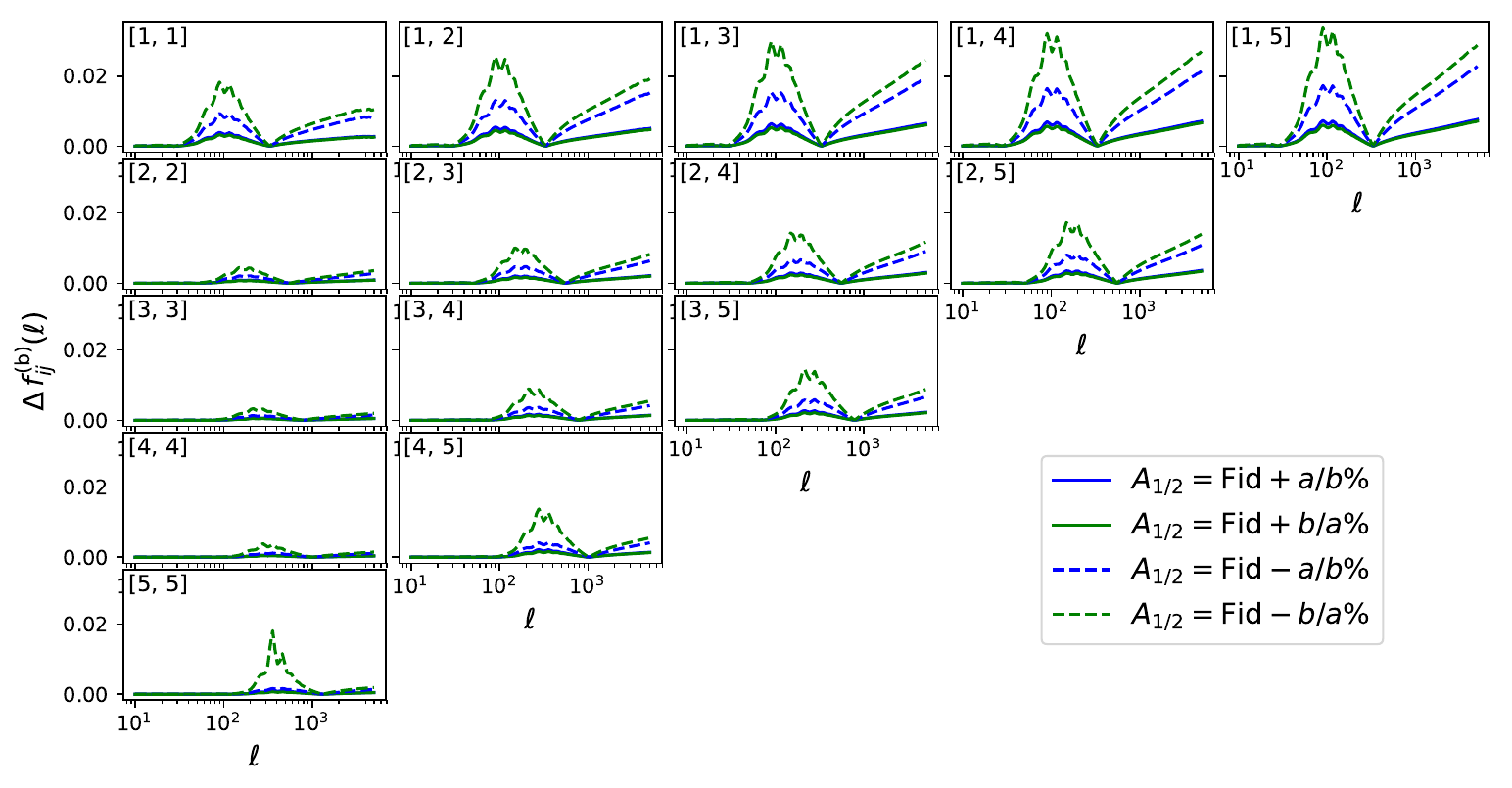}
    \label{fig:6a}
    \includegraphics[width=0.8\linewidth]{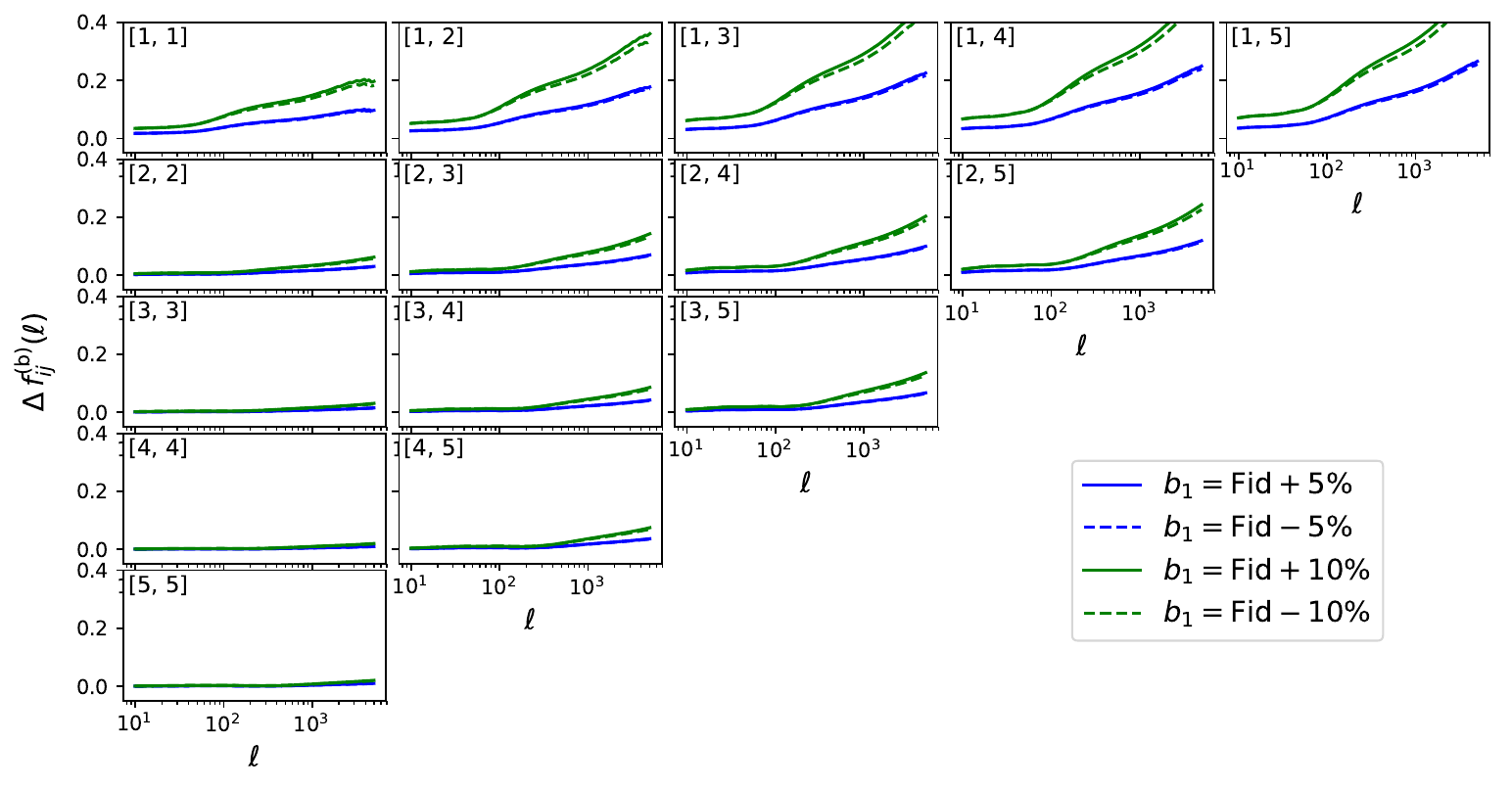}
    \label{fig:6b}
    
    \caption{Residual statistical error $\Delta f_{ij}^{(\mathrm{b})}$ due to uncertainty in IA parameters and the galaxy bias parameter have been shown. The figure on the top represents $\Delta f_{ij}^{(\mathrm{b})}$ for the uncertainties in $A_1 \ \&\ A_2$  while the rest of the IA parameters and galaxy bias parameter are set to their fiducial value. The figure on the bottom represents $\Delta f_{ij}^{(\mathrm{b})}$ for the uncertainties in $b_1$  while the rest of the IA parameters are set to their fiducial value. $A_{1/2} = \mathrm{Fid} + a/b \%$ can be read as $A_1 = A_1^{\mathrm{Fid}} + a\% \ \&\ A_2 = A_2^{\mathrm{Fid}} + b\%$, where $a\ \&\ b$ refer to $50\ \&\ 60$, respectively. }
    \label{fig:6}
\end{figure}

After combining Equations \ref{eq:35} and \ref{eq:36}, we get $\Delta f_{ij}^{(\mathrm{a})} = f_{ij}^{\text{thresh}}$. When the IA signal is sufficiently large and satisfies $f^{\mathrm{I}}_{ij} > f_{ij}^{\text{thresh}}$, the IG contamination can be detected and eliminated using the SC method from the observed lensing signal. It renders a systematic error in lensing measurement with amplitude $f^{\mathrm{I}}_{ij}$ into a statistical error $\Delta f_{ij}^{(\mathrm{a})} = f_{ij}^{\text{thresh}}$ which is calculated and shown in the Figure \ref{fig:6} for various choices of bin-combinations.  

The measurement error in $C_{ij}^{\mathrm{GG}}$ is described by the rms fluctuation induced by the cosmic variance in the GG correlation and shot noise due to random galaxy ellipticities with the assumption that no other systematic error exists. The lower limit of the fractional measurement error on $C_{ij}^{\mathrm{GG}}$ (when $i\neq j$) is  
\begin{eqnarray}\label{eq:37}
    \big(e_{ij}^{\text{min}}\big)^2 = \frac{(C_{ij}^{\mathrm{GG}})^2  + \big(C_{ii}^{\mathrm{GG}} + C_{ii}^{\mathrm{GG, N}} \big)\big(C_{jj}^{\mathrm{GG}} + C_{jj}^{{GG, N}} \big)} {2\ell \Delta\ell f_{\text{sky}}(C_{ij}^{\mathrm{GG}})^2}.
\end{eqnarray}

When $\Delta f_{ij}^{(\mathrm{a})}$ is smaller than $e^{\text{min}}_{ij}$, the residual error after the SC will then be negligible, which indicates a SC with a little cosmological information loss. We find that $\Delta f_{ij}^{(\mathrm{a})} < e^{\text{min}}_{ij}$ is, in general, valid for all the combinations of the redshift bins. Figure \ref{fig:9} shows the numerical verification. 


\subsection{\texorpdfstring{Uncertainty in determining $R_i(\ell)$}{Uncertainty in determining Ri}}\label{sec:4.2}
\par\noindent

From the definition of $R_i(\ell)$ (Equation \ref{eq:28}), it is evident that any uncertainty in $R_i(\ell)$ must be caused by the uncertainties in the IA parameters and galaxy bias parameters.  
The variations and percentage deviations of $R_i(\ell)$ for all 5 redshift bins when the uncertainties are included in the IA and bias parameters are shown in Figure \ref{fig:5}. 

The individual pair of rows represent the variation and the percentage deviations of $R_i(\ell)$ from its fiducial value, respectively when a single parameter is varied. The individual column represents the 5 different redshift bins. The first pair of rows correspond to when the $\pm 50\%$ and $\pm 60\%$ uncertainties are considered for IA amplitudes ($A_1$ \& $A_2$), simultaneously. The second pair of rows corresponds to when the $\pm 80\%$ and $\pm 90\%$ uncertainties are considered for $\eta_1$ \& $\eta_2$, simultaneously. And, the last pair of rows correspond to when the $\pm 5\%$ and $\pm 10\%$ uncertainties are considered for bias parameter $b_1$.

We observe that $R_i$ is more sensitive to the bias parameter $b_1$ compared to the IA parameters. And the dependence on IA amplitude parameters ($A_1$ and $A_2$) is stronger compared to the other two IA parameters ($\eta_1$ and $\eta_2$). 
As we observe the dependence on $\eta_1$ and $\eta_2$ are relatively weaker compared to other free parameters, any error in the SC due to the uncertainties of these two parameters will be subdominant. It is safe to ignore the error propagated through the uncertainties in $R_i(\ell)$ caused by these two IA parameters. We also observe the dependencies are a little stronger for the higher redshift bins compared to the lower redshift bins. This fact can be attributed to the redshift dependencies of the IA model which is encapsulated in $C_1(z)$ and $C_2(z)$.     

The uncertainty in $R_i(\ell)$ propagates through the scaling relation and gives rise to an uncertainty in the SC. These uncertainties induce a residual statistical error 
\begin{eqnarray}\label{eq:38}
    \Delta f_{ij}^{(\mathrm{b})} = f^{\mathrm{I}}_{ij} \frac{\Delta R_i(\ell)}{R_i(\ell)},
\end{eqnarray}
where $\Delta R_i(\ell)$ is the uncertainty in $R_i(\ell)$. The statistical error $\Delta f_{ij}^{(\mathrm{b})}$ due to the uncertainties in IA amplitude and bias parameters are shown in Figure \ref{fig:5}. The residual error due to the uncertainties in IA amplitude parameters is smaller than the residual error due to the uncertainty in the bias parameter for all combinations of redshift bins. 

In general, the residual error due to the uncertainties in the bias parameter is comparable to the minimum statistical error in the $C^{\mathrm{GG}}_{ij}$ measurement, $e_{ij}^{\mathrm{min}}$ (Figure \ref{fig:9}). So, the uncertainty in the bias parameter that causes the uncertainties in determining $R_i(\ell)$ does not have a significant impact on the reduction of the performance of the SC method. In conclusion, it is safe to assume that the dependencies on the IA and bias parameters have no significant impact on the error budget.

Not only the uncertainties of the IA parameters but also the uncertainties of the cosmological parameters affect the calculation of $R_i(\ell)$. 
The associated uncertainties in the cosmological parameters are well below $1\%$. As long as the associated uncertainty is smaller than $1\%$, the induced error will be sub-dominant to the systematic error discussed later in Section \ref{sec:4.3}.   

\begin{figure*}
    \centering
    \includegraphics[width=1.0\linewidth, height=4.25cm]{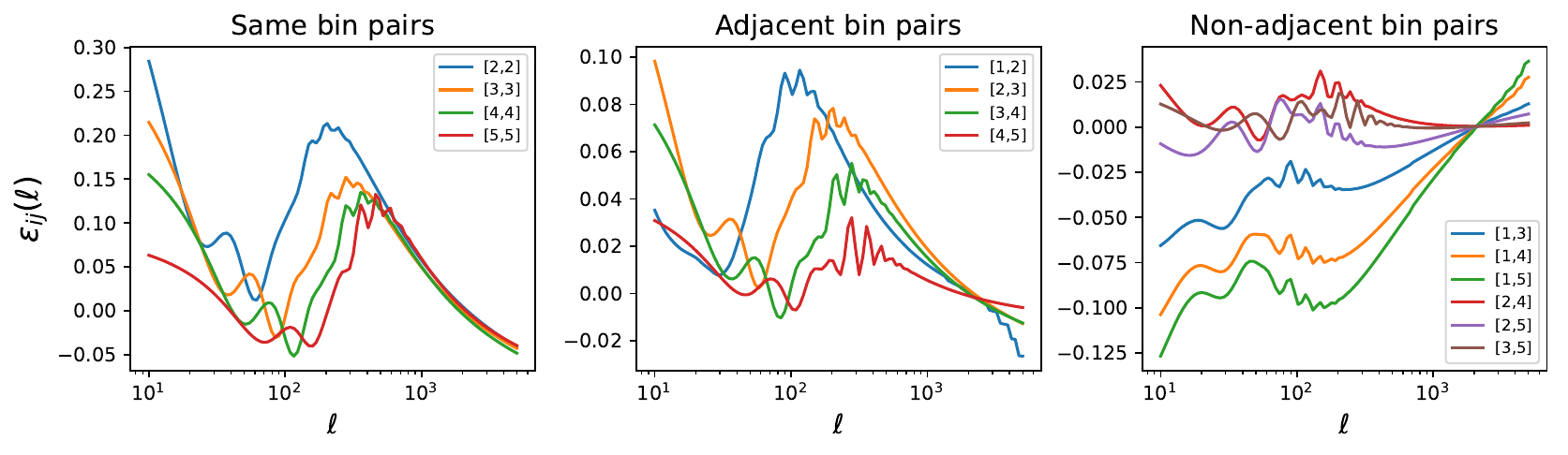}
    \caption{$\varepsilon_{ij}$ which quantifies the accuracy of the scaling relation are shown for - (a) the auto-correlatingbin pairs, i.e., $j=i$; (b) the adjacent bin pairs, i.e., $j=i+1$; (c) the non-adjacent bin pairs, i.e., $j>i+1$. The scaling relation is accurate well within 20$\%$ so the IG contamination can be suppressed by a factor of 5 or larger for auto-correlating bin pairs. Whereas, for the non-adjacent bin pairs, the scaling relation is accurate within 10$\%$ which implies suppression of the IG contamination by a factor of 10 or larger can be achieved. Different colours represent different bin combinations as mentioned in the plot. } 
    \label{fig:7}
\end{figure*}

\subsection{Accuracy of the new scaling relation}\label{sec:4.3}
\par\noindent

The most important step in the SC method is to calculate the IG contamination ($C_{ij}^{\mathrm{IG}}$) using the observed $C_{ii}^{\mathrm{Ig}}$ and other observables from the surveys through the scaling relation. To validate the performance of the SC method it is necessary to check the accuracy of the scaling relation. The accuracy can be quantified by 
\begin{eqnarray}\label{eq:39}
    \varepsilon_{ij}(\ell) \equiv \frac{R_i(\ell) C_{ij}^{\mathrm{IG}}(\ell)} {W_{ij} \Delta_i C_{ii}^{\mathrm{Ig}}(\ell)} - 1 ,
\end{eqnarray}
$\varepsilon_{ij}(\ell)$ quantifies the residual systematic error $\delta f_{ij}  =  \varepsilon_{ij} f^{\mathrm{I}}_{ij}$. The numerical results are shown in Figure \ref{fig:7}. 

We observe $|\varepsilon_{ij}|$ for the non-adjacent bin pairs ($i, j>i+1$) is less than $2.5\%$ and $10\%$ without and with bin 1, respectively. It signifies that IG contamination can be suppressed by a factor of $40$ and $10$ or larger for the bin pairs without and with bin 1, respectively if other errors are negligible. For the adjacent bin pairs ($i, j=i+1$), there is a drop in accuracy and $|\varepsilon_{ij}|$ for such bin pairs are within $10\%$ which signifies that IG contamination can be suppressed by a factor of $10$ or larger if other errors are negligible. For the auto-correlating bins ($i, j=i$), accuracy decreases and $|\varepsilon_{ij}|$ for such bin pairs are within $20\%$ (above $\ell=20$) which signifies that IG contamination can be suppressed by a factor of $5$.
The lensing kernel $W_j(z)$ varies quickly for the adjacent bins, which are close in redshift. This fast redshift evolution degrades the accuracy of the SC and hence results in a larger $|\varepsilon_{ij}|$. For the non-adjacent pairs of bins, $W_j(z)$ varies slowly due to the larger lens-source distance, resulting in better accuracy. So, the SC for the adjacent bins is not as accurate as for the non-adjacent bins. Our results agree with the results of (\cite{Zhang_2010b}, \cite{Troxel_2012a}, \cite{Yao_2017}, \cite{Yao_2019}). The lensing signal is generally weak for the bins with lower redshift which may result in poor accuracy of the scaling relation for such bins. We observe that the scaling relation is not very accurate for the bin pairs that involve bin 1. 

We observe a significant degradation in the performance of the scaling relation for all combinations of bin pairs if an effective galaxy bias $\big(b_i^{\rm eff} \equiv \sqrt{C_{ii}^{\rm gg}(\ell)/C_{ii}^{\rm mm}(\ell)}\big)$ is used instead of $R_i$. Our results are evidence of the improved accuracy of the SC method due to better IA modelling. 



\section{Other sources of errors}\label{sec:5}
\par\noindent

Beyond the three sources of error mentioned earlier, additional factors contribute to uncertainties in the IG measurement. We qualitatively explore magnification bias and catastrophic photo-z errors based on the discussion of \cite{Zhang_2010b}, \cite{Troxel_2012a}. However, based on simplified estimates, we find that none of these factors can entirely invalidate the SC technique.

\subsection{\texorpdfstring{Catastrophic photo-$z$ errors}{Catastrophic photo-z errors}}\label{sec:5.1}
\par\noindent

In this study, our numerical computations exclusively consider a Gaussian photo-$z$ probability density function (PDF). However, in observational data, the photo-$z$ PDF exhibits greater complexity due to non-negligible outliers. This presence of significant outliers introduces a catastrophic error, affecting the SC technique in two distinct ways by reducing the accuracies of the $Q$ estimation, and the scaling relation. It has been considered that the true galaxy distribution in a given photo-$z$ bin is sufficiently narrow and smooth in the derivation of the scaling relation. Thus the catastrophic error generally leads to degradation in performance of the scaling relation. 

Catastrophic error introduces the bias in $Q$ estimation mainly through its impact on $\eta$. Stage IV photometric surveys will have the outlier rate $f_{\mathrm{cat}}$ to $\sim 0.1\%$ accuracy for the induced systematic errors to be subdominant. Following \cite{Zhang_2010b}, we choose $|z-z^{\mathrm{P}}|>\Delta$ as the criteria of the catastrophic error and $f_{\mathrm{cat}} = \int_0^{z^{\mathrm{P}}-\Delta} p(z|z^{\mathrm{P}}) dz + \int_{z^{\mathrm{P}}+\Delta}^\infty p(z|z^{\mathrm{P}}) dz$. Using $f_{\mathrm{cat}} << 1$ and Equation \ref{eq:C3}, we observe that the induced bias $\delta Q \simeq O(f_{\mathrm{cat}})$. However, if $|f_{\mathrm{cat}}| < 0.1\%$, can be achieved, the resulting error in $Q$ remains below 1\%. Consequently, this induced error does not significantly impact the SC process. Moreover, despite the presence of substantial catastrophic errors (as documented by studies such as \cite{Schneider_2006}, \cite{Newman_2008}, \cite{Zhang_2010c}), we can statistically infer the photo-$z$ probability density function (PDF) through both self- and cross-calibrations of photo-$z$ errors. Importantly, since $Q$ can be predicted based on the photo-z PDF, we can mitigate potential biases in $Q$, even when the actual $|f_{\mathrm{cat}}| > 0.1\%$.

The catastrophic error also affects the scaling relation. It biases both $C^{\mathrm{IG}}$, through the term $W_j$ and $n_i$ in Equation \ref{eq:25}, and $C^{\mathrm{Ig}}$, through the term $n_i^2$ in Equation \ref{eq:27}. Although part of this effect cancels out when taking the ratio of the two, a residual error remains at the order of $O(f_{\mathrm{cat}})$. However, based on the magnitude estimation, it appears that the bias induced by catastrophic errors is likely subdominant compared to the major systematic error $\delta f^{(\mathrm{c})}_{ij}$ in the scaling relation.

\begin{figure*}
    \begin{minipage}{0.49\linewidth}
        \includegraphics[width=1\linewidth]{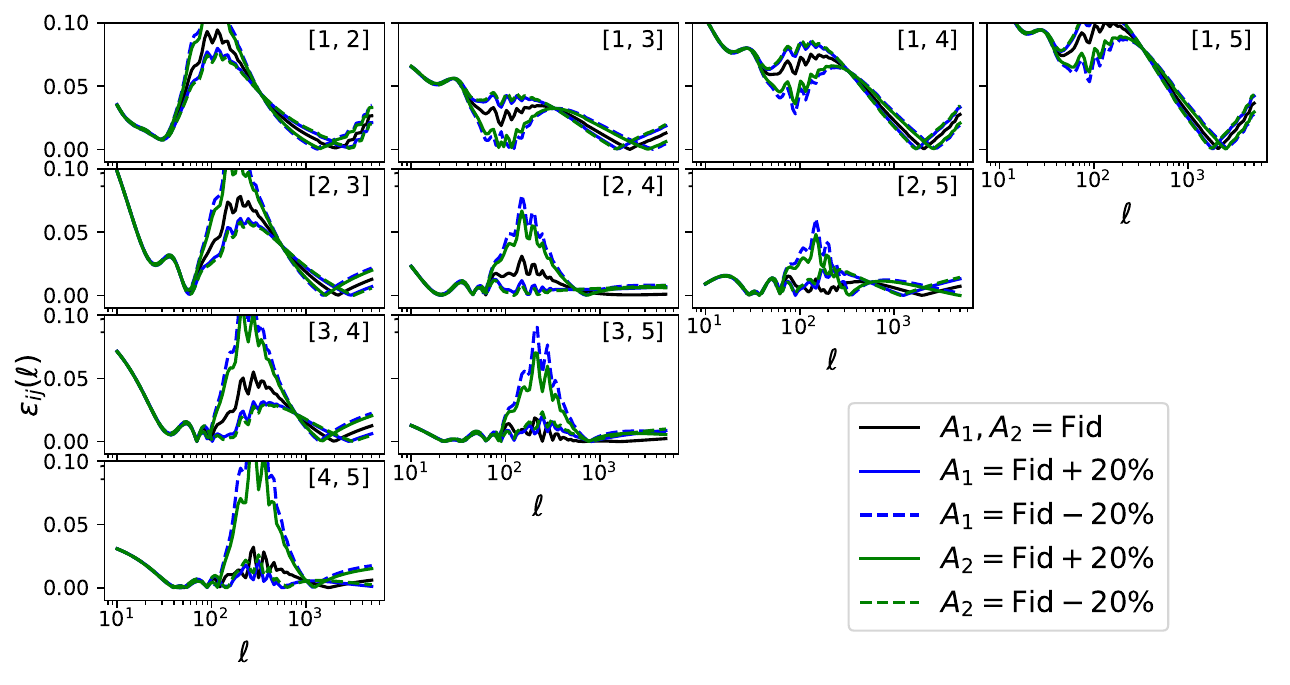}
    \end{minipage}
    \hfill
    \begin{minipage}{0.49\linewidth}
        \includegraphics[width=\linewidth]{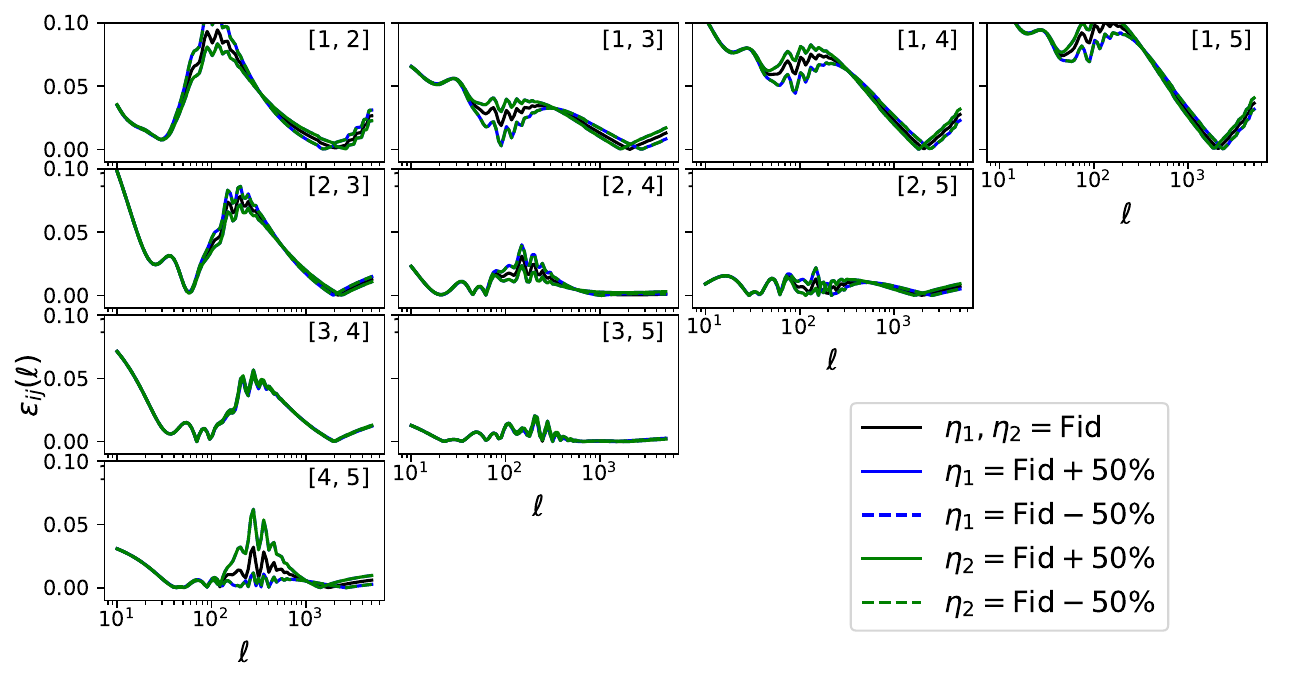}
    \end{minipage}
    \centering
    \begin{minipage}{0.5\linewidth}
        \includegraphics[width=\linewidth]{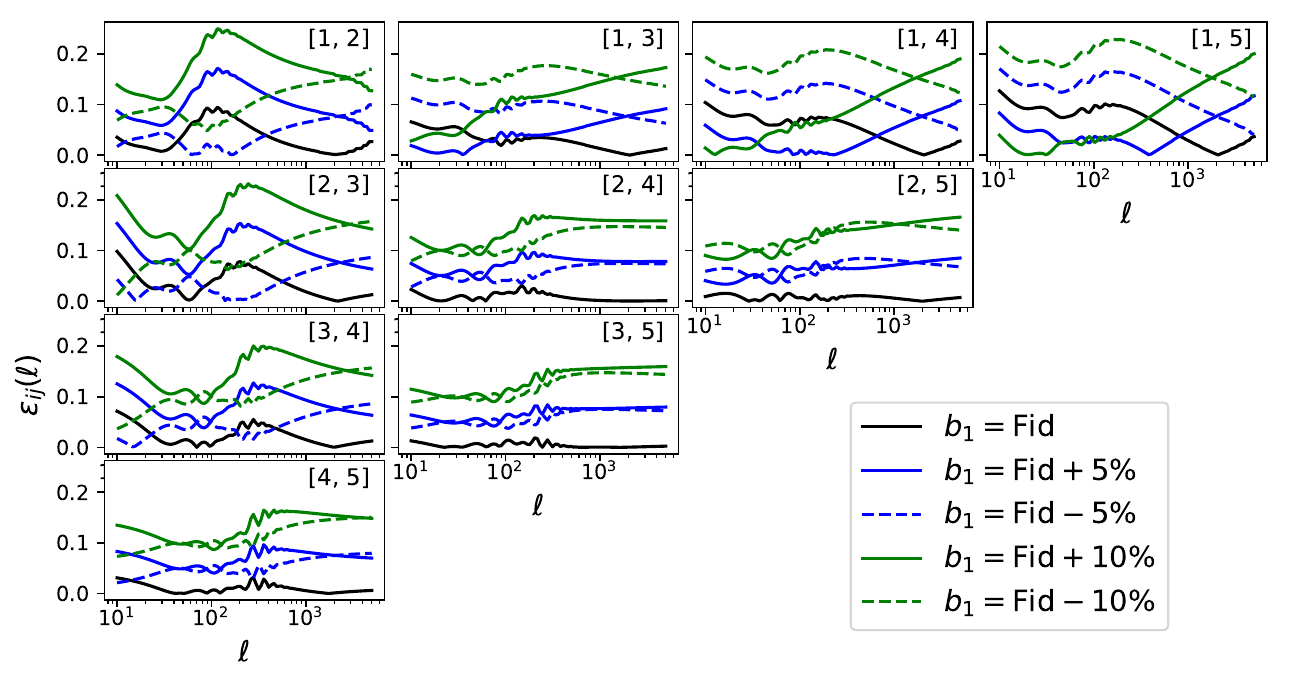}
    \end{minipage}
    \caption{Variation in the accuracy of the scaling relation when the uncertainty in IA and bias parameters are propagated through $R_i(\ell)$, keeping the rest of the IA parameters fixed to their fiducial values. The upper left and the upper right plot correspond to when the uncertainties in the IA amplitude ($A_1 \ \&\ A_2$) and the other two IA parameters ($\eta_1 \ \& \ \eta_2$) are considered. The lower plot corresponds to when the uncertainties in the bias parameter ($b_1$) are considered. Different colours represent the accuracy of the scaling relation for a particular amount of uncertainties in the IA/bias parameters which are mentioned in the legend.}
    \label{fig:8}
\end{figure*}

\subsection{\texorpdfstring{Photo-$z$ bias}{Photo-z bias}}\label{sec:5.4}
\par\noindent

We aim to delve deeper into the relationship between the quality of photo-z and SC. The shift in the photo-$z$ i.e. photo-$z$ bias can affect the results of the SC significantly. \cite{Yao_2017} studied the effect of photo-$z$ marginalization in forecasting the cosmological parameters after self-calibrating the IG contamination from the cosmic shear signal. They observed that the increase in contour size for SC is primarily due to photo-$z$ marginalization whereas the additional measurement error in the case is insignificant. Given that SC utilizes photo-$z$ information via the $Q$ estimation and the scaling relation, it is logical to anticipate that SC's performance will be more susceptible to the quality of photo-$z$.

They experimented with various photo-$z$ priors to contrast the contours of the SC and marginalization cases, as depicted in Figure 6 in \cite{Yao_2017}. The figure demonstrates that as the photo-z prior weakens, the contours enlarge for both the SC and marginalization cases. However, the contour size escalates slightly more rapidly with the photo-$z$ priors in the SC case. This indicates that the statistical error in the SC case has a moderately higher dependency on the photo-$z$ quality than in the marginalization case.

In this study, we have not considered the effects of photo-$z$ parameters in obtaining the final results however they will be marginalized when we will perform cosmological inference using the SC method being applied to real data. With the LSST Y1 level accuracy in the photo-$z$ we expect the errors associated with these will be negligible.  

\subsection{Magnification bias}\label{sec:5.2}
\par\noindent

Gravitational lensing not only warps the shapes of galaxies but also affects their spatial distribution, leading to a phenomenon known as magnification bias or cosmic magnification. In essence, it modifies the observed galaxy density, denoted as $\delta_{\rm g}^{\mathrm{L}} = \delta_{\rm g} + g(F)\kappa$. The function $g(F) = 2(-d \mathrm{ln} N(>F)/d \mathrm{ln} F -1)$, which can be measured in principle, depends on the logarithmic slope of the unlensed galaxy luminosity function $N(>F)$.

Though magnification bias affects the measurement of cosmic shear and galaxy clustering, the effects are not significant in the context of this paper. The magnification bias has a relatively larger effect on galaxy-galaxy lensing and modifies Equation \ref{eq:7} to
\begin{eqnarray}\label{eq:40}
    C_{ii}^{\mathrm{g}\gamma} &=& C_{ii}^{\mathrm{gG}} + C_{ii}^{\mathrm{gI}} + g_i(C_{ii}^{\mathrm{GG}} + C_{ii}^{\mathrm{IG}}) \\\nonumber
    &=& \big[C_{ii}^{\mathrm{gG}} + g_i C_{ii}^{\mathrm{IG}} \big] + \big[C_{ii}^{\mathrm{gI}} + g_i C_{ii}^{\mathrm{GG}} \big]
\end{eqnarray}
where $g_i$ is the averaged $g(F)$ over galaxies in the $i$th redshift bin, and $g(F)$ is of order unity. However, since it changes sign from the bright end of the luminosity function to the faint end, we expect the averaged $g_i<1$ for sufficiently deep surveys.

Our goal is to measure $C_{ii}^{\mathrm{Ig}}$ or $C_{ii}^{\mathrm{gI}}$ with new contamination from the magnification bias. We can apply the same weighting of the estimator Equation \ref{eq:21} here. On the one hand, both $C_{ii}^{\mathrm{Ig}}$ and $C_{ii}^{\mathrm{GG}}$ are unchanged by this weighting. Both $C_{ii}^{\mathrm{IG}}$ and $C_{ii}^{\mathrm{Gg}}$ are reduced by virtually the same $1-Q$. These behaviours mean that the estimator Equation \ref{eq:21} eliminates the combination $C_{ii}^{\mathrm{Gg}} +g_i C_{ii}^{\mathrm{IG}}$ and measures the combination $C_{ii}^{\mathrm{Ig}} + g_i C_{ii}^{\mathrm{GG}}$ in which the term $g_i C_{ii}^{\mathrm{GG}}$ ii contaminates the $I-g$
measurement.

We cannot remove this contamination with any certainty due to measurement errors on $g_i$, $C_{ii}^{\mathrm{GG}}$. The direct estimation of the errors involved is lengthy, so we will instead determine the accuracy to which these measurements must be made for the contribution due to magnification bias to be negligible concerning other errors in the IG SC(\cite{Zhang_2010b}, \cite{Troxel_2012a}). 

If $g_i$ and $C_{ii}^{\mathrm{GG}}$ measurements have error $\delta g_i$ and $\delta C_{ii}^{\mathrm{GG}}$, respectively, then the induced fractional error in the $C_{ij}^{\mathrm{GG}}$ measurement is

\begin{eqnarray}\label{eq:41}
    \frac{W_{ij} \Delta_i}{R_i} \frac{\delta g_i C_{ii}^{\mathrm{GG}} + g_i \delta C_{ii}^{\mathrm{GG}}}{C_{ij}^{\mathrm{GG}}} < \frac{W_{ij} \Delta_i}{R_i} \bigg( |\delta g_i| + \bigg| g_i \frac{\delta C_{ii}^{\mathrm{GG}}}{C_{ii}^{\mathrm{GG}}} \bigg | \bigg) 
    = O(10^{-3}) \bigg( \bigg| \frac{\delta g_i}{0.1}\bigg | + \bigg| \frac{g_i \delta C_{ii}^{\mathrm{GG}}/C_{ii}^{\mathrm{GG}}}{10\%}\bigg | \bigg)
\end{eqnarray}

The relationship outlined above holds under specific conditions: $C_{ii}^{\mathrm{GG}} < C_{ij}^{\mathrm{GG}}$ (where $i<j$), $R_i = O(1)$, and $W_{ij}\delta_i = O(10^{-2})$. 

For a $g_i$ which is large enough to be non-negligible, we need only require accuracy in its measurement of $\delta g_i = 0.1$ and measurement accuracy $C_{ii}^{\mathrm{GG}}$ of 10
per cent the magnification bias induced error will be $O(10^{-3})$, which is safely negligible by a factor of 10 compared to the minimum measurement error $e_{ij}^{\mathrm{min}}$ in the cosmic shear or to the other residual errors of the SC technique. As discussed by \cite{Zhang_2010b}, this level of accuracy can likely be accomplished by direct measurement of $g_i$ under the approximation $C_{ii}^{\mathrm{\gamma \gamma}} \approx C_{ii}^{\mathrm{GG}}$, if the lensing contamination $C_{ii}^{\mathrm{II}}$ < 10 per cent. However, if the II contamination is greater than 10 per cent of the lensing signal, more detailed methods must be employed to achieve a great enough accuracy in the $g_i$ measurement for it to be safely negligible, some of which are discussed by \cite{Zhang_2010b}.

\begin{figure*}
    \centering
    \includegraphics[width=1.0\linewidth, height=8.cm]{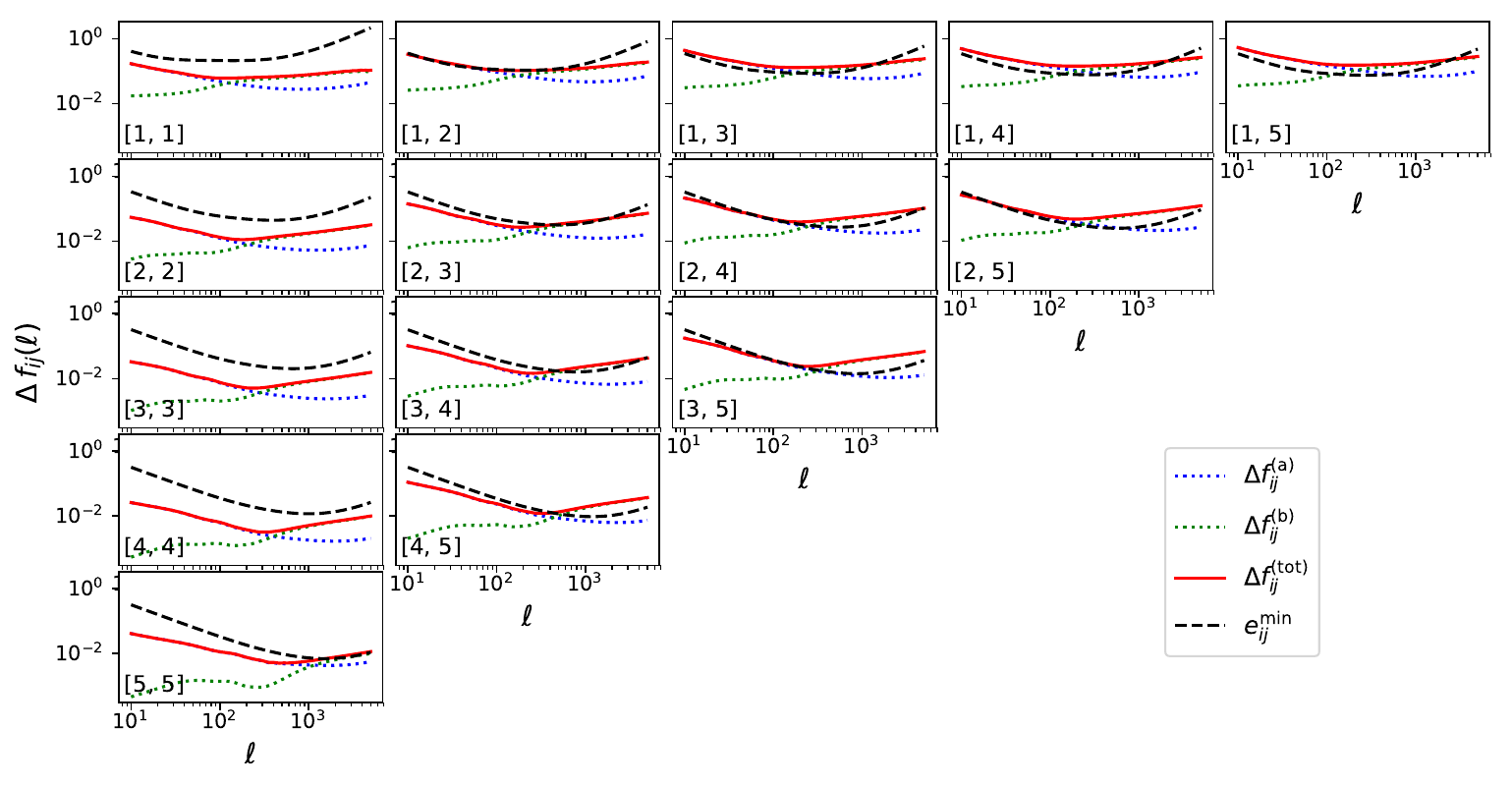}
    \caption{This figure represents the comparison between the residual statistical errors and minimum fractional error in the measurement of $C_{ij}^{\mathrm{GG}}$. The black dashed curves correspond to the minimum fractional statistical error $e_{ij}^{\text{min}}$ which is caused by the cosmic variance in the lensing field and the random shape fluctuation in the $C_{ij}^{\mathrm{GG}}$ measurement. The blue and green dotted curves represent the $\Delta f_{ij}^{\mathrm{(a)}}$ and $\Delta f_{ij}^{\mathrm{(b)}}$, respectively. The red curves correspond to $\Delta f_{ij}^{\mathrm{(tot)}} = \sqrt{\big(\Delta f_{ij}^{\mathrm{(a)}}\big)^2 + \big(\Delta f_{ij}^{\mathrm{(b)}}\big)^2}$. 
    } 
    \label{fig:9}
\end{figure*}

\section{Uncertainty in the accuracy of scaling relation}\label{sec:6}
\par\noindent

The choices and uncertainties in the IA \& galaxy bias parameters and cosmological parameters propagate through various components of the scaling relation which might affect the performance of the SC. So we explore the possible dependencies of the accuracy of the scaling relation on the uncertainties of these parameters, in this section.

\subsection{Uncertainties in IA parameters} \label{sec:6.1}
\par\noindent

The SC techniques require the evaluation of $R_{i}(\ell)$ which varies with the choices of the IA parameters. The dependencies on the IA parameters in determining $R_{i}(\ell)$ have been studied in Section \ref{sec:3.2}. We want to obtain these parameters by fitting the TATT model and nonlinear galaxy bias with the observed galaxy-clustering and gI correlations. There is a possibility of a mismatch between the obtained values of any parameter from its true value due to some systematic errors or computational errors, etc., in our pipeline. We assumed that these uncertainties should not be more than 20\% for IA amplitude parameters ($A_1\ \&\ A_2$) and 10\% for the bias parameter $b_1$ and not more than 50\% for the other IA parameters($\eta_1\ \&\ \eta_2$).

The accuracy of the scaling relation has been calculated when the uncertainties in IA and bias parameters, propagated through $R_i(\ell)$ are taken into account. The numerical results are shown in Figure \ref{fig:8}. The upper left and the upper right plot correspond to when the uncertainties in the IA amplitude ($A_1 \ \&\ A_2$) and the other two IA parameters ($\eta_1 \ \& \ \eta_2$) are considered. The lower plot corresponds to when the uncertainties in the bias parameter ($b_1$) are considered. Different colours represent the accuracy of the scaling relation for a particular amount of uncertainties in the IA/bias parameters which are mentioned in the legend.


Though it is well evident that when the uncertainties in IA and bias parameters are propagated through $R_i(\ell)$ into the scaling relation it can cause significant deviations in $\varepsilon_{ij}$ yet the values of $\varepsilon_{ij}$ do not exceed 0.2 for all possible combinations of redshift bins except the auto-correlating bin pairs. Hence, the detection of the IG correlation by a factor of 5 or better is still achievable when a moderate amount of uncertainties in both the IA and bias parameters are considered.

\subsection{Uncertainties in cosmological parameters}\label{sec:6.2}
\par\noindent

The SC techniques require the evaluation of $W_{ij}$ and $Q$. Both evaluations involve the cosmology-dependent lensing kernel $W_{\mathrm{L}}(z_{\mathrm{L}}, z_{\mathrm{G}}) \propto \Omega_{\rm m} (1+z_{\mathrm{L}})\chi_{\mathrm{L}}(1 - \chi_{\mathrm{L}}/\chi_{\mathrm{G}})$. Fortunately, the cosmological parameters have already been measured to $1\%$ accuracy by \cite{Planck_2018}. So, the uncertainties in cosmology can at most bias the SC at a percent level of accuracy, which is negligible compared to the identified $\sim 5-10\%$ error in the scaling relation in Section \ref{sec:4.3}.

\begin{figure*}
    \centering
    \includegraphics[width=1.0\linewidth, height=8.cm]{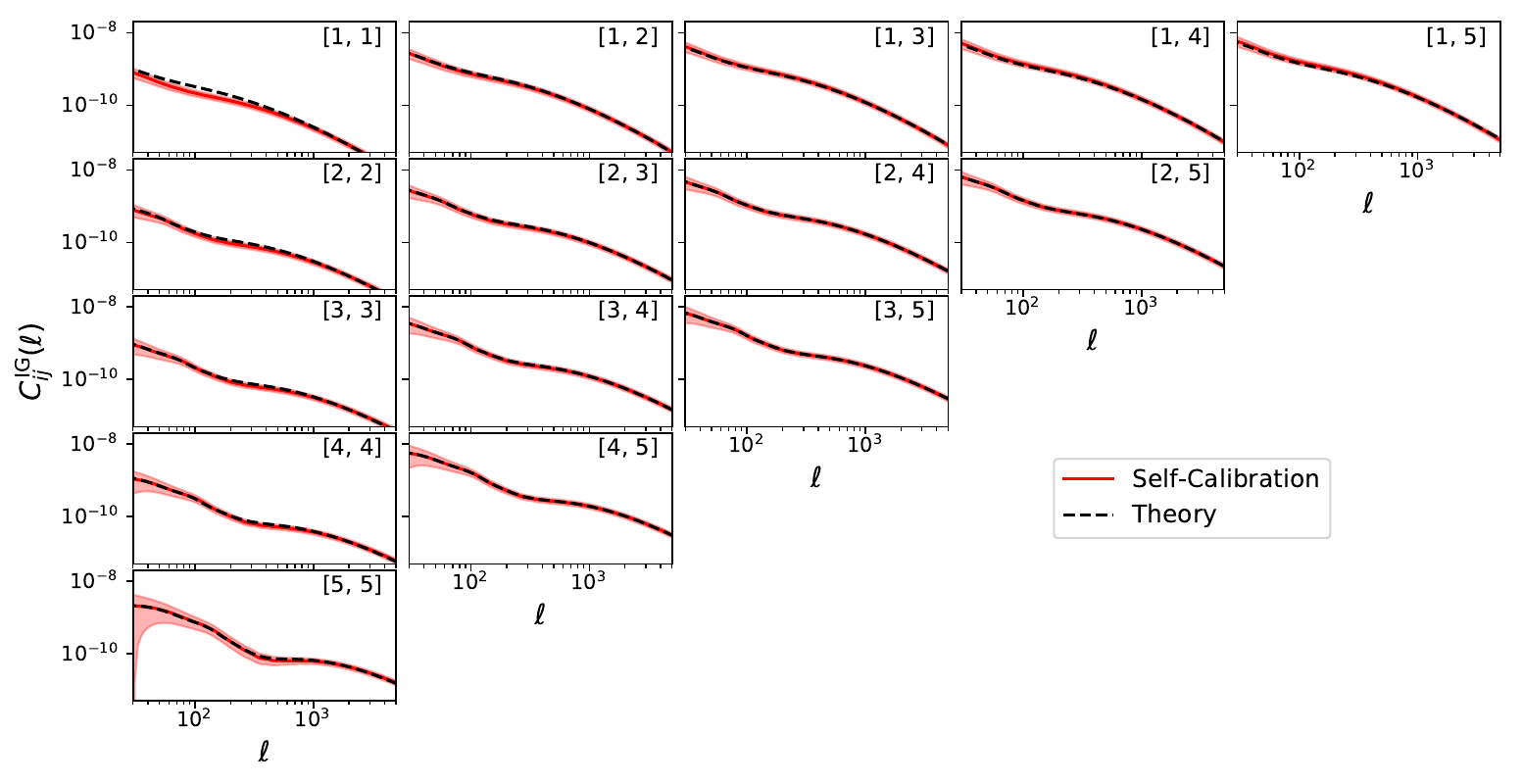}
    \caption{This figure represents the final results of the SC method. The red solid curves and the black dashed curves represent the IG correlation determined through SC and the predicted value from the theory. The shaded regions represent the error bars computed through the error propagation relation (\ref{eq:42}).  } 
    \label{fig:10}
\end{figure*}

\section{Summary of the residual errors \& error propagation}\label{sec:7}
\par\noindent

We discuss whether the residual statistical error associated with the measurement error in $C^{\mathrm{Ig}}_{ii}$ or the residual systematic error associated with the accuracy of the scaling relation dominates the error budget of the SC method. 

If $f_{ij}^{\mathrm{I}} < f^{\text{thresh}}_{ij}$,  the IA is too weak to be detected in the galaxy galaxy lensing correlation. Thus, the SC technique is not applicable. However, this also usually means the IA is negligible in the cosmic shear measurement. Thus, there is no need to correct the IG contamination in this case. However, there are important exceptions to the above conclusion. When one photo-z bin is at sufficiently low redshift, the IG contamination is undetectable by our method, but the systematic error it induces is non-negligible. In these cases, other methods (e.g., \cite{Joachimi_2008}, \cite{Joachimi_2009}, \cite{Zhang_2010b}) shall be applied to correct for IA.

SC is applicable when $f_{ij}^{\mathrm{I}} > f^{\text{thresh}}_{ij}$. If $f_{ij}^{\mathrm{I}} < \Delta f_{ij}^{(\mathrm{a})}/ \varepsilon_{ij}$, the residual statistical error induced by the uncertainty in $C_{ii}^{\mathrm{Ig}}$ measurement dominates. However, this residual error, $\Delta f_{ij}^{(\mathrm{a})} = f_{ij}^{\text{thresh}}$, is usually negligible, compared to the minimum statistical error $e^{\mathrm{min}}_{ij}$ in shear measurement ($\Delta f_{ij} < e^{\text{min}}_{ij}$). If $f_{ij}^{\mathrm{I}} > \Delta f_{ij}/ \varepsilon_{ij}$, the residual systematic error arising from the imperfect scaling relation dominates. The fractional residual error in the lensing–lensing measurement is $\delta f_{ij} = \varepsilon_{ij} f^{\mathrm{I}}_{ij} \approx 0.1f^{\mathrm{I}}_{ij}$. This error will still be sub-dominant to the lensing statistical fluctuation if $f^{\mathrm{I}}_{ij} < e^{\text{min}}_{ij}/\varepsilon_{ij}$. If this is not the case, SC can work to suppress the IG contamination by a factor of 10.

Figure \ref{fig:10} represents the comparison between the residual statistical errors and minimum fractional error in the measurement of $C_{ij}^{\mathrm{GG}}$. The black dashed curves correspond to the minimum fractional statistical error $e_{ij}^{\text{min}}$ which is caused by the cosmic variance in the lensing field and the random shape fluctuation in the $C_{ij}^{\mathrm{GG}}$ measurement. The blue and green dotted curves represent the $\Delta f_{ij}^{\mathrm{(a)}}$ and $\Delta f_{ij}^{\mathrm{(b)}}$, respectively. The red curves correspond to $\Delta f_{ij}^{\mathrm{(tot)}} = \sqrt{(\Delta f_{ij}^{\mathrm{(a)}})^2 + (\Delta f_{ij}^{\mathrm{(b)}})^2}$. We observe that $\Delta f_{ij}^{\mathrm{(a)}} < e_{ij}^{\text{min}}$ for all combinations of redshift bins. $\Delta f_{ij}^{\mathrm{(b)}}$ corresponds to the 10\% uncertainty in bias parameter and we observe $\Delta f_{ij}^{\mathrm{(b)}} \sim e_{ij}^{\text{min}}$ for all combinations of redshift bins. As the bias parameter can be obtained with much better accuracy than 10\%, it is safe to assume that $\Delta f_{ij}^{\mathrm{(tot)}} < e_{ij}^{\text{min}}$. Hence, the residual errors do not have a significant impact in degrading the accuracy of the SC method.   

The measurement error in $C_{ii}^{\mathrm{Ig}}$, i.e., $\Delta C_{ii}^{\mathrm{Ig}}$ and the uncertainty in $R_i$ propagate through the scaling relation \ref{eq:33} and contribute to the error budget of the measurement of $C_{ii}^{\mathrm{IG}}$. The error in $C_{ii}^{\mathrm{IG}}$ resulting from the SC method can be quantified through the equation
\begin{eqnarray}\label{eq:42}
    \Delta C_{ij}^{\mathrm{IG}} = C_{ij}^{\mathrm{IG}} \times \bigg[ \bigg(\frac{\Delta C_{ii}^{\mathrm{Ig}}}{C_{ii}^{\mathrm{Ig}}}\bigg)^2 + \bigg(\frac{\Delta R_i}{R_i}\bigg)^2\bigg]
\end{eqnarray}
where we considered that the uncertainty in $R_i$ is only arising from the uncertainty in the galaxy bias parameter. The numerical results are shown in Figure \ref{fig:10}. The solid red curves and the dashed black curves represent the measurement of $C_{ii}^{\mathrm{IG}}$ using SC and its theoretical value, respectively for all possible combinations of redshift bin pairs. The 1$\sigma$ contours are also shown for all of these cases.

\section{Conclusion and Discussion}\label{sec:8}
\par\noindent

The SC technique for the detection and the removal of the IG correlation from lensing surveys consists of two key ingredients. (1) It extracts the galaxy number density-galaxy intrinsic ellipticity correlation from the observed galaxy–galaxy lensing correlation from the same redshift bin in a given lensing survey with photo-$z$ measurement, i.e. from the second lensing observable ($C_{ii}^{\mathrm{g}\gamma}$). (2) It converts this Ig (or gI) measurement into a measure of the IG correlation through a generic scaling relation. The SC technique only has a moderate requirement for photo-$z$ accuracy and results in a little loss of cosmological information. 

We tailored a SC method that can extract and mitigate the IG contamination from the lensing power spectrum in the nonlinear regime. To be more specific, the IG contamination can be extracted at the level of nonlinear galaxy bias and TATT. In the original derivation of the SC scaling relation, which was developed for the linear regime, the ratio $P_{\mathrm{gI}}/P_{\mathrm{mI}}$ was considered to be the linear galaxy bias. This is generally true irrespective of the IA model if one uses the linear galaxy bias model. But, in the nonlinear regime, the simplified approximation doesn't hold. So, we calculate the exact ratio $P_{\mathrm{gI}}/P_{\mathrm{mI}}$ assuming the TATT model for the IA of galaxies and the non-linear galaxy bias model which contains the matter density field and the tidal field up to the second order. We only use this theoretical ratio in the scaling relation to obtain the IG correlation from the observed Ig correlation. 

The scaling relation for the non-adjacent bin pairs ($i, j>i+1$) is accurate within $2.5\%$ and $10\%$ without and with bin 1, respectively. It signifies that IG contamination can be suppressed by a factor of $40$ and $10$ or larger for the bin pairs without and with bin 1, respectively if other errors are negligible. For the adjacent bin pairs ($i, j=i+1$), there is a drop in accuracy and the scaling relation for such bin pairs is accurate within $10\%$ which signifies that IG contamination can be suppressed by a factor of $10$ or larger if other errors are negligible. For the auto-correlating bins ($i, j=i$), there is a drop in accuracy and $|\varepsilon_{ij}|$ for such bin pairs are within $20\%$ (above $\ell=20$) which signifies that IG contamination can be suppressed by a factor of $5$. The SC is more accurate if the bin separation is higher. The lower strength of the shear signal in the bins with lower redshift bins results in lower accuracy of the scaling relation.

We observe that $\Delta f_{ij}^{\mathrm{(a)}} < e_{ij}^{\text{min}}$ for all combinations of redshift bins. $\Delta f_{ij}^{\mathrm{(b)}}$ corresponds to the 10\% uncertainty in the bias parameter and we observe $\Delta f_{ij}^{\mathrm{(b)}} \sim e_{ij}^{\text{min}}$ for all combinations of redshift bins. As the bias parameter can be obtained with much better accuracy than 10\%, it is safe to assume that $\Delta f_{ij}^{\mathrm{(tot)}} < e_{ij}^{\text{min}}$. Hence, the residual errors do not have a significant impact in degrading the accuracy of the SC method.   

We also observe that the residual systematic error associated with the scaling relation is much lower for the bin pairs with higher redshifts. The residual statistical uncertainty $\Delta f_{ij}^{(\mathrm{a})}$ is slightly lower for the tomographic bins with higher redshift compared to the bins with lower redshift. Hence, the errors associated with the SC method are significantly lower for the higher redshift bins. Thus the IG contamination can be extracted more precisely or accurately for the bin pairs with higher redshift and also for the non-adjacent ones. For all possible combinations of the redshift bins, we observe $\Delta f_{ij}^{(\mathrm{a})}$ is smaller than the lower limit of the fractional measurement error in $C_{ij}^{\mathrm{GG}}$, $e^{\text{min}}_{ij}$. This implies the residual error after the SC will be negligible, which results in little cosmological information loss. 


The uncertainties in the IA and galaxy bias parameters have a small impact on the accuracy of the scaling relation. We consider the uncertainties are 20\% for IA amplitude parameters ($A_1\ \&\ A_2$) and 10\% for the bias parameter $b_1$ and not more than 50\% for the other IA parameters($\eta_1\ \&\ \eta_2$. 
When uncertainties in these parameters are studied in a group, keeping the other parameters unchanged, we observe a little drop in the accuracy of the scaling relation. Even after considering these uncertainties, we show that the scaling relation is accurate within 20\% for all choices of bin combinations except the auto-correlating bins.

Similar to the two-point IA correlations, the three-point correlations, IGG, IIG, and III) cause a serious contaminant to the 3-point cosmic shear measurements. The SC for these three-point IA correlations was first proposed by \cite{Troxel_2012a}, \cite{Troxel_2012b}. In a recent study \cite{Pyne_2021}, the SC technique has been applied for both the IA power spectrum and the IA bispectrum for the Euclid-like surveys. We will revisit the SC of the three-point correlation in a nonlinear regime in the near future.  

The proposed SC method is also applicable to the nonlinear models, e.g., the halo alignment model and the effective field theory (EFT) model of IA. The ratio $P_{\mathrm{gI}}/P_{\mathrm{mI}}$ can be determined more accurately if we consider a more general model of galaxy bias. Not only this method will benefit upcoming Stage-IV imaging surveys such as Rubin LSST, Roman etc., but also the \texttt{FAST-PT} implementation of the gI correlation can be used for 3x2 point analysis in nonlinear regime.

\section*{Acknowledgements}
\par\noindent

This paper has undergone internal review in the LSST Dark Energy Science Collaboration. The internal reviewers were Jonathan Blazek and Christos Georgiou and we thank them for their valuable comments and suggestions. The DESC acknowledges ongoing support from the Institut National de 
Physique Nucl\'eaire et de Physique des Particules in France; the 
Science \& Technology Facilities Council in the United Kingdom; and the
Department of Energy and the LSST Discovery Alliance
in the United States.  DESC uses resources of the IN2P3 
Computing Center (CC-IN2P3--Lyon/Villeurbanne - France) funded by the 
Centre National de la Recherche Scientifique; the National Energy 
Research Scientific Computing Center, a DOE Office of Science User 
Facility supported by the Office of Science of the U.S.\ Department of
Energy under Contract No.\ DE-AC02-05CH11231; STFC DiRAC HPC Facilities, 
funded by UK BEIS National E-infrastructure capital grants; and the UK 
particle physics grid, supported by the GridPP Collaboration.  This 
work was performed in part under DOE Contract DE-AC02-76SF00515.

We thank Cristhian Garcia-Quintero for his valuable suggestions. We thank Jonathan Blazek for cross-checking each term of the gI correlation, computed independently using in \texttt{FAST-PT} during the ECHO-IA LILAC workshop held in May 2024 at the Center for Astrophysics (CFA) at Harvard University.

We acknowledge that this material is based upon work supported in part by the Department of Energy, Office of Science, under Award Numbers DE-SC0022184 and DE-SC0024787 and also in part by the U.S. National Science Foundation under grants AST2327245 and AST2206563.

\section*{Author contributions}
\par\noindent

Avijit Bera is the lead author of the paper who carried out most of the analyses and wrote the paper. Leonel Medina-Varela contributed by coding, especially the $\eta$ integrals to the main analysis pipeline. He helped with paper writing and reviewing, and also with discussion and guidance of the main analysis. Vinu Sooriyaarachchi contributed to the first implementation of the gI correlation in \texttt{FAST-PT}. Mustapha Ishak set the initial plan of the project. He mentored and coordinated the work throughout the project's early to the final stages. He also discussed the final result, and mentored during the paper writing. Carter Williams assisted in calculating and validating each term of gI cross-correlation.


\bibliography{citation}{}
\bibliographystyle{plain}

\appendix

\section{Modeling galaxy number density-galaxy intrinsic ellipticity power spectrum in nonlinear regime}\label{AppA}
\par\noindent

We calculate the number density - galaxy intrinsic ellipticity (gI) power spectrum (PS) up to one loop order for TA and TT models separately and then we will consider the complete model using \texttt{FAST-PT} (\cite{McEwen_2016}, \cite{Fang_2017}). We only consider the E mode contribution of intrinsic ellipticity to calculate the gI PS as the B mode contribution is suppressed at the leading order. It is inevitable to consider the B mode contribution for II correlations as it is not suppressed in the smallest scale. In evaluating the PS corresponding to the correlations note that we will be separating the cut-off dependent contributions (see Appendix A of \cite{Blazek_2015}) and absorbing them into the IA parameters as will be discussed in \cite{Blazek_2019}. 

The galaxy bias model that includes the second order contribution from matter density and tidal fields is  
\begin{eqnarray}
    \delta_{\rm g}(\mathbf{x})= b_1\delta_{\rm m}(\mathbf{x}) + \frac{b_2}{2}\big( \delta_{\rm m}(\mathbf{x})^2 \big) + \frac{b_s}{2}\big( s(\mathbf{x})^2 \big) + ...,
\end{eqnarray}
where $\delta_{\rm m}$ and $s$ are the nonlinear density and tidal field, respectively. In our study, we do not consider $b_{3\rm NL}$. The matter density field can be expanded in terms of linear matter density and relevant gravity kernels in standard perturbation theory (SPT) \cite{Bernardeau_2002}
\begin{eqnarray}
    \delta_{\rm m} = \delta_{\rm m}^{(1)} + \delta_{\rm m}^{(2)} + \delta_{\rm m}^{(3)} + ...,
\end{eqnarray}
where $\delta_{\rm m}^{(1)}$ is the linear density field; $\delta_{\rm m}^{(2)}$ and $\delta_{\rm m}^{(3)}$ are the second and third-order contributions. It is convenient to work in Fourier space and the power spectrum (PS) are defined as the ensemble average over the Fourier space fields as  
\begin{eqnarray}
    \langle A(\mathbf{k})B(\mathbf{k'})\rangle = (2\pi)^3 \delta_{D}^{(3)}(\mathbf{k} + \mathbf{k'}) P_{AB}(k).
\end{eqnarray}
In Fourier space, $\delta_{\rm m}^{(2)}$ and $\delta_{\rm m}^{(3)}$ are described by
\begin{eqnarray}
    \delta_{\rm m}^{(2)}(\mathbf{k}) &=& \int \frac{d^3\mathbf{k}_1}{(2\pi)^3} F_{2}(\mathbf{k}_1,\mathbf{k}_2) \delta_{\rm m}^{(1)}(\mathbf{k}_1) \delta_{\rm m}^{(1)}(\mathbf{k}_2),\\
    \delta_{\rm m}^{(3)}(\mathbf{k}) &=& \int \frac{d^3\mathbf{k}_1}{(2\pi)^3} \int \frac{d^3\mathbf{k}_2}{(2\pi)^3}  F_{3}(\mathbf{k}_1,\mathbf{k}_2,\mathbf{k}_3) \delta_{\rm m}^{(1)}(\mathbf{k}_1) \delta_{\rm m}^{(1)}(\mathbf{k}_2) \delta_{\rm m}^{(1)}(\mathbf{k}_3),
\end{eqnarray}
where $\mathbf{k}_2 = \mathbf{k} - \mathbf{k}_1$. The second and third-order density kernels are  
\begin{eqnarray}
    F_2(\mathbf{q},\mathbf{k}-\mathbf{q}) &=& \frac{7\mu +3\alpha -10\alpha \mu^2}{14\alpha(1+\alpha^2-2\alpha\mu)},\\
    F_3(\mathbf{k},\mathbf{q},-\mathbf{q}) &=& \frac{1}{3024q^2} \Bigg[\frac{12}{\alpha^2}-158+100\alpha^2-42\alpha^4 + \frac{3}{\alpha^3}(7\alpha^2+2) \mathrm{ln} \bigg(\frac{\alpha+1}{|\alpha-1|} \bigg)\Bigg],
\end{eqnarray}
where $\alpha=q/k$ and $\mu=\hat{k}\cdot\hat{q}$.

In configuration space, a tidal tensor $s_{ij}(\mathbf{x})$ is defined in terms of $\delta_{\rm m}(\mathbf{x})$ 
\begin{eqnarray}
    s_{ij}(\mathbf{x}) = \Big[\nabla_i \nabla_j \nabla^{-2} - \frac{1}{3}\delta_{ij}\Big]\delta_{\rm m}(\mathbf{x}).
\end{eqnarray}

In Fourier space, the normalized tidal tensor is 
\begin{eqnarray}
    s_{ij}(\mathbf{k}) = \Big[\hat{k}_i \hat{k}_j  - \frac{1}{3}\delta_{ij}\Big]\delta_{\rm m}(\mathbf{k}). 
\end{eqnarray}
    
The squared tidal tensor is  
\begin{eqnarray}
    s^2(\mathbf{k}) = \int \frac{d^3\mathbf{k_1}}{(2\pi)^3} S_{2}(\mathbf{k}_1,\mathbf{k}_2) \delta_{\rm m}(\mathbf{k}_1) \delta_{\rm m}(\mathbf{k}_2),
\end{eqnarray}
where $S_{2}(\mathbf{k}_1,\mathbf{k}_2) = \mu_{12}^2-\frac{1}{3}$ and $\mu_{12}=\hat{k}_1\cdot\hat{k}_2$.

Following the TATT formalism, the intrinsic ellipticity or shape of the galaxies can be expanded up to the second order in terms of linear matter density field 
\begin{eqnarray}
    \gamma^{\mathrm{I}}_{ij}(\mathbf{x}) = C_1s_{ij} + C_2\bigg(s_{ik}s_{kj} - \frac{1}{3}\delta_{ij} s^2 \bigg) + C_{1\delta}\delta_{\rm m} s_{ij} + 
    C_{2\delta}\delta_{\rm m}\bigg(s_{ik}s_{kj} - \frac{1}{3}\delta_{ij} s^2 \bigg) + C_{t} t_{ij} + ..,
\end{eqnarray}

In configuration space the components of $\gamma$ are 
\begin{eqnarray}
    (\gamma_+,\gamma_{\times}) = (C_1+C_{1\delta}\delta_{\rm m})(s_{xx}-s_{yy}, 2s_{xy}) +   
    (C_2+C_{2\delta}\delta_{\rm m})(s_{xk}s_{xk}-s_{yk}s_{yk}, 2s_{xk}s_{yk})
    + ...,
\end{eqnarray}

In Fourier space, $\gamma$ can be decomposed into a curl-free (E) and a divergence-free (B) components (\cite{Mackey_2002}, \cite{Kamionkowski_1998})
\begin{eqnarray}
     \gamma_E(\mathbf{k}) &=& p(\hat{k}) ^{-1} \big[f_E(\hat{k}) \gamma_+(\mathbf{k}) + f_B(\hat{k}) \gamma_\times(\mathbf{k}) \big],\\
     \gamma_{\mathrm{B}}(\mathbf{k}) &=& p(\hat{k}) ^{-1} \big[-f_B(\hat{k}) \gamma_+(\mathbf{k}) + f_E(\hat{k}) \gamma_\times(\mathbf{k}) \big],
\end{eqnarray}
where the angular operators are defined as $f_E(\hat{k}) = \hat{k}_x^2-\hat{k}_y^2$, and $f_B(\hat{k}) = 2 \hat{k}_x \hat{k}_y$. The projection operator is defined as $p(\hat{k}) = 1-\hat{k}_z^2$ which removes the unobservable line of sight shear. 

E and B mode decomposition of $\gamma$ are  
\begin{eqnarray}
    \label{gamma_EB}
    \gamma_{(E,B)}(\mathbf{k}) &=& C_1 f_{(E,B)}(\hat{k})\delta_{\rm m}(k) + C_{1\delta} \int \frac{d^3\mathbf{k_1}}{(2\pi)^3} f_{(E,B)}(\hat{k}_1) \delta_{\rm m}(\mathbf{k}_1) \delta_{\rm m}(\mathbf{k}_2) \\\nonumber
    && +\ C_2 \int \frac{d^3\mathbf{k_1}}{(2\pi)^3} h_{(E,B)}(\hat{k}_1,\hat{k}_2) \delta_{\rm m}(\mathbf{k}_1) \delta_{\rm m}(\mathbf{k}_2)\\\nonumber
    && +\ C_{2\delta} \int \frac{d^3\mathbf{k_1}}{(2\pi)^3} h_{(E,B)}(\hat{k}_1,\hat{k}_2) \delta_{\rm m}(\mathbf{k}_1) \delta_{\rm m}(\mathbf{k}_2)\delta_{\rm m}(\mathbf{k}_3),
\end{eqnarray}
where $\mathbf{k}_1+\mathbf{k}_2+\mathbf{k}_3 = \mathbf{k}$, with $\mathbf{k}_3 = 0$ except in the last term.

Equation \ref{gamma_EB} can be also expressed in a compact way 
\begin{eqnarray}
    \gamma_{(E,B)}(\mathbf{k}) = C_1 f_{(E,B)}(\delta_{\rm m}) + C_{1\delta} \delta_{\rm m} f_{(E,B)}(\delta_{\rm m}) + C_2 h_{(E,B)}(\delta_{\rm m},\delta_{\rm m})
    +C_{2\delta} \delta_{\rm m} h_{(E,B)}(\delta_{\rm m},\delta_{\rm m}),
\end{eqnarray}
where the angular operators are defined as 
\begin{eqnarray}
    h_E(\hat{u},\hat{v}) &=& \hat{u}\cdot\hat{v} (\hat{u}_x\hat{v}_x-\hat{u}_y\hat{v}_y) -\frac{1}{3}(\hat{u}_x^2 + \hat{v}_x^2 - \hat{u}_y^2 - \hat{v}_y^2),\\\nonumber
    h_B(\hat{u},\hat{v}) &=& \hat{u}\cdot\hat{v} (\hat{u}_x\hat{v}_y + \hat{u}_y\hat{v}_x) -\frac{2}{3}(\hat{u}_x\hat{u}_y + \hat{v}_x\hat{v}_y),
\end{eqnarray}

and the projection operator $p(\hat{u}) = 1-\hat{u}_z^2$ removes the unobservable line of sight shear.

\subsection{Tidal alignment}
\par\noindent

Tidal alignment contribution to  the gI correlation $\langle \delta_{\mathrm{g}} |\gamma^{\rm I}_{\rm ta} \rangle$ up to one loop order
\begin{eqnarray}
    \langle \delta_{\rm g} | \gamma^{\rm I}_{\mathrm{ta}} \rangle &=& 
    b_1 C_1 \langle \delta_{\rm m}| f_E(\delta_{\rm m}) \rangle + 
    b_1 C_{1\delta} \langle \delta_{\rm m}| \delta_{\rm m} f_E(\delta_{\rm m}) \rangle +
    \frac{b_2}{2}C_1  \langle \delta_{\rm m}^2| f_E(\delta_{\rm m}) \rangle\\\nonumber 
    && +\ \frac{b_2}{2} C_{1\delta}\langle \delta_{\rm m}^2| \delta_{\rm m} f_E(\delta_{\rm m}) \rangle + 
    \frac{b_{\mathrm{s}}}{2}C_{1\delta} \langle s^2| f_E(\delta_{\rm m}) \rangle + 
    \frac{b_{\mathrm{s}}}{2}C_{1\delta} \langle s^2| \delta_{\rm m} f_E(\delta_{\rm m}) \rangle.
\end{eqnarray}

This correlation corresponds to the gI PS
\begin{eqnarray}
    \langle \delta_{\rm g}|\gamma^{\rm I}_{\mathrm{ta}} \rangle = (2\pi)^3 P_{\rm gI}|_{\rm ta}(k) \delta^3_D(\mathbf{k}+\mathbf{k}'),
\end{eqnarray}
where $P_{\rm gI}|_{\mathrm{ta}}(k) $ is the tidal alignment contribution of gI PS and expanding this up to one loop order term
\begin{eqnarray}\nonumber
   P_{\rm gI}|_{\mathrm{ta}}(k) &=&  b_1 C_1 p(\hat{k}) P_\delta(k) + b_1 C_{1\delta} P_{0|0E}(k,\mu_k)  + \frac{b_2}{2}C_1 p(\hat{k}) P_{00|E}(k,\mu_k) \\
    && +\ \frac{b_2}{2} C_{1\delta} P_{00|0E}(k,\mu_k) +
     \frac{b_{\mathrm{s}}}{2}C_{1\delta} P_{SS|E}(k,\mu_k) + 
     \frac{b_{\mathrm{s}}}{2}C_{1\delta} P_{SS|0E}(k,\mu_k).
\end{eqnarray}

The detailed explanation of each of these terms are shown in below 
\begin{itemize}
    \item $\langle \delta_{\rm m}|f_E(\delta_{\rm m}) \rangle$
    
    Expanding up to one loop order term, we get 
    \begin{eqnarray}\label{eqA1}
    \langle \delta_{\rm m}|f_E(\delta_{\rm m}) \rangle = 
    \langle \delta_{\rm m}^{(1)}| f_E(\delta_{\rm m}^{(1)}) \rangle + 
    \langle \delta_{\rm m}^{(2)}| f_E(\delta_{\rm m}^{(2)}) \rangle +
    2\langle \delta_{\rm m}^{(1)}|f_E(\delta_{\rm m}^{(3)}) \rangle,
\end{eqnarray}
and the corresponding PS is
\begin{eqnarray}
    \langle \delta_{\rm m}| f_E(\delta_{\rm m}) \rangle = (2\pi)^3 p(\hat{k}) P_{\delta}(k) \delta^3_D(\mathbf{k}+\mathbf{k}'),
\end{eqnarray}
where $P_{\delta}(k)$ is the matter PS and up to one loop order, the corresponding PS is  
\begin{eqnarray}
    P_{\delta}(k) = P_{\rm L}(k) + 2I_{00}(k) + 6k^2 J_{00}(k) P_{\rm L}(k),
\end{eqnarray}
where $P_{\rm L}(k)$ is the linear matter PS and defined as $\langle \delta_{\rm m}^{(1)}| \delta_{\rm m}^{(1)}\rangle = (2\pi)^3  P_{\rm L}(k) \delta^3_D(\mathbf{k}+\mathbf{k}')$, and 
\begin{eqnarray}\label{eqA4}
    I_{00}(k) &=& \int \frac{d^3q}{(2\pi)^3} P_{\rm L}(q)P_{\rm L}(|\mathbf{k}-\mathbf{q}|) F_2^2(\mathbf{k}-\mathbf{q}, \mathbf{q}),\\\label{eqA5}
    J_{00}(k) &=& \frac{1}{k^2}\int \frac{d^3q}{(2\pi)^3} P_{\rm L}(q) F_3(\mathbf{k},\mathbf{q}, -\mathbf{q}).
\end{eqnarray}
    \item $\langle \delta_{\rm m}|\delta_{\rm m} f_{E}(\delta_{\rm m}) \rangle$
    
    Expanding up to one loop order term, we get
    \begin{eqnarray}
    \langle \delta_{\rm m}|\delta_{\rm m} f_{E}(\delta_{\rm m}) \rangle = \langle \delta_{\rm m}^{(2)}|\delta_{\rm m}^{(1)} f_{E}(\delta_{\rm m}^{(1)}) \rangle + 
    \langle \delta_{\rm m}^{(1)}|\delta_{\rm m}^{(2)} f_{E}(\delta_{\rm m}^{(1)}) \rangle +
    \langle \delta_{\rm m}^{(1)}|\delta_{\rm m}^{(1)} f_{E}(\delta_{\rm m}^{(2)}) \rangle,
\end{eqnarray}
and the corresponding PS is
\begin{eqnarray}\label{eq:A.28}
    P_{0|0E}(k,\mu_k) = A_{0|0E}(k,\mu_k) + B_{0|0E}(k,\mu_k) + C_{0|0E}(k,\mu_k),
\end{eqnarray}
where 
\begin{eqnarray}\label{eq:A.29}
    A_{0|0E}(k,\mu_k) &=&  2 \int \frac{d^3q}{(2\pi)^3} f_E(\hat{q}) F_2(\mathbf{q}_2,\mathbf{q}) P_{\rm L}(q)P_{\rm L}(q_2),  \\\label{eq:A.30}
    B_{0|0E}(k,\mu_k) &=&  2 P_{\rm L}(k) \int \frac{d^3q}{(2\pi)^3} f_E(\hat{q}) F_2(-\mathbf{k}, \mathbf{q}) P_{\rm L}(q), \\\label{eq:A.31}
    C_{0|0E}(k,\mu_k) &=& 2 P_{\rm L}(k) \int \frac{d^3q}{(2\pi)^3} f_E(\hat{q}_2) F_2(-\mathbf{k}, \mathbf{q}) P_{\rm L}(q),   
\end{eqnarray}
where $\mathbf{q}_2 = \mathbf{k}-\mathbf{q}$. $ B_{0|0E}(k,\mu_k)$ and $ C_{0|0E}(k,\mu_k)$ can be simplified as  
\begin{eqnarray}\label{eq:A.32}
    B_{0|0E}(k,\mu_k) &=&  \bigg( \frac{8}{105} \sigma^2 p(\hat{k}) P_{\rm L}(k) \bigg),\\ \nonumber
    C_{0|0E}(k,\mu_k) &=& \bigg( \frac{10}{21} \sigma^2 p(\hat{k}) P_{\rm L}(k) \bigg) + 2 P_{\rm L}(k) \int \frac{d^3q}{(2\pi)^3} P_{\rm L}(q) \bigg[ f_E(\hat{q}_2) F_2(-\mathbf{q}, \mathbf{k}) - \frac{5}{21} p(\hat{k})\bigg],  \\
\end{eqnarray}
where $\sigma^2 = \int \frac{d^3q}{(2\pi)^3} P_{\rm L}(q)$. In the last integral, we separated the $k\rightarrow 0$ contribution to the integral and placed it in bracket. This contribution is similar to $ B_{0|0E}(k,\mu_k)$ and both of those are proportional to the linear $C_1$ term. $C_1$ parameter can be renormalized by absorbing the cutoff-dependent terms. 

    \item $\langle \delta_{\rm m}^2| f_{E}(\delta_{\rm m}) \rangle$

     Expanding up to one loop order term, we get
    \begin{eqnarray}\label{eq:A.34}
    \langle \delta_{\rm m}^2| f_{E}(\delta_{\rm m}) \rangle = 
     \langle \delta_{\rm m}^{(1)} \cdot \delta_{\rm m}^{(1)} | f_{E}(\delta_{\rm m}^{(2)}) \rangle +
    2 \langle \delta_{\rm m}^{(2)} \cdot \delta_{\rm m}^{(1)} | f_{E}(\delta_{\rm m}^{(1)}) \rangle, 
\end{eqnarray}
and the corresponding PS is
\begin{eqnarray}\label{eq:A.35}
    P_{00|E}(k,\mu_k) = A_{00|E}(k,\mu_k) + 2B_{00|E}(k,\mu_k),
\end{eqnarray}
where 
\begin{eqnarray}\label{eq:A.36}
    A_{00|E}(k,\mu_k) &=&  2 \int \frac{d^3q}{(2\pi)^3} F_2(\mathbf{q}_2,\mathbf{q}) P_{\rm L}(q)P_{\rm L}(q_2),  \\\label{eq:A.37}
    B_{0|0E}(k,\mu_k) &=&  2 P_{\rm L}(k) \int \frac{d^3q}{(2\pi)^3} F_2(-\mathbf{k}, \mathbf{q}) P_{\rm L}(q).   
\end{eqnarray}

$B_{00|E}(k,\mu_k)$ can be simplified as 
\begin{eqnarray}\label{eq:A.38}
    B_{00|E}(k,\mu_k) = \frac{34}{21} \sigma^2 P_{\rm L}(k).
\end{eqnarray}

    \item $\langle \delta_{\rm m}^2| \delta_{\rm m} f_{E}(\delta_{\rm m}) \rangle$

    Expanding up to one loop order term, we get
    \begin{eqnarray}\label{eq:A.39}
    \langle \delta_{\rm m}^2| \delta_{\rm m} f_{E}(\delta_{\rm m}) \rangle = 
    \langle \delta_{\rm m}^{(1)} \cdot \delta_{\rm m}^{(1)} | \delta_{\rm m}^{(1)} f_{E}(\delta_{\rm m}^{(1)}) \rangle, 
\end{eqnarray}
and the corresponding PS is
\begin{eqnarray}\label{eq:A.40}
    P_{00|0E}(k,\mu_k) =  2 \int \frac{d^3q}{(2\pi)^3} f_E(\hat{q})  P_{\rm L}(q)P_{\rm L}(q_2).
\end{eqnarray}

    \item $\langle s^2| f_{E}(\delta_{\rm m}) \rangle$
    
    Expanding up to one loop order term, we get
    \begin{eqnarray}\label{eq:A.41}
 \langle s^2| f_{E}(\delta_{\rm m}) \rangle=\langle s^{2^{(1)(1)}}| f_{E}(\delta_{\rm m}^{(2)}) \rangle + 2\langle s^{2^{(2)(1)}}| f_{E}(\delta_{\rm m}^{(1)}) \rangle,
    \end{eqnarray}
where we have used the following notations
\begin{eqnarray} \label{eq:A.42}
  s^{2^{(2)(1)}}&=&\int\frac{d^{3}\mathbf{k}}{(2\pi)^{3}}S_{2}(\mathbf{k_{1}},
    \mathbf{k_{2}})\delta^{(2)}_{\rm m}(\mathbf{k_{1}})\delta^{(1)}_{\rm m}(\mathbf{k_{2}}),\\ \label{eq:A.43}
    s^{2^{(1)(1)}}&=&\int\frac{d^{3}\mathbf{k}}{(2\pi)^{3}}S_{2}(\mathbf{k_{1}},
    \mathbf{k_{2}})\delta^{(1)}_{\rm m}(\mathbf{k_{1}})\delta^{(1)}_{\rm m}(\mathbf{k_{2}}).
\end{eqnarray}

Then we have the relevant PS as follows
\begin{eqnarray}\label{eq:A.44}
  P_{SS|E}(k,\mu_k) = A_{SS|E}(k,\mu_k) + B_{SS|E}(k,\mu_k),   
\end{eqnarray}
  where the individual contributions are,
\begin{eqnarray}\label{eq:A.45}
A_{SS|E}(k,\mu_k) &=& 2p(\hat{k})\int\frac{d^{3}q}{(2\pi)^{3}}S_{2}(\mathbf{q_{2},\mathbf{q}})F_{2}(\mathbf{q_{2},\mathbf{q}})P_{\rm L}(q)P_{\rm L}(q_{2}), \\ \label{eq:A.46}
B_{SS|E}(k,\mu_k) &=& 4p(\hat{k})P_{\rm L}(k)\int\frac{d^{3}q}{(2\pi)^{3}}\left[S_{2}(\mathbf{q_{2},\mathbf{q}})F_{2}(\mathbf{k},\mathbf{-q})-\frac{34}{136}p(\hat{k})\right]P_{\rm L}(q).
\end{eqnarray}

    \item $\langle s^2| \delta_{\rm m} f_{E}(\delta_{\rm m}) \rangle$
    
    Expanding up to one loop order term, we get
\begin{eqnarray}\label{eq:A.47}
   \langle s^2| \delta_{\rm m} f_{E}(\delta_{\rm m}) \rangle=\langle s^{2^{(1)(1)}}| \delta_{\rm m}^{(1)} f_{E}(\delta_{\rm m}^{(1)}) \rangle,
\end{eqnarray}
and the corresponding PS is 
\begin{eqnarray}\label{eq:A.48}
    P_{SS|0E}(k,\mu_{k})=2\int\frac{d^{3}q}{(2\pi)^{3}}f_{E}(\hat{q})S_{2}(\mathbf{q_{2},\mathbf{q}})P_{\rm L}(q)P_{\rm L}(q_{2}).
\end{eqnarray}

\end{itemize}

\subsection{Tidal torquing}
\par\noindent

Tidal torquing contribution to the gI correlation $\langle \delta_{\rm g} |\gamma^{\rm I}_{\rm tt} \rangle$  up to one loop order
\begin{eqnarray}\label{eq:A.49}
    \langle \delta_{\rm g} | \gamma^{\rm I}_{\mathrm{tt}} \rangle &=& 
    b_1 C_2 \langle \delta_{\rm m}| h_{E}(\delta_{\rm m},\delta_{\rm m}) \rangle + 
    b_1 C_{2\delta} \langle \delta_{\rm m}| \delta_{\rm m} h_{E}(\delta_{\rm m},\delta_{\rm m}) \rangle +
    \frac{b_2}{2}C_2  \langle \delta_{\rm m}^2| h_{E}(\delta_{\rm m},\delta_{\rm m}) \rangle\\\nonumber 
    && +\ \frac{b_2}{2} C_{2\delta}\langle \delta_{\rm m}^2| \delta_{\rm m} h_{E}(\delta_{\rm m},\delta_{\rm m}) \rangle + 
    \frac{b_s}{2}C_{2\delta} \langle s^2| h_{E}(\delta_{\rm m},\delta_{\rm m}) \rangle + 
    \frac{b_s}{2}C_{2\delta} \langle s^2| \delta_{\rm m} h_{E}(\delta_{\rm m},\delta_{\rm m}) \rangle.
\end{eqnarray}

The corresponding PS is 
\begin{eqnarray}\label{eq:A.50}
    \langle \delta_{\rm g} | \gamma^{\rm I}_{\mathrm{tt}} \rangle = (2\pi)^3 P_{\rm gI}|_{\rm tt}(k) \delta^D(\mathbf{k}+\mathbf{k}'),
\end{eqnarray}
where $P_{\rm gI}|_{\mathrm{tt}}(k)$ is the tidal torquing contribution of gI PS and we will consider up to one loop order.  $\langle \delta_{\rm m}| \delta_{\rm m} h_{E}(\delta_{\rm m},\delta_{\rm m}) \rangle$,  $\langle \delta_{\rm m}^2| \delta_{\rm m} h_{E}(\delta_{\rm m},\delta_{\rm m}) \rangle$ and $\langle s^2| \delta_{\rm m} h_{E}(\delta_{\rm m},\delta_{\rm m}) \rangle$ do not contribute to one loop order, and $P_{\rm gI}|_{\mathrm{tt}}(k)$ can be expressed as 
\begin{eqnarray}\label{eq:A.51}
    P_{\rm gI}|_{\mathrm{tt}}(k) = b_1 C_2 P_{0|E2}(k,\mu_k) +  \frac{b_2}{2}C_2  P_{00|E2}(k,\mu_k) + 
     \frac{b_s}{2}C_{2\delta}  P_{SS|E2}(k,\mu_k).
\end{eqnarray}

The detailed explanation of each of these terms are given below 

\begin{itemize}
    \item $\langle \delta_{\rm m}| h_{E}(\delta_{\rm m},\delta_{\rm m}) \rangle$

    Expanding up to one loop order term, we get
    \begin{eqnarray}\label{eq:A.52}
 \langle \delta_{\rm m}| h_{E}(\delta_{\rm m},\delta_{\rm m}) \rangle=\langle \delta_{\rm m}^{(2)}| h_{E}(\delta_{\rm m}^{(1)},\delta_{\rm m}^{(1)}) \rangle+2\langle \delta_{\rm m}^{(1)}| h_{E}(\delta_{\rm m}^{(2)},\delta_{\rm m}^{(1)}) \rangle,
\end{eqnarray}
and the corresponding PS is 
\begin{eqnarray}\label{eq:A.53}
     P_{0|E2}(k,\mu_k) =  A_{0|E2}(k,\mu_k) +   B_{0|E2}(k,\mu_k),
\end{eqnarray}
where
\begin{eqnarray}\label{eq:A.54}
    A_{0|E2}(k,\mu_k) &=&  2 \int \frac{d^3q}{(2\pi)^3}   P_{\rm L}(q)P_{\rm L}(q_2)F_{2}(\mathbf{q},\mathbf{q_{2}})h_{E}(\hat{q},\hat{q_{2}}),  \\ \label{eq:A.55}
    B_{0|E2}(k,\mu_k) &=&  4 P_{\rm L}(k) \int \frac{d^3q}{(2\pi)^3}P_{\rm L}(q)[F_2( \mathbf{q},-\mathbf{k}) h_E(\hat{q},\hat{q_{2}})-\frac{29}{630}p(\hat{k})]. 
\end{eqnarray}

Note that here the $k\rightarrow0$ contribution to the $ B_{0|E2}$ term, which will be renormalized into the definition of the $C_{1}$ parameter, has explicitly been removed and the same contribution from the $A_{0|E2}$ term is zero.

    \item $\langle \delta_{\rm m}^2 | h_{E}(\delta_{\rm m},\delta_{\rm m}) \rangle$

    Expanding up to one loop order term, we get
\begin{eqnarray}\label{eq:A.56}
 \langle \delta_{\rm m}^2 | h_{E}(\delta_{\rm m},\delta_{\rm m}) \rangle=\langle \delta_{\rm m}^{(1)}\cdot \delta_{\rm m}^{(1)}| h_{E}(\delta_{\rm m}^{(1)},\delta_{\rm m}^{(1)}) \rangle,
\end{eqnarray}
and the corresponding PS is 
\begin{eqnarray}\label{eq:A.57}
    P_{00|E2}=2\int\frac{d^{3}q}{(2\pi)^{3}} h_E(\hat{q},\hat{q_{2}}) P_{\rm L}(q)P_{\rm L}(q_{2}).
\end{eqnarray}

    \item $\langle s^2| h_{E}(\delta_{\rm m},\delta_{\rm m}) \rangle$

    Expanding up to one loop order term, we get
    \begin{eqnarray}\label{eq:A.58}
\langle s^2| h_{E}(\delta_{\rm m},\delta_{\rm m}) \rangle=\langle s^{2^{(1)(1)}}| h_{E}(\delta_{\rm m}^{(1)},\delta_{\rm m}^{(1)}) \rangle,  
\end{eqnarray}
and the corresponding PS is
\begin{eqnarray}\label{eq:A.59}
    P_{SS|E2}=2\int\frac{d^{3}q}{(2\pi)^{3}} h_E(\hat{q},\hat{q_{2}}) S_{2}(\mathbf{q},\mathbf{q_{2}}) P_{\rm L}(q)P_{\rm L}(q_{2}).
\end{eqnarray}

\end{itemize}

\subsection{Tidal alignment and tidal torquing complete model}
\par\noindent

We provide the final expression of the gI correlations up to one loop order for the complete TATT model. We have only considered the E-mode contribution of IA shear to calculate the gI correlations as the B-mode contribution is suppressed in the leading order
\begin{eqnarray}\nonumber
     P_{\rm gI}(k,\mu_k) &=&  b_1 C_1 p(\hat{k}) P_\delta(k) + 
     b_1 C_{1\delta} P_{0|0E}(k,\mu_k)  + 
     \frac{b_2}{2}C_1 p(\hat{k}) P_{00|E}(k,\mu_k) \\\nonumber
     && +\ \frac{b_2}{2} C_{1\delta} P_{00|0E}(k,\mu_k) +
     \frac{b_s}{2}C_{1\delta} P_{SS|E}(k,\mu_k) + 
     \frac{b_s}{2}C_{1\delta} P_{SS|0E}(k,\mu_k) \\
     && +\ b_1 C_2 P_{0|E2}(k,\mu_k) +  \frac{b_2}{2}C_2  P_{00|E2}(k,\mu_k) + 
     \frac{b_s}{2}C_{1\delta}  P_{SS|E2}(k,\mu_k).
     \label{P_gI}
\end{eqnarray}

\begin{figure}
    \centering
        \includegraphics[width=\linewidth]{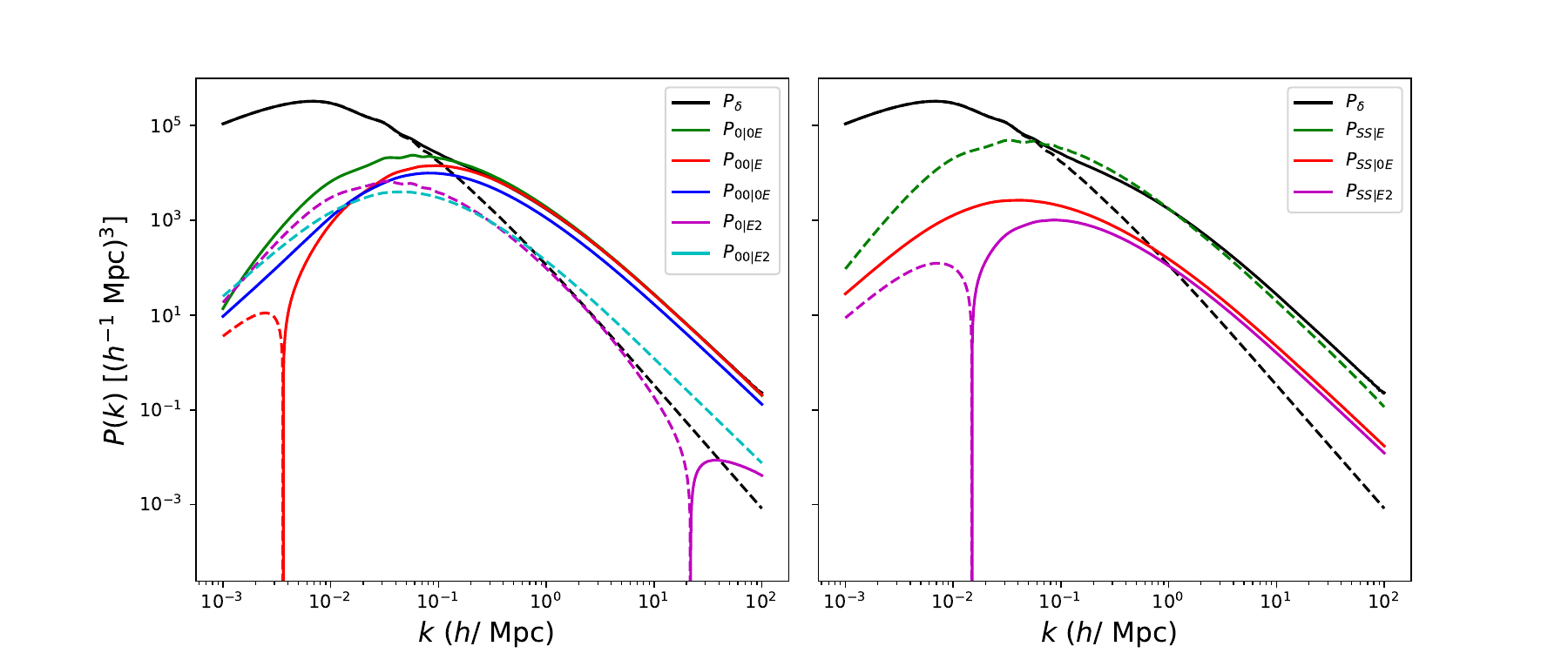}
    \caption{The individual components, evaluated at $z=0$ of the gI PS from is shown. We assume transverse modes ($\mu_k = 0$). Negative values are indicated with dashed lines. Left panel: contributions from matter density field. Right panel: contributions from tidal tensor field. For reference, in both panels the leading tidal alignment contribution $P_{\delta}$ is shown, with the solid line and the dashed line for $P_{\rm L}$.}
    \label{fig:P_gI_full}
\end{figure}

\begin{figure}
    \centering
        \includegraphics[width=0.6\linewidth]{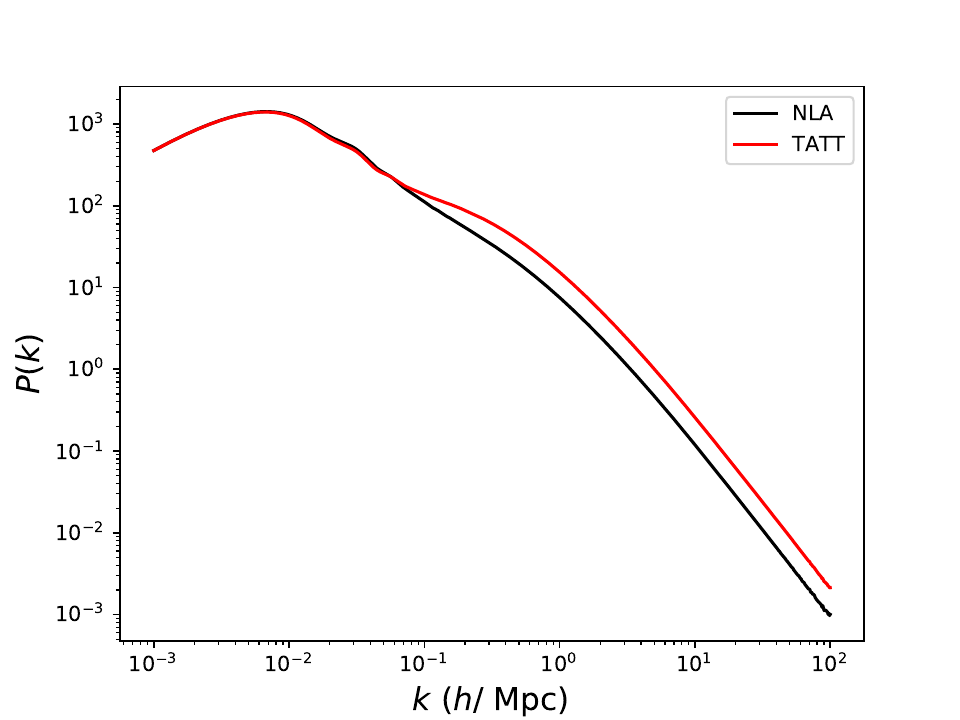}
    \caption{A comparsion plot of gI PS, evaluated at $z=0$ for NLA and TATT model. The pre-factors in Equation \ref{P_gI} are included, with
    $C_1 = C_{1\delta} = -1$ and $C_2 = 5$. We assume transverse modes ($\mu_k = 0$). }
    \label{fig:P_gI_nla_tatt}
\end{figure}

All the individual component of the gI PS are shown in Figure \ref{fig:P_gI_full}. A comparison plot of the $P_{\rm gI}$ for both the nonlinear alignment model (NLA) and the TATT has been shown in Figure \ref{fig:P_gI_nla_tatt}. The pre-factors in Equation \ref{P_gI} are included, with
$C_1 = C_{1\delta} = -1$ and $C_2 = 5$. We assume transverse modes ($\mu_k=0$).

\section{\texttt{FAST-PT} implementation}\label{AppB}
\par\noindent

The \texttt{FAST-PT} implementation by extending the gI PS in nonlinear regime is available on \texttt{GitHub}\footnote{\href{https://github.com/Vinukcs/FAST-PT/blob/ia-nl-bias}{https://github.com/Vinukcs/FAST-PT/blob/ia-nl-bias}}, while the necessary theoretical derivations of the implemented terms are shown below.

\subsection{Derivation of the Legendre Polynomial expansion for each of the correlated terms of the gI power spectrum} \label{AppB1}
\par\noindent

In this subsection, we provide the Legendre Polynomial expansion for each of the correlated terms of the gI PS in terms of the polar angles. Following the convention of \cite{McEwen_2016}, we have $\mu_{12} \equiv \mathbf{q}_1 \cdot \mathbf{q}_2/(q_1q_2) = \hat{\mathbf{q}}_1 \cdot \hat{\mathbf{q}}_2$, which is the cosine of the angle between
$q_1$ and $q_2$. Also, $\mu_{1} \equiv \hat{\mathbf{q}}_1 \cdot \hat{\mathbf{k}}$ and $\mu_{2} \equiv \hat{\mathbf{q}}_2 \cdot \hat{\mathbf{k}}$.

\begin{itemize}

    \item $P_{0|0E}(k,\mu_{k})$

    The $P_{0|0E}(k,\mu_{k})$ term (Equation \ref{eq:A.28}) consists of $A_{0|0E}(k,\mu_{k})$, $B_{0|0E}(k,\mu_{k})$and $C_{0|0E}(k,\mu_{k})$. Both of these terms incorporate the kernel $ f_{E}(\hat{\mathbf{q}}_{1}) F_2(\mathbf{q}_1, \mathbf{q}_2)$ which can be expressed as
    \begin{eqnarray}
     f_{E}(\hat{\mathbf{q}})F_2(\mathbf{q}_{1}, \mathbf{q}_{2}) &=& -\frac{5}{14} - \frac{1}{7}\mu_{12}^2 + \frac{15}{14}\mu_2^2 + \frac{3}{7}\mu_{12}^2\mu_2^2 
     - \frac{1}{4}\frac{q_1}{q_2} \mu_{12}
     + \frac{3}{4}\frac{q_1}{q_2} \mu_{12}\mu_{2}^2 \\ \nonumber
     && - \frac{1}{4}\frac{q_2}{q_1} \mu_{12}
     + \frac{3}{4}\frac{q_2}{q_1} \mu_{12}\mu_{2}^2,
\end{eqnarray}
and its corresponding Legendre polynomials expansion is
\begin{eqnarray}
     f_{E}(\hat{\mathbf{q}})F_2(\mathbf{q}_{1}, \mathbf{q}_{2}) &=& \frac{17}{21}\mathcal{P}_{2}(\mu_{2}) + \frac{1}{2}\frac{q_1}{q_2}\mathcal{P}_{2}(\mu_{2})\mathcal{P}_{1}(\mu_{12}) + \frac{1}{2}\frac{q_2}{q_1}\mathcal{P}_{2}(\mu_{2})\mathcal{P}_{1}(\mu_{12}) \\ \nonumber
     && + \frac{4}{21}\mathcal{P}_{2}(\mu_{2})\mathcal{P}_{2}(\mu_{12}).
\end{eqnarray}

    \item $P_{00|E}(k,\mu_{k})$

    The $P_{00|E}(k,\mu_{k})$ term (Equation \ref{eq:A.35}) consists of $A_{00|E}(k,\mu_{k})$ and $B_{00|E}(k,\mu_{k})$. 
    
    The $A_{00|E}(k,\mu_{k})$ term (Equation \ref{eq:A.36}) contains the scalar kernel $F_2(\mathbf{q}_{1}, \mathbf{q}_{2})$. The corresponding kernel in terms of the angles, as described in \cite{McEwen_2016} is 
\begin{eqnarray}
    F_2(\mathbf{q}_1, \mathbf{q}_2) = \frac{5}{7} + \frac{1}{2}\frac{q_1}{q_2}\mu_{12} + \frac{1}{2}\frac{q_2}{q_1}\mu_{12} + \frac{6}{21}\mu_{12}^2,
\end{eqnarray}
and its corresponding Legendre polynomials expansion is
\begin{eqnarray}
    F_2(\mathbf{q}_1, \mathbf{q}_2) = \frac{17}{21}\mathcal{P}_{0}(\mu_{12}) + \frac{1}{2}\frac{q_1}{q_2}\mathcal{P}_{1}(\mu_{12}) + \frac{1}{2}\frac{q_2}{q_1}\mathcal{P}_{1}(\mu_{12}) + \frac{4}{21}\mathcal{P}_{2}(\mu_{12}).
\end{eqnarray}

The $B_{00|E}(k,\mu_k)$ term (Equation \ref{eq:A.37})can be simplified and absorbed into the definition of bias parameter $b_{1}$ with the bias renormalization (see Appendix A of  \cite{McEwen_2016}).

\item $P_{00|0E}(k,\mu_{k})$

For the $P_{00|0E}(k,\mu_{k})$ term (Equation \ref{eq:A.40}), it is sufficient to expand the kernel $f_{E}(\hat{\mathbf{q}}_{1})$ in terms of a single angle, as described in \cite{McEwen_2016}. This results in the function
\begin{eqnarray}
    f_{E}(\hat{\mathbf{q}}) = \frac{3}{2} \mu_2^2 - \frac{1}{2},
\end{eqnarray}
and its corresponding Legendre polynomials expansion is
\begin{eqnarray}
    f_{E}(\hat{\mathbf{q}}) = \mathcal{P}_{2}(\mu_{2}).
\end{eqnarray}

    \item $P_{SS|E}(k,\mu_{k})$

    The $P_{SS|E}(k,\mu_{k})$ term (Equation \ref{eq:A.44}) consists of $A_{SS|E}(k,\mu_{k})$ and $B_{SS|E}(k,\mu_{k})$.

    The $A_{SS|E}(k,\mu_{k})$ term (Equation \ref{eq:A.45}) contains the kernel $F_2(\mathbf{q}_{1}, \mathbf{q}_{2}) S_2(\mathbf{q}_{1}, \mathbf{q}_{2})$ which can be expressed as 
\begin{eqnarray}
     S_2(\mathbf{q}_1, \mathbf{q}_2) F_2(\mathbf{q}_1, \mathbf{q}_2) = -\frac{5}{21} + \frac{13}{21}\mu_{12}^2 + \frac{2}{7}\mu_{12}^4 - \frac{1}{6}\frac{q_1}{q_2}\mu_{12} - \frac{1}{6}\frac{q_2}{q_1}\mu_{12} + \frac{1}{2}\frac{q_1}{q_2}\mu_{12}^3 + \frac{1}{2}\frac{q_2}{q_1}\mu_{12}^3 ,
\end{eqnarray}
and its corresponding Legendre polynomials expansion is
\begin{eqnarray}
    S_2(\mathbf{q}_1, \mathbf{q}_2) F_2(\mathbf{q}_1, \mathbf{q}_2) &=& \frac{8}{315}\mathcal{P}_{0}(\mu_{12}) + \frac{254}{441}\mathcal{P}_{2}(\mu_{12}) + \frac{16}{245}\mathcal{P}_{4}(\mu_{12})+ \frac{2}{15} \frac{q_1}{q_2}\mathcal{P}_{1}(\mu_{12})\\ \nonumber
    && +\ \frac{2}{15} \frac{q_2}{q_1}\mathcal{P}_{1}(\mu_{12}) + \frac{1}{5} \frac{q_1}{q_2}\mathcal{P}_{3}(\mu_{12}) + \frac{1}{5} \frac{q_2}{q_1}\mathcal{P}_{3}(\mu_{12}).
\end{eqnarray}


The $B_{SS|E}(k,\mu_{k})$ term is similar in structure to the $J_{00}(k)$ contribution in the one-loop order SPT correction to the matter PS. Taking $\alpha=\frac{q_{1}}{k}$ and $\mu_{1} \equiv \hat{\mathbf{q}}_1 \cdot \hat{\mathbf{k}}$, Equation \ref{eq:A.46} may be expressed in the following form
\begin{eqnarray}\nonumber 
    B_{SS|E}(k,\mu_k) &=& 4p(\hat{k})P_{\rm L}(k)\int\frac{k^{3}\alpha^{3}d\alpha d\mu_1}{(2\pi)^{2}} \bigg[ \left(\frac{(\alpha-\mu_1)^{2}}{1+\alpha^{2}-2\alpha \mu_1}-\frac{1}{3}\right) \\ \label{eq:B.9}
    && \ \ \ \ \ \ \ \ \ \ \ \ \ \ \ \times \left(\frac{5}{7}-\frac{1}{2}\mu_1\left(\alpha+\frac{1}{\alpha}\right)+\frac{2}{7}\mu_1^{2}\right) - \frac{34}{126} \bigg] P_{\rm L}(q_{1}).
\end{eqnarray}

Analytically evaluating the angular part of the integral yields
\begin{eqnarray}\label{eq:B.8}
    B_{SS|E}(k,\mu_k) &=& 4p(\hat{k})P_{\rm L}(k)k^{3}\int\frac{\alpha^{2}d\alpha}{(2\pi)^{2}} \Big[ Z(\alpha) -\frac{34}{126} \Big]  P_{\rm L}(q_{1})
\end{eqnarray}
where
\begin{eqnarray}\nonumber
Z(\alpha)&=&\frac{1}{2016\alpha^{5}} \times \bigg[ 4\alpha(45-165\alpha^{2}+379\alpha^{4}+45\alpha^{6})+45(-1+\alpha^{2})^{4} \mathrm{log}\left[(-1+\alpha)^{2}\right] \\ 
&& \ \ \ \ \ \ \ \ \ \ \ \ \ \ -\ 90(-1+\alpha^{2})^{4}  \mathrm{log}\left[1+\alpha\right] \bigg] \label{eq:B.11}
\end{eqnarray}
and the Taylor expansions for $\alpha\rightarrow0$ and $\alpha\rightarrow \infty$ are respectively given by
\begin{eqnarray}
    \alpha^{2}Z(\alpha)=-\frac{16}{21}\alpha^{2} + \frac{16}{49}\alpha^{4} - \frac{16}{441}\alpha^{6} - \frac{16}{4851}\alpha^{8} -\frac{16}{21021} \alpha^{10}- \frac{16}{63063}\alpha^{12}+\mathcal{O}\left(\alpha^{13}\right)
\end{eqnarray}
and
\begin{eqnarray}\nonumber
    \alpha^{2}Z(\alpha) &=& - \frac{16}{21} + \frac{16}{49}\alpha^{-2} - \frac{16}{441}\alpha^{-4} -\frac{16}{4851} \alpha^{-6}- \frac{16}{21021}\alpha^{-8}\\ 
    &&\ -\ \frac{16}{63063}\alpha^{-10}-\frac{16}{153153}\alpha^{-12}+\mathcal{O}\left(\alpha^{-13}\right)
\end{eqnarray}

The integral in Equation \ref{eq:B.9} is directly implemented in \texttt{FAST-PT} via discrete convolutions.

    \item $P_{SS|0E}(k,\mu_{k}$)

    The $P_{SS|0E}(k,\mu_{k}$) term (Equation \ref{eq:A.48}) contains the kernel $ f_{E}(\hat{\mathbf{q}}_{1})S_2(\mathbf{q}_{1}, \mathbf{q}_{2})$ which can be expressed as
\begin{eqnarray}
     f_{E}(\hat{\mathbf{q}})S_2(\mathbf{q}_{1}, \mathbf{q}_{2}) = \frac{1}{4} - \frac{3}{4}\mu_{12}^2 - \frac{3}{4}\mu_2^2 + \frac{9}{4}\mu_{12}^2\mu_2^2,
\end{eqnarray}
and its corresponding Legendre polynomials expansion is
\begin{eqnarray}
     f_{E}(\hat{\mathbf{q}})S_2(\mathbf{q}_{1}, \mathbf{q}_{2}) = \frac{2}{3} \mathcal{P}_{2}(\mu_{12})\mathcal{P}_{2}(\mu_{2}).
\end{eqnarray}

    \item $P_{0|E2}(k,\mu_{k}$)

The $P_{0|E2}(k,\mu_{k}$) term (Equation \ref{eq:A.53}) consists of the terms $A_{0|E2}(k,\mu_{k}$) and $B_{0|E2}(k,\mu_{k}$). Both of these terms incorporate the kernel $ h_{E}(\hat{\mathbf{q}}_{1}, \hat{\mathbf{q}}_{2}) F_2(\mathbf{q}_1, \mathbf{q}_2)$ which can be expressed as
\begin{eqnarray}\nonumber
    h_{E}(\hat{\mathbf{q}}_{1}, \hat{\mathbf{q}}_{2}) F_2(\mathbf{q}_1, \mathbf{q}_2) &=& 
    \frac{5}{21} 
    - \frac{11}{42}\mu_{12}^2  
    - \frac{1}{7}\mu_{12}^4
    - \frac{5}{14}\mu_{1}^2
    - \frac{1}{7}\mu_{1}^2\mu_{12}^2
    + \frac{5}{14}\mu_{1}\mu_{2}\mu_{12}
    + \frac{3}{7}\mu_{1}\mu_{2}\mu_{12}^3 \\ \nonumber
    && - \frac{5}{14}\mu_{2}^2
    - \frac{1}{7}\mu_{2}^2\mu_{12}^2
    + \frac{1}{6}\frac{q_1}{q_2}\mu_{12}
    - \frac{1}{4}\frac{q_1}{q_2}\mu_{12}^3
    - \frac{1}{4}\frac{q_1}{q_2}\mu_{12}^3 \mu_{1}^2
    - \frac{1}{4}\frac{q_1}{q_2}\mu_{12}^3 \mu_{2}^2 \\ \nonumber
    && + \frac{3}{4}\frac{q_1}{q_2}\mu_{12}^2 \mu_{1}\mu_{2}
    + \frac{1}{6}\frac{q_2}{q_1}\mu_{12}
    - \frac{1}{4}\frac{q_2}{q_1}\mu_{12}^3
    - \frac{1}{4}\frac{q_2}{q_1}\mu_{12}^3 \mu_{1}^2
    - \frac{1}{4}\frac{q_2}{q_1}\mu_{12}^3 \mu_{2}^2 \\
    && + \frac{3}{4}\frac{q_2}{q_1}\mu_{12}^2 \mu_{1}\mu_{2},
\end{eqnarray}
and its corresponding Legendre polynomials expansion is
\begin{eqnarray}\nonumber
    h_{E}(\hat{\mathbf{q}}_{1}, \hat{\mathbf{q}}_{2}) F_2(\mathbf{q}_1, \mathbf{q}_2) &=&
    -\frac{31}{210} \mathcal{P}_{0}(\mu_{12})
    -\frac{17}{63}\mathcal{P}_{2}(\mu_{1})
    -\frac{47}{147}\mathcal{P}_{2}(\mu_{12})
    -\frac{4}{63}\mathcal{P}_{2}(\mu_{1})\mathcal{P}_{2}(\mu_{12}) \\ \nonumber
    && + \frac{93}{70}\mathcal{P}_{1}(\mu_{1})\mathcal{P}_{1}(\mu_{2})\mathcal{P}_{1}(\mu_{12})
    -\frac{17}{63}\mathcal{P}_{2}(\mu_{2})
     -\frac{4}{63}\mathcal{P}_{2}(\mu_{2})\mathcal{P}_{2}(\mu_{12}) \\ \nonumber
    && + \frac{6}{35}\mathcal{P}_{1}(\mu_{1})\mathcal{P}_{1}(\mu_{2})\mathcal{P}_{3}(\mu_{12})
    - \frac{8}{245} \mathcal{P}_{4}(\mu_{12}) 
    - \frac{3}{20}\frac{q_1}{q_2}\mathcal{P}_{1}(\mu_{12}) \\ \nonumber
    && - \frac{1}{6}\frac{q_1}{q_2} \mathcal{P}_{2}(\mu_{1})  \mathcal{P}_{1}(\mu_{12})
    + \frac{1}{4}\frac{q_1}{q_2}\mathcal{P}_{1}(\mu_{1}) \mathcal{P}_{1}(\mu_{2}) 
    + \frac{1}{2}\frac{q_1}{q_2}\mathcal{P}_{1}(\mu_{1}) \mathcal{P}_{1}(\mu_{2}) \mathcal{P}_{2}(\mu_{12}) \\ \nonumber
    && - \frac{1}{6}\frac{q_1}{q_2} \mathcal{P}_{2}(\mu_{2})  \mathcal{P}_{1}(\mu_{12})
    - \frac{1}{10}\frac{q_1}{q_2} \mathcal{P}_{3}(\mu_{12})
    - \frac{3}{20}\frac{q_2}{q_1}\mathcal{P}_{1}(\mu_{12}) \\\nonumber
    && - \frac{1}{6}\frac{q_2}{q_1} \mathcal{P}_{2}(\mu_{1})  \mathcal{P}_{1}(\mu_{12})
    + \frac{1}{4}\frac{q_2}{q_1}\mathcal{P}_{1}(\mu_{1}) \mathcal{P}_{1}(\mu_{2}) 
    + \frac{1}{2}\frac{q_2}{q_1}\mathcal{P}_{1}(\mu_{1}) \mathcal{P}_{1}(\mu_{2}) \mathcal{P}_{2}(\mu_{12}) \\ 
    && - \frac{1}{6}\frac{q_2}{q_1} \mathcal{P}_{2}(\mu_{2})  \mathcal{P}_{1}(\mu_{12})
    - \frac{1}{10}\frac{q_2}{q_1} \mathcal{P}_{3}(\mu_{12}).
\end{eqnarray}

    \item $P_{00|E2}(k,\mu_{k}$)

The $P_{00|E2}(k,\mu_{k}$) term (Equation \ref{eq:A.57}) contains the kernel $ h_{E}(\hat{\mathbf{q}}_{1}, \hat{\mathbf{q}}_{2})$ which can be expressed as
\begin{eqnarray}
    h_{E}(\hat{\mathbf{q}}_{1}, \hat{\mathbf{q}}_{2}) = 
    \frac{1}{3} + \frac{3}{2}\mu_{12}\mu_1\mu_2 
    - \frac{1}{2}\mu_{12}^2
    - \frac{1}{2}\mu_1^2
    - \frac{1}{2}\mu_2^2,
\end{eqnarray}
and its corresponding Legendre polynomials expansion is
\begin{eqnarray}\nonumber
    h_{E}(\hat{\mathbf{q}}_{1}, \hat{\mathbf{q}}_{2}) = - \frac{1}{6}\mathcal{P}_{0}(\mu_{12}) + \frac{3}{2}\mathcal{P}_{1}(\mu_{12})\mathcal{P}_{1}(\mu_{1})\mathcal{P}_{1}(\mu_{2}) - \frac{1}{3}\mathcal{P}_{2}(\mu_{12}) - \frac{1}{3}\mathcal{P}_{2}(\mu_{1}) - \frac{1}{3}\mathcal{P}_{2}(\mu_{2}).\\
\end{eqnarray}

    \item $P_{SS|E2}(k,\mu_{k}$)

The $P_{SS|E2}(k,\mu_{k}$) term (Equation \ref{eq:A.59}) contains the kernel $ h_{E}(\hat{\mathbf{q}}_{1}, \hat{\mathbf{q}}_{2}) S_{2}(\mathbf{q}_{1},\mathbf{q}_{2})$ which can be expressed as
\begin{eqnarray}\nonumber
    h_{E}(\hat{\mathbf{q}}_{1}, \hat{\mathbf{q}}_{2}) S_{2}(\mathbf{q}_{1},\mathbf{q}_{2}) &=& 
    - \frac{1}{9}
    - \frac{1}{2}\mu_{12}^4 
    + \frac{3}{2}\mu_{12}^3\mu_{1}\mu_{2}
    - \frac{1}{2}\mu_{12}^2\mu_1^2 
    - \frac{1}{2}\mu_{12}^2\mu_2^2 
    + \frac{1}{2}\mu_{12}^2\\ 
    && -\ \frac{1}{2}\mu_{12}\mu_{1}\mu_{2}
    + \frac{1}{6}\mu_{1}^2
    + \frac{1}{6}\mu_{2}^2,
\end{eqnarray}
and its corresponding Legendre polynomials expansion is
\begin{eqnarray}\nonumber
    h_{E}(\hat{\mathbf{q}}_{1}, \hat{\mathbf{q}}_{2}) S_{2}(\mathbf{q}_{1},\mathbf{q}_{2}) &=& 
    - \frac{2}{45} \mathcal{P}_{0}(\mu_{12})
    + \frac{2}{5}\mathcal{P}_{1}(\mu_{12})\mathcal{P}_{1}(\mu_{1})\mathcal{P}_{1}(\mu_{2}) 
    - \frac{4}{35}\mathcal{P}_{4}(\mu_{12}) 
    - \frac{11}{63}\mathcal{P}_{2}(\mu_{12}) \\ \nonumber
    && + \frac{3}{5}\mathcal{P}_{3}(\mu_{12})\mathcal{P}_{1}(\mu_{1})\mathcal{P}_{1}(\mu_{2}) 
    - \frac{2}{9}\mathcal{P}_{2}(\mu_{12}){P}_{2}(\mu_{1}) 
    - \frac{2}{9}\mathcal{P}_{2}(\mu_{12}){P}_{2}(\mu_{2}). \\
\end{eqnarray}

\end{itemize}

\subsection{Coefficients for the Lengendre polynomial expansion for the different convolution integrals}
\label{AppB2}
\par\noindent

In this section we summarize the coefficients obtained in Appendix \ref{AppB1} and organize them in tables where $\alpha$, $\beta$, $l$ and $A_l^{\alpha\beta}$ are such that we can write the expansion as
\begin{eqnarray}
K(\mu_{12},\mu_1,\mu_2,q_{1},q_{2}) = \sum_{l_{1},l_{2},l,\alpha,\beta}A_{l_{1}l_{2}l}^{\alpha,\beta}\mathcal{P}_{l}(\mu_{12})\mathcal{P}_{l_{1}}(\mu_1)\mathcal{P}_{l_{2}}(\mu_2)q_{1}^{\alpha}q_{2}^{\beta}.
\end{eqnarray}

These tables are essentially the format in which the \texttt{FAST-PT} framework accepts inputs for the convolution integrals that we are computing with it, the different functions expanded here are the integration kernels shown in the text above for the different IA operators.

\FloatBarrier

\begin{longtable}{|c c c c|}
    \hline
    $\alpha$ & $\beta$ & $l$ & $A_{l}^{\alpha\beta}$ \\ [1ex] 
    \hline\hline
    0 & 0 & 0 & $\frac{17}{21}$ \\[1ex]
    \hline
    1 & -1 & 1 & $\frac{1}{2}$ \\[1ex]
    \hline
    -1 & 1 & 1 & $\frac{1}{2}$ \\[1ex]
    \hline
    0 & 0 & 2 & $\frac{4}{21}$ \\[1ex]
    \hline
    \caption{Coefficients for the Legendre polynomial expansion of $F_2(\mathbf{q}_{1}, \mathbf{q}_2)$}
\end{longtable}

\begin{longtable}{|c c c c c c|} 
    \hline
    $\alpha$ & $\beta$ & $l$ &$l_{1}$ & $l_{2}$ & $A_{l_{1}l_{2}l}^{\alpha\beta}$ \\ [1ex] 
    \hline\hline
    0 & 0 & 0 & 0 & 2 & 1 \\[1ex]
    \hline
    \caption{Coefficients for the Legendre polynomial expansion of $f_{E}(\hat{\mathbf{q}})$} 
\end{longtable}

\begin{longtable}{|c c c c c c|} 
    \hline
    $\alpha$ & $\beta$ & $l$ &$l_{1}$ & $l_{2}$ & $A_{l_{1}l_{2}l}^{\alpha\beta}$ \\ [1ex] 
    \hline\hline
    0 & 0 & 0 & 2 & 0 & $\frac{17}{21}$ \\[1ex]
    \cline{3-6}
    &  & 2 & 0 & 2 & $\frac{4}{21}$ \\[1ex]
    \hline
    1 & -1 & 1 & 0 & 2 & $\frac{1}{2}$ \\[1ex]
    \hline
    -1 & 1 & 1 & 0 & 2 & $\frac{1}{2}$ \\[1ex]
    \hline
    \caption{Coefficients for the Legendre polynomial expansion of $f_{E}(\hat{\mathbf{q}})F_{2}(\mathbf{q_{1},\mathbf{q}_{2}})$}
\end{longtable}

\begin{longtable}{|c c c c c c|}
    \hline
    $\alpha$ & $\beta$ & $l$ &$l_{1}$ & $l_{2}$ & $A_{l_{1}l_{2}l}^{\alpha\beta}$ \\ [1ex]
    \hline\hline
    0 & 0 & 2 & 0 & 2 & $\frac{2}{3}$ \\[1ex]
    \hline
    \caption{Coefficients for the Legendre polynomial expansion of $f_{E}(\hat{\textbf{q}})S_{2}(\textbf{q}_{1},\textbf{q}_{2})$}
\end{longtable}

\newpage
\begin{longtable}{|c c c c c c|} 
    \hline
    $\alpha$ & $\beta$ & $l$ &$l_{1}$ & $l_{2}$ & $A_{l_{1}l_{2}l}^{\alpha\beta}$ \\ [1ex] 
    \hline\hline
    0 & 0 & 4 & 0 & 0 & $\frac{16}{245}$ \\[1ex]
    \cline{3-6}
    &  & 2 & 0 & 0 & $\frac{254}{441}$ \\[1ex]
    \cline{3-6}
    &  & 0 & 0 & 0 & $\frac{8}{315}$ \\[1ex]
    \hline
    1 & -1 & 3 & 0 & 0 & $\frac{1}{5}$ \\[1ex]
    \cline{3-6}
    &  & 1 & 0 & 0 & $\frac{2}{15}$ \\[1ex]
    \hline
    -1 & 1 & 3 & 0 & 0 & $\frac{1}{5}$ \\[1ex]
    \cline{3-6}
    &  & 1 & 0 & 0 & $\frac{2}{15}$ \\[1ex]
    \hline
    \caption{Coefficients for the Legendre polynomial expansion of $S_{2}(\mathbf{q_{1},\mathbf{q}_{2}})F_{2}(\mathbf{q_{1},\mathbf{q}_{2}})$}
\end{longtable}


\begin{longtable}{|c c c c c c|} 
    \hline
    $\alpha$ & $\beta$ & $l$ &$l_{1}$ & $l_{2}$ & $A_{l_{1}l_{2}l}^{\alpha\beta}$ \\ [1ex] 
    \hline\hline
    0 & 0 & 1 & 1 & 1 & $\frac{3}{2}$ \\[1ex]
    \cline{3-6}
    &  & 2 & 0 & 0 & $\frac{-1}{3}$ \\[1ex]
    \cline{3-6}
    &  & 0 & 2 & 0 & $\frac{-1}{3}$ \\[1ex]
    \cline{3-6}
    &  & 0 & 0 & 2 & $\frac{-1}{3}$ \\[1ex]
    \cline{3-6}
    &  & 0 & 0 & 0 & $-\frac{1}{6}$ \\[1ex]
    \hline
    \caption{Coefficients for the Legendre polynomial expansion of $h_{E}(\textbf{q}_{1},\textbf{q}_{2})$}
\end{longtable}

\begin{longtable}{|c c c c c c|} 
    \hline
    $\alpha$ & $\beta$ & $l$ &$l_{1}$ & $l_{2}$ & $A_{l_{1}l_{2}l}^{\alpha\beta}$ \\ [1ex] 
    \hline\hline
    0 & 0 & 1 & 1 & 1 & $\frac{2}{5}$ \\[1ex]
    \cline{3-6}
    &  & 4 & 0 & 0 & $\frac{-4}{35}$ \\[1ex]
    \cline{3-6}
    &  & 3 & 1 & 1 & $\frac{3}{5}$ \\[1ex]
    \cline{3-6}
    &  & 2 & 2 & 0 & $\frac{-2}{9}$ \\[1ex]
    \cline{3-6}
    &  & 2 & 0 & 2 & $\frac{-2}{9}$ \\[1ex]
    \cline{3-6}
    &  & 2 & 0 & 0 & $\frac{-11}{63}$ \\[1ex]
    \cline{3-6}
    &  & 0 & 0 & 0 & $\frac{-2}{45}$ \\[1ex]
    \hline
    \caption{Coefficients for the Legendre polynomial expansion of $h_{E}(\textbf{q}_{1},\textbf{q}_{2})S_{2}(\textbf{q}_{1},\textbf{q}_{2})$}
\end{longtable}

\newpage
\begin{longtable}{|c c c c c c|} 
    \hline
    $\alpha$ & $\beta$ & $l$ &$l_{1}$ & $l_{2}$ & $A_{l_{1}l_{2}l}^{\alpha\beta}$ \\ [1ex] 
    \hline\hline
    0 & 0 & 0 & 0 & 0 & $\frac{-31}{210}$ \\[1ex]
    \cline{3-6}
    &  & 0 & 2 & 0 & $\frac{-17}{63}$ \\[1ex]
    \cline{3-6}
    &  & 2 & 0 & 0 & $\frac{-47}{147}$ \\[1ex]
    \cline{3-6}
    &  & 2 & 2 & 0 & $\frac{-4}{63}$ \\[1ex]
    \cline{3-6}
    &  & 1 & 1 & 1 & $\frac{93}{70}$ \\[1ex]
    \cline{3-6}
    &  & 0 & 0 & 2 & $\frac{-17}{63}$ \\[1ex]
    \cline{3-6}
    &  & 2 & 0 & 2 & $\frac{-4}{63}$ \\[1ex]
    \cline{3-6}
    &  & 3 & 1 & 1 & $\frac{6}{35}$ \\[1ex]
    \cline{3-6}
    &  & 4 & 0 & 0 & $\frac{-8}{245}$ \\[1ex]
    \hline
    1 & -1 & 1 & 0 & 0 & $\frac{-3}{20}$ \\[1ex]
    \cline{3-6}
    &  & 1 & 2 & 0 & $\frac{-1}{6}$ \\[1ex]
    \cline{3-6}
    &  & 0 & 1 & 1 & $\frac{1}{4}$ \\[1ex]
    \cline{3-6}
    &  & 2 & 1 & 1 & $\frac{1}{2}$ \\[1ex]
    \cline{3-6}
    &  & 1 & 0 & 2 & $\frac{-1}{6}$ \\[1ex]
    \cline{3-6}
    &  & 3 & 0 & 0 & $\frac{-1}{10}$ \\[1ex]
    \hline
    -1 & 1 & 1 & 0 & 0 & $\frac{-3}{20}$ \\[1ex]
    \cline{3-6}
    &  & 1 & 2 & 0 & $\frac{-1}{6}$ \\[1ex]
    \cline{3-6}
    &  & 0 & 1 & 1 & $\frac{1}{4}$ \\[1ex]
    \cline{3-6}
    &  & 2 & 1 & 1 & $\frac{1}{2}$ \\[1ex]
    \cline{3-6}
    &  & 1 & 0 & 2 & $\frac{-1}{6}$ \\[1ex]
    \cline{3-6}
    &  & 3 & 0 & 0 & $\frac{-1}{10}$ \\[1ex]
    \hline
    \caption{Coefficients for the Legendre polynomial expansion of $h_{E}(\mathbf{q}_{1},\mathbf{q}_{2}) F_{2}(\mathbf{q_{2},\mathbf{q}_{1}})$}
\end{longtable}

\end{document}